\RequirePackage[l2tabu, orthodox]{nag}
\documentclass[journal,10pt,onecolumn]{IEEEtran}

\usepackage{authblk}
\usepackage{subcaption}

\usepackage{lmodern}
\usepackage[T1]{fontenc}
\usepackage[utf8]{inputenc}

\usepackage{geometry}
\usepackage[lined,boxed,algosection]{algorithm2e}
\SetAlCapSkip{0.5em}
\IncMargin{1em}

\usepackage{color}
\usepackage{xcolor}
\definecolor{dark-red}{rgb}{0.4,0.15,0.15}
\definecolor{dark-blue}{rgb}{0.15,0.15,0.4}
\definecolor{medium-blue}{rgb}{0,0,0.5}

\definecolor{mycomment}{rgb}{0.3,0.7,0.8}
\definecolor{mygray}{rgb}{0.5,0.5,0.5}
\definecolor{lightgray}{rgb}{0.95,0.95,0.95}
\definecolor{mymauve}{rgb}{0.58,0,0.82}

\definecolor{mediumlightgray}{rgb}{0.8,0.8,0.8}

\ifxetex
	\usepackage[setpagesize=false, % page size defined by xetex
				unicode=false, % unicode breaks when used with xetex
				xetex]{hyperref}
\else
	\usepackage[unicode=true]{hyperref}
\fi
\hypersetup{breaklinks=true,
		bookmarks=true,
		colorlinks,
		linkcolor={dark-blue},
		citecolor={dark-blue},
		urlcolor={medium-blue},
		pdfauthor={Eddie Schoute},
		pdftitle={Shortcuts to Quantum Network Routing},
		pdfborder={0 0 0}
}
\usepackage{amssymb,amsmath,amsthm}
\usepackage[capitalise]{cleveref}

\usepackage[colorinlistoftodos]{todonotes}
\usepackage{listings}

\usepackage{tikz}
\usetikzlibrary{graphs,calc,decorations.pathreplacing,decorations.pathmorphing,shapes.misc,patterns,fit}
\tikzset{%
	inner sep=0.2em
}

\usepackage{xifthen}

\DeclareMathOperator*{\argmin}{arg\,min}
\DeclareMathOperator*{\argmax}{arg\,max}

\newcommand\doubleplus{\ensuremath{\mathbin{+\mkern-10mu+}}}

\newtheorem{theorem}{Theorem}[section]
\newtheorem{lemma}[theorem]{Lemma}
\newtheorem{definition}[theorem]{Definition}
\newtheorem{proposition}[theorem]{Proposition}
\newtheorem{corollary}[theorem]{Corollary}

\newcommand{\basis}{\noindent {\bfseries Basis:} }
\newcommand{\ih}{\noindent {\bfseries Induction hypothesis:} }
\newcommand{\is}{\noindent {\bfseries Induction:} }

\newenvironment{subproof}[1][\proofname]{%
  \begin{proof}[#1]%
}{%
  \end{proof}%
}
\newenvironment{fancyproof}[1][\proofname]{%
  \begin{proof}[#1]%
}{%
  \end{proof}%
}

\newcommand{\ly}{{l}}
\DeclareMathOperator{\la}{{label}}

\newcommand{\edge}{\ensuremath{\mathtt{\sim}}}
\DeclareMathOperator{\polylog}{polylog}

\usepackage{mathtools}
\mathtoolsset{centercolon}
\DeclarePairedDelimiter{\set}{\lbrace}{\rbrace}
\DeclarePairedDelimiter{\abs}{\lvert}{\rvert}
\DeclareMathOperator{\gcdd}{gcd_2}

\begin{document}
	\title{Shortcuts to quantum network routing}
	\author[1,2]{Eddie Schoute
		\thanks{Correspondence to:
			\href{mailto:eschoute@umd.edu}{eschoute@umd.edu}
				and \href{mailto:s.d.c.wehner@tudelft.nl}{s.d.c.wehner@tudelft.nl}
		}
	}
	\affil[1]{Joint Center for Quantum Information and Computer Science (QuICS), University of Maryland}
	\affil[2]{QuTech, Delft University of Technology, Lorentzweg 1, 2628 CJ Delft, Netherlands}
	\author[3]{Laura Man\v{c}inska}
	\affil[3]{School of Mathematics, University of Bristol, Bristol, UK}
	\author[4]{Tanvirul Islam}
	\affil[4]{Centre for Quantum Technologies, National University of Singapore, 114375 Singapore}
	\author[5]{Iordanis Kerenidis}
	\affil[5]{IRIF, Univ Paris Diderot, CNRS, Paris France}
	\author[2]{Stephanie~Wehner}

\maketitle
	\begin{abstract}%
		\noindent

A quantum network promises to enable long distance quantum communication, and assemble small quantum devices into a large quantum computing cluster. Each network node can thereby be seen as a small few qubit quantum computer. Qubits can be sent over direct physical links connecting nearby quantum nodes, or by means of teleportation over pre-established entanglement amongst distant network nodes. Such pre-shared entanglement effectively forms a shortcut - a virtual quantum link - which can be used exactly once. 

Here, we present an abstraction of a quantum network that allows ideas from computer science to be applied to the problem of routing qubits, and manage entanglement in the network. Specifically, we consider a scenario in which each quantum network node can create EPR pairs with its immediate neighbours over a physical connection, and perform entanglement swapping operations in order to create long distance virtual quantum links. 
We proceed to discuss the features unique to quantum networks, which call for the development of new routing techniques. As an example, we present two simple hierarchical routing schemes for a quantum network of $N$ nodes for a ring and sphere topology. For these topologies we present efficient routing algorithms requiring $O(\log N)$ qubits to be stored at each network node, $O(\polylog N)$ time and space to perform routing decisions, and $O(\log N)$ timesteps to replenish the virtual quantum links in a model of entanglement generation.
	\end{abstract}

\tableofcontents

%\listoftodos

% Use include instead of input, to generate separate .aux files and reduce conflicts (in dropbox).
% It does add a \clearpage, but we can switch to \input for the final version.

\section{Introduction}\label{sec:introduction}
	Quantum communication offers unparalleled advantages over classical communication.
	Possibly the most well-known application is quantum key distribution~\cite{bb84,e91}
	that allows two parties to establish an encryption key with security guarantees that are provably impossible to attain classically.
	Nevertheless, quantum communication offers a wide range of other applications ranging from cryptographic protocols,
	efficient communication protocols \cite{R99,buhrman2001quantum,bar2004exponential,gavinsky2007exponential,regev2011quantum},
	applications in distributed systems~\cite{byzantine},  better clock synchronization~\cite{clocks}, to extending the baseline of telescopes~\cite{telescope}.
	While quantum key distribution is already commercially available at short distances, sending quantum bits (qubits) 
	over long distances remains an outstanding challenge requiring the construction of a quantum 
	repeater (see~\cite{repeaterReview} for a review).
	Just as in classical communication networks, however, we do not only desire to bridge long distances.
	We would also like to enable communication to be routed efficiently to the correct destination even when many network nodes communicate simultaneously.
	This is also highly relevant for quantum communication at short distances, when smaller quantum computers on chip are connected by a network
	to form a larger quantum computing device.

	\subsection{Quantum networks: Basic properties}
\begin{figure}
		\centering
		\begin{tikzpicture}

			\foreach \x / \y / \angle in {0/0/45, 0/3/315, 3/0/135, 3/3/225}
			{%
				\node[circle, draw, minimum size=1.5cm] (\x\y) at (\x,\y) {};
				\node[fill, blue, circle, minimum size=0.5cm] (\x\y primary) at ($(\x,\y)+(\angle:0.4)$) {};
				\node[pattern color=gray, pattern=vertical lines, circle, minimum size=0.5cm] (\x\y second) at ($(\x,\y)+(\angle+120:0.4)$) {};
				\node[pattern color=gray, pattern=vertical lines, circle, minimum size=0.5cm] (\x\y tert) at ($(\x,\y)+(\angle+240:0.4)$) {};
			}

			\draw[dotted, very thick, blue] (00primary) -- (33primary);
			\draw[dotted, very thick, blue] (03primary) -- (30primary);

			\draw (00) to[bend right] (30);
			\draw (30) to[bend right] (33);
			\draw (33) to[bend right] (03);
			\draw (03) to[bend right] (00);

			\draw[dashed, red] (00.south east) to[bend right] (30);
			\draw[dashed, red] (30.north east) to[bend right] (33);
			\draw[dashed, red] (33.north west) to[bend right] (03);
			\draw[dashed, red] (03.south west) to[bend right] (00);

			% Legend
			\node[circle, draw, label=right:Network node, minimum size = 1em] (networknode) at (5,3) {};
			\node[circle, pattern color=gray, pattern=vertical lines, label=right:Unused Qubit memory, minimum size = 1em] (emptyqubit) at ($(networknode) - (0,1.5em)$) {};
			\node[circle, fill, blue, label=right:Used Qubit memory, minimum size = 1em] (qubit) at ($(emptyqubit) - (0,1.5em)$) {};

			\draw ($(qubit) - (0.5em, 1.5em)$) -- +(1em,0) node[label=right:Physical quantum communication link] (physicalquantum) {};
			\draw[dashed, draw=red, text=black] ($(physicalquantum) - (1em, 1.5em)$) -- +(1em,0) node[label=right:Physical classical communication link] (physicalclassical) {};
			\draw[dotted, very thick, draw=blue, text=black] ($(physicalclassical) - (1em, 1.5em)$) -- +(1em,0) node[label=right:Virtual link via entanglement (VQL)] (virtuallink) {};
		\end{tikzpicture}
		\caption{Quantum network. The goal of a quantum network is to send qubits from one node to another, or equivalently, to establish entanglement between them. Each node in a quantum network is a small quantum computer that can store and operate on few qubits. Network nodes can be connected by standard classical communication channels, as well as quantum ones.
		In addition, entanglement can be established between nodes and be used to send qubits using quantum teleportation. Entanglement thus forms a virtual quantum link which can however only be used once.}\label{fig:quantumNetwork}
	\end{figure}
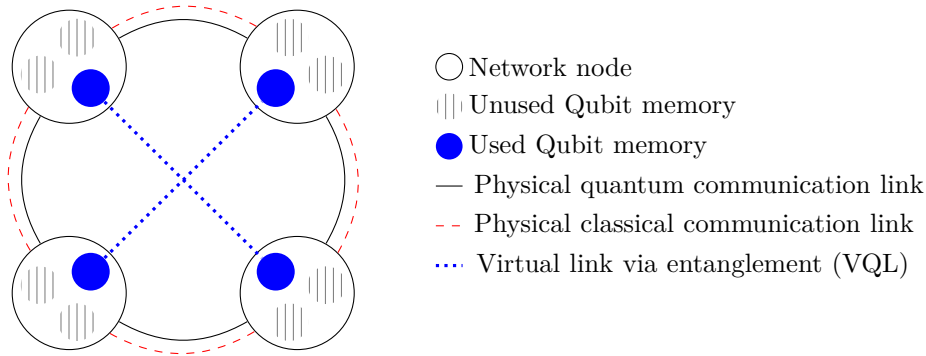

	We focus on the fundamental issue of routing quantum information to the right destination in a quantum network~\cite{kimble2008QuantumInternet} , or equivalently, manage entanglement in the network.
	To establish a connection to classical networking,
	let us briefly explain what a quantum network looks like (see \cref{fig:quantumNetwork} for an overview, and \cref{fig:NetworkBox} for a detailed summary of the general model). 
	In essence, each node in a quantum network is a small quantum computer that can 
	store and operate on a few
	qubits. According to current technology, if we ask for nodes that can both manipulate as well as store qubits, then 
	the number of qubits in each node is less than 10.
	This may seem like an extremely small number, but 
	it is important to note that 
	-- unlike for the purpose of quantum computation -- most quantum network applications can be executed using even fewer	
	qubits - usually just one. However, having more qubits at each node offers the opportunity to perform error-correction, and will be useful when 
	considering the routing protocols below. We also emphasize that for quantum communication protocols we generally do not require universal quantum computation, but
	it is sufficient if each node can perform more elementary operations. 

	Quantum network nodes can exchange classical control information over standard classical communication channels. This may be by means of a direct physical connection or via, for example, the internet. 
	In addition, nodes which are physically close - at a maximum distance of around 250kms - may be optically connected by direct quantum communication channels such as telecom fibers in a useful way.
	Direct quantum communication over longer distances is made challenging by the fact that quantum error-correction requires many qubits, qubits cannot be copied, and we cannot amplify signals in the way 
	a classical repeater can. Quantum repeaters thus work in a fundamentally different way.

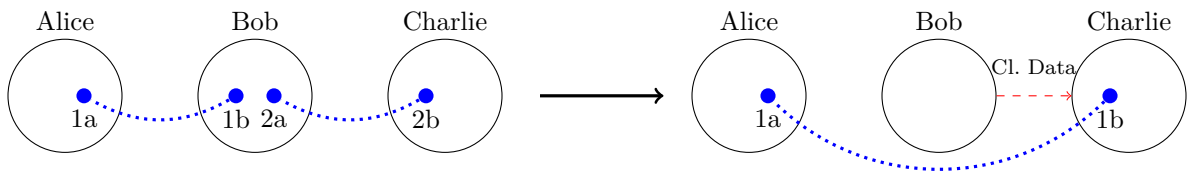
\begin{figure}
	\centering
	\begin{tikzpicture}
		\node[circle,draw, minimum size=1.5cm] (alice1) at (0,0) {};
		\node (a1) at (0,1) {Alice};
		\node[circle,draw, minimum size=1.5cm] (bob1) at (2.5,0) {};
		\node (b1) at (2.5,1) {Bob};
		\node[circle,draw, minimum size=1.5cm] (charlie1) at (5,0) {};
		\node (c1) at (5,1) {Charlie};

		\node[fill, circle, blue] (alice11) at (0.25,0) {};
		\node (node11a) at (0.25, -0.3) {1a};
		\node[fill,circle, blue] (bob11) at (2.25,0) {}
			edge[very thick, dotted, blue, bend left] (alice11);
		\node (node11b) at (2.25,-0.3) {1b};
		\node[fill,circle, blue] (bob12) at (2.75,0) {};
		\node (node12a) at (2.75,-0.3) {2a};
		\node[fill,circle, blue] (charlie11) at (4.75,0) {}
			edge[very thick, dotted, blue, bend left] (bob12);
		\node (node12b) at (4.75,-0.3) {2b};

		\draw[->, very thick] (6.25,0) -- (7.875,0);

		\node[circle,draw, minimum size=1.5cm] (alice2) at (9,0) {};
		\node (a2) at (9,1) {Alice};
		\node[circle,draw, minimum size=1.5cm] (bob2) at (11.5,0) {};
		\node (b2) at (11.5,1) {Bob};
		\node[circle,draw, minimum size=1.5cm] (charlie2) at (14,0) {};
		\draw[dashed, red, ->] (bob2) -- (charlie2) node[midway,above=6pt,black] {\footnotesize Cl. Data};
		\node (c2) at (14,1) {Charlie};

		\node[fill,circle, blue] (alice21) at (9.25,0) {};
		\node (node21a) at (9.25, -0.3) {1a};
		\node[fill,circle, blue] (charlie21) at (13.75,0) {}
			edge[very thick, dotted, blue, bend left=45] (alice21);
		\node (node21b) at (13.75, -0.3) {1b};
	\end{tikzpicture}
	\caption{%
		Entanglement swapping: Two network nodes, Alice and Charlie establish entanglement via an 
		intermediary node Bob (a quantum repeater~\cite{cirac:repeater,repeaterReview}). 
		In the beginning, Alice and Bob share an entangled link, and Bob and Charlie share an entanglement - virtual quantum link (VQL). 
		Each end of the entangled links requires 1 qubit of quantum memory (1a,1b,2a,2b).
		Bob subsequently performs a procedure known as entanglement swapping~\cite{artur:entSwap}:
		he teleports~\cite{teleport} qubit
		(1b) to Charlie using the entangled link (2a,2b).
		This requires classical communication from Bob to Charlie,
		and the entangled link (2a,2b) is consumed by this process.
		Then, Charlie performs a final correction operation.
		The final state is a VQL between Alice and Charlie (qubits (1a,1b)).
		The use of teleportation in this way is also known as entanglement swapping.
		Note that the number of operations to create a VQL between two distant nodes is thus equal to the path length.
		In practice, all operations are subject to noise, meaning that the final entanglement is generally not perfect. 
		One way to overcome this challenge is by first producing multiple entangled links, and then applying a procedure known as entanglement distillation~\cite{durSurvey}.
		Since here we take a rather abstract view of counting required resources, we merely note that this means that long distance VQL carry a higher cost to be established than shorter ones.
		In this first simple stage, we will take all VQLs to have equal cost, but more realistic protocols will need to take such costs into consideration.
	}
	\label{fig:entanglementSwap}
\end{figure}

	An important concept for quantum networks is the relation between sending a qubit from network node $A$ to node $B$, and creating entanglement between the two nodes. 
	If $A$ is capable of sending a qubit to $B$, then such entanglement can be established by $A$ preparing two entangled qubits, and sending one of them to $B$. 
	In turn if $A$ and $B$ already share entanglement, then they can send a qubit using quantum teleportation~\cite{teleport} - even if they are not directly connected by a quantum communication channel. Deterministic quantum teleportation has been shown to be technologically feasible~\cite{pfaff2014unconditional}. Quantum teleportation requires classical communication and consumes the entanglement in the process. It forms a key ingredient for realizing quantum repeaters as illustrated in \cref{fig:entanglementSwap}.
	We may thus think of shared entanglement (together with a classical communication channel) as forming a virtual connection - a shortcut - in the network. In analogy to the notion of virtual circuits
	in classical networking, we will refer to such a link as a \emph{virtual quantum link (VQL)}. Importantly, such VQLs are rather unusual from a classical perspective in that they 
	can be used only once, and require one qubit of quantum memory at each end point to 
	be maintained. As explained in \cref{fig:entanglementSwap}, long distance VQLs are more time consuming to create~\cite{dur1999quantum} 
	and thus carry a higher cost when used. In this first stage of studying quantum networks, we will however assume
	that all costs are equal.
	
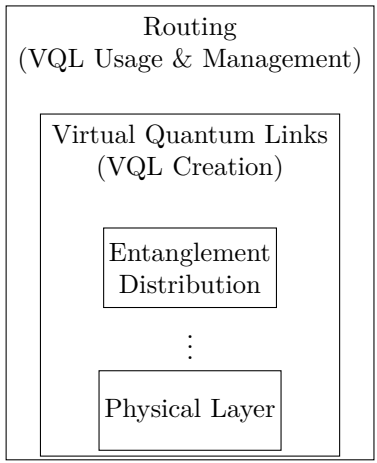
\begin{figure}
		\begin{minipage}[c]{0.3\textwidth}
			\begin{tikzpicture}
				\node[draw, minimum size=3em] (physical) at (0,0) {Physical Layer};
				\node (ellipsis) [above=0.2em of physical] {\vdots};
				\node[draw] (entanglement) [above=0.2em of ellipsis, align=center, minimum size=3em] {Entanglement\\Distribution};

				\node (qlinks) [above=1.5em of entanglement, align=center] {Virtual Quantum Links\\(VQL Creation)};
				\node[draw, fit=(qlinks) (entanglement) (physical)] {};

				\node (qrouting) [above=1.5em of qlinks, align=center] {Routing\\(VQL Usage \& Management)};
				% Hack in extra space for the surrounding square
				\node (spacer) [below=-0.3em of physical] {};
				\node[draw, fit=(qrouting) (qlinks) (entanglement) (physical) (spacer)] {};
			\end{tikzpicture}
		\end{minipage}
		\begin{minipage}[c]{0.67\textwidth}
			\caption{In a quantum network, the underlying physical layer, followed by (iterations of)
				error-correction and entanglement swapping and distribution can be used to create virtual quantum
				links (VQL). We here adopt such a perspective of virtualization, both to allow techniques from
				classical network routing to be applied, as well as to focus on protocols that use simple(r)
				quantum operations.
				This modifies previous models that mirror the OSI model for classical networks, whose lower layers
				can be understood as a procedure to generate VQLs~\cite{vanMeterOSI}.
			}
			\label{fig:osiModel}
		\end{minipage}
	\end{figure}

	While the requirements of a VQL for one qubit of memory at each node appears benign from a classical perspective, it is rather significant in a quantum network. 
	First of all, due to current technological limitations each network node can store only very few qubits.
	What's more, however, such quantum storage is typically rather noisy, meaning that each qubit has a limited lifetime. 
	The latter can in theory be overcome by performing error-correction at each network node at the expense of using additional qubits. Therefore, we will take the simplifying perspective that
	while each network node may only store very few qubits, they have an arbitrarily long lifetime.
	Such an abstraction is useful to devise new protocols (see also \cref{fig:osiModel}),
	while of course not being a realistic assumption for the forseeable future.
	A more realistic model would thus be to expire VQLs after a certain period of time.
	In that case the latency of the classical communication channels becomes crucial.
	If the classical communication required by routing or indeed any protocol takes too much time,
	then the VQL is already gone. This concern is less 
	relevant for routing quantum information between on-chip quantum processors, but plays
	an important role in long-distance quantum networks.

	Let us briefly summarize the network model we consider, before discussing how such a model enables classical networking concepts to be applied to the study of quantum ones. \cref{fig:NetworkBox} provides a succinct overview of the relevant parameters. 

	\begin{itemize}
	\item Every node in the quantum network can store $Q$ qubits and perform the operations given below. We remark that as in classical networks, quantum network nodes can of course be more diverse. One could, 
	for example, consider a situation in which the end-points are very simple nodes with no storage capabilities, whereas the network has more powerful routers that can store and operate on many of them. 
	It is also possible to consider nodes that can just perform entanglement swapping operations (see \cref{fig:entanglementSwap}), 
	but have no storage\footnote{The storage requirement is - in theory - overcome by demanding that both qubits arrive at precisely the 
	same time, and the node immediately performs a swapping operation. In practise, however, this can realistically only be done with a significant failure probability at each node without storage.}. 
	For simplicity, we will here assume all nodes are equivalent. 
	\item Every node can establish entanglement with neighboring nodes to which it is directly connected by a quantum communication channel such as optical fiber. There are many quantum protocols for generating entanglement, and we will take such a protocol as a given.
	\item Every node can perform quantum teleportation (i.e., a Bell measurement). In particular, this means that each node can act as a repeater in \cref{fig:entanglementSwap} to establish VQLs in the 
	network. Being able to perform teleportation is also sufficient to use a VQL to later transmit qubits. Note that these are operations which are much simpler than full quantum computation.
	\item Every VQL has a cost, modeling the time it takes to establish this VQL, and lifetime. As outlined above, we will here work in a simplfied model in which all costs are one, and idealized error-correction allows an arbitrary lifetime.
	\item Both the entanglement generation and swapping operation take one time unit. At each time step, a network node can either make entanglement with one of its neighbours or perform a swap. 
	We remark that different experimental proposals for quantum repeaters lead to different timings and that the time for entanglement generation generally depends also on the physical distance between the nodes
	as well as how noisy the operations are. A procedure known as entanglement distillation can, for example, be used to reduce noise but requires several imperfect entangled links to be generated first which makes
	it time-consuming (and difficult).
	\item All network nodes can communicate classically within one time unit.
		Since we work in a simplified model where error-correction is performed to extend the lifetime of qubits,
		we will ignore the latency of classical communication.
		However, our protocols are designed with an eye towards lifting this requirement as we will explain below.
	\end{itemize}

	Evidently, such an abstraction hides much of the complexity of realizing quantum networks.
	However, just as layers of abstractions used in classical networking~\cite{tanenbaum},
	it allows us to break the problem
	down into simpler components - layers - which simplifies further study. On the one hand, it allows us to apply ideas and concepts from classical computer science 
	to the problem of routing qubits in a quantum network. Specifically, our model of abstraction leads to a layer in a ``quantum OSI-model'' (\cref{fig:osiModel}), that allows an embedding in classical techniques. The properties of our model are summarized in~\cref{fig:NetworkBox}.
	On the other hand, it also motivates routing protocols which are based on only very simple components: namely means to generate (and maintain)
	entangled links and perform swapping operations. This is in contrast to works which study maximizing flows over quantum channels which generally require complex quantum computing operations on 
	potentially very many qubits (see related work below). 

	Given the abstraction proposed here, routing in a quantum network can be understood as routing on VQLs
	that can be assigned deliberately and dynamically, carry a certain cost,
	can only be used once, and which may expire after a given time. 
	We remark that communication in a quantum network can in principle be understood entirely as transformations performed on the graph of VQLs: Given existing VQLs, 
	to send a qubit from two nodes $A$ and $B$ we consume VQLs to create a new entangled link
	-- a new VQL -- directly between $A$ and $B$,
	followed by teleportation of the qubit over said VQL. 
	That is, we have transformed the network topology given by a set of VQLs into a new one
	in which $A$ and $B$ are connected.
	Such transformations can be very useful even if the VQLs are themselves noisy, leading to rather beautiful transformations on the VQL network
	topology~\cite{cirac:percol}. As we will discuss in \cref{sec:discuss} at the end, using multi-partite entanglement, a quantum network can also exhibit a more involved form of VQLs as hyper-edges where more than two nodes hold part of the entanglement. The problem of routing in a quantum network using VQLs is thus equivalent to managing entanglement in the network.

	\begin{figure}
	\noindent\fbox{%
	\parbox{\textwidth+4pt}{%
		\begin{tabular}{|l|l|p{3.6cm}|}
		\hline
		{\bfseries Property} & {\bfseries General Network} & {\bfseries Our Network}\\
		\hline
		Size & $N$ & $N$\\
		\hline 
		Size of quantum memory & $Q$ & $Q=O(\log N)$ \\
		\hline 
		Lifetime of qubits & fixed time $t_{LT}$ & $t_{LT}=\infty$ \\
		\hline 
		Time to establish entanglement & fixed time $t_{EE}$ & $t_{EE} = 1$\\
		\hline 
		Type of entanglement & arbitrary $k$-partite entanglement & bipartite EPR (VQL)\\
		\hline 
		Time for swapping operation & fixed time $t_\text{SWAP}$ & $t_\text{SWAP} = 1$\\
		\hline 
		Time for measurement & fixed time $t_{M}$ & $t_{M} = 1$\\
		\hline 
		Parallelism of operations at node & parallel or sequential & parallel \\
		\hline 
		Topology of the physical communication links & arbitrary & ring or sphere\\
		\hline 
		Classical communication delays & from node $i$ to $j$ $t^{i\rightarrow j}_{COM}$ & $t^{i \rightarrow j}_{COM} = 1, \forall i,j$\\
		\hline
		Size of classical memory & $S$ & $S_{Ring}=O(\log N)$, $S_{Sphere}=O(\log^6 N)$\\
		\hline
		\end{tabular}%
		\\
		\begin{description}
			\item[{\bf Size}] \hfill \\ The number of nodes in the network.
			\item[{\bf Size of quantum memory}] \hfill \\ The number of qubits that each network node can store simultaneously.
			\item[{\bf Lifetime of qubits}] \hfill \\ The units of time that a qubit can be kept in storage.
			\item[{\bf Time to establish entanglement}] \hfill \\
				The units of time that it takes to establish the relevant type of entanglement between
				network nodes that are directly connected by a physical quantum communication link such as fiber
				(typically, bipartite EPR pairs corresponding to VQLs between physically adjacent network nodes).
			\item[{\bf Type of entanglement}] \hfill \\
				Generally, bipartite EPR pairs corresponding to immediate VQLs. Quantum information allows more complicated entangled states which can later be converted into VQLs using measurements 
				(see Discussion), yet these are very difficult to create in practice.
			\item[{\bf Time for swapping operation}] \hfill \\
				Time that it takes a network node to perform an entanglement swapping operation (see Figure~\ref{fig:entanglementSwap}).
			\item[{\bf Time for measurement}] \hfill \\
				Time to perform a measurement at each node.
				Often the same as the time needed to swap entanglement.
				Only relevant when considering more general forms of entanglement which require
				measurements to be converted to VQLs.
			\item[{\bf Parallelism of operations at node}] \hfill \\ Determines whether the node can simultaneously create entanglement with more than one other network node connected by physical quantum communication channels.
				In many systems, such as NV centers in diamond~\cite{ronald:prx}
				only one VQL can be created by one node at the same time meaning such operations are sequential.
			\item[{\bf Topology of the physical communication links}] \hfill \\
				Determines how the quantum nodes are connected by direct physical quantum communication channels over which they can easily communicate.
			\item[{\bf Classical communication delays}] \hfill \\
				Determines the time needed to send classical control information from node $i$
				to another node $j$ over standard classical communication networks.
				The time for quantum communication over physical links is
				- along with other delays - absorbed into the time to establish entanglement.
			\item[{\bf Size of classical memory}] \hfill \\
				The number of classical bits the nodes have to store about the network.
				We enforce that the nodes only have a local view of the network and not the entire network topology.	
		\end{description}
	}\\ %
	}
	\caption{Abstraction of an $N$ node quantum network}
	\label{fig:NetworkBox}
	\end{figure}

	\subsection{Requirements and routing protocols}

	Having established a suitable model, we are motivated to consider how to best assign VQLs in order to route quantum
	information, or equivalently manage entanglement, as effectively as possible. That is, we are now facing a rather classical problem of resource allocation
	and management. In this first work, we will adopt the above model where additional simplifications, such as taking all 
	VQLs to have arbitrary lifetime and ignoring the latency of classical communication enables us to prove analytical results. 
	However, we will design our protocol to have specific features that will likely lead to a good performance even if we were
	to relax such additional simplifications.

	The design of good routing protocols can exploit several features: First, we can of course decide ourselves
	what the topology of the network of VQLs should look like. 
	Since VQLs can in principle be established between any pair of nodes, a naive approach
	would be to establish a fully connected graph of VQLs. Recall, however, that each VQL needs one qubit of quantum memory 
	at each end point, and hence this naive approach would require $\Omega(N)$ qubits of memory per node for a network of $N$ nodes. On the other hand, one might also consider establishing VQLs on the fly whenever it is clear
	that we would like to form a long distance entangled link between two nodes. While this may be feasible in the regime of entirely parallel, and perfect, quantum operations creating entanglement ahead of time has significant advantages:
	first of all, we can create high-quality (low noise) VQLs by performing entanglement distillation which is too time consuming to perform on the fly. Similarly, we will need a large number of swapping operations on the entanglement thus
	created, while our objective here is to minimize the final number of - in practise very noisy - swapping operations. Finally, however, distinguishing between a background process that creates, manages and maintains entangled links - and the actual transmission of quantum information, allows many parties in the network to communicate simultaneously on the already pre-shared entanglement.
	
	Our goal will thus be to minimize the number of memory qubits needed at each node, while, at the same time, have the path of VQLs between pairs of nodes to be short.
	This means that the number entanglement swapping operations (equal to the path length, see \cref{fig:entanglementSwap})
	to form the final VQL for transmission between $A$ and $B$ should be small.
	While this may be less important in the regime of perfect operations in the abstract model of \cref{fig:osiModel},
	it is clear that it becomes highly relevant if such operations are noisy.
	Hence, we will pay special attention to this here.

	The question of routing is then how we can efficiently find a path between a sender $A$ and receiver $B$.
	To minimize local latency we desire to reduce the computation time and classical memory
	needed at each node for storing and processing routing information. 
	More significantly, however, we desire a decentralized routing algorithm in which each node can make a routing decision 
	without having to communicate with all other nodes or storing the complete topology of the network. While of lesser importance
	in our abstract model, this will become relevant when counting the time needed for classical communication: the longer the classical
	communication takes, the longer the lifetime of the qubit in the node's quantum memory needs to be. 

	Here, we exhibit routing algorithms that fulfil these goals for simple network structures,
	namely the ring and the sphere. 
	The first is motivated by a ring topology in a metropolitan area, whereas the latter is inspired by considering satellites around Earth.
	Note that both topologies only use short physical links, as is necessary. 
	We remark, nevertheless, that the essential ideas present in our approach for the ring and the sphere, could also be carried over to other structures such as a 
	grid as we will discuss in Section~\ref{sec:discuss}.
	Specifically, we present
	the following results, summarized in Table~\ref{tab:contrib}.
	\begin{itemize}
	\item A way to assign virtual quantum links using entanglement between $N$ network nodes distributed equally around a ring or sphere, such that the quantum memory required at
	each node is $O(\log N)$. 
	\item A decentralized routing protocol on the resulting network of VQLs for which we prove analytically that it is efficient in all resources,
		i.e.\ the routing time, the classical memory of each node, and the time needed to replenish the consumed VQLs are all $O(\polylog N)$. 
	We show that, in fact, our algorithm finds the shortest path along VQLs connecting a sender $A$ and a receiver
	$B$.
	%\footnote{Using classical routing terminology, this means the stretch factor, i.e., the ratio between the routed path and the optimal one is $1$.}
    Crucially, our algorithm uses only local information for routing, since each node only stores information about neighbouring nodes, hence having to store very little information about the graph as a whole.
    Contrasting with, for example, Dijkstra's algorithm, where nodes have to store information about the entire graph layout.
	Our routing algorithm can be understood as a hierarchical routing scheme tailored to the specific demands above. As we will discuss below, our assignment of VQLs and routing
	algorithm still works even if part of the network is missing, i.e., the nodes are not equally distributed, but denser in certain areas than in others.
	\item We perform an initial simulation to study how our network of VQLs and routing procedure performs under load, that is, if many network nodes attempt to communicate simultaneously. We see that our routing algorithm is not optimized for multiple routing requests. We provide a discussion and ideas how to overcome this issue in Section \ref{simulations}.
	
	\end{itemize}

	%\addtocounter{footnote}{1}
	\begin{table}[t!]
		\centering
		\begin{tabular}{l|c|c|c|c}
			& Complete VQLs  & Ring & Sphere \\ \hline
			Diameter & $O(1)$ & $O(\log N)$ & $O(\log N)$ \\
			Quantum Memory &  $O(N)$ & $O(\log N)$ & $O(\log N)$ \\
			Entangl. Distribution &  $O(N)$ & $O(\log N)$ & $O(\log N)$ \\
			Routing (time/node)\protect\footnotemark[\value{footnote}] & $O(1)$ & $O(\log N)$ & $O(\log N)$\\
			Routing (space/node) & $O(N)$ & $O(\log N)$ & $O(\log^6 N)$ \\
			$\hookrightarrow$ Label size (node) & $O(1)$ & $O(1)$ & $O(\log N)$
		\end{tabular}
		\caption{%
			A comparison of a communication protocol 
			that distributes VQLs between all nodes in the network,
			and our contributions for the ring the sphere networks.
			The structure of the network in our contribution allows for decreasing the diameter
			(maximum path length between any two nodes) and keeping the quantum memory size feasible.
			We route using a decentralised routing algorithms
			with a hierarchical labelling scheme (similar to Internet Protocol~\cite{postel1981ip})
			that saves only possible shortest paths in a logarithmic size collection.
			Redistributing entanglement is also simplified by VQLs that can be rebuilt concurrently (see \cref{fig:NetworkBox} for a more general model).
		}
		\label{tab:contrib}
	\end{table}
	%\footnotetext{The time complexity of the ring algorithm is per path.}

\subsection{Related Work}

Routing messages to the right destination in a network has seen enormous attention in the classical literature, and the term routing is also used for a number of different concepts in 
quantum networks. While we are not aware of any work taking the perspective employed above, we briefly elucidate\footnote{Due to the vast amount of literature, our references will be non-exhaustive and we focus on providing entry-points.} the relation of our work to the classical and quantum
literature. It is highly likely that concepts from both domains will form an important ingredient in designing quantum networks as we will discuss further in 
Section~\ref{sec:discuss}.

\subsubsection{Related work in quantum information}
Let us first consider routing in quantum networks. The first is at the so-called physical layer (\cref{fig:osiModel}) in which a physical information carrier such as a photon
is directed to the correct network node. This forms an essential ingredient in establishing entanglement between network nodes, which has seen a number of experimental
implementations (see e.g.~\cite{andreas:elementary,andreas:cavityReview}). 
In contrast, we here are concerned with a much higher level protocol in which a means to generate entanglement is a given, and which
is independent of the exact physical implementation of the qubits. 

Another line of quantum research considers quantum network coding (see e.g.~\cite{Bruss:codingRouter,hayashi:networkCode,Leung:networkCode,kobayashi:coding,pirandola:capacities}). 
We remark that in this context, the term routing is in the quantum literature sometimes taken synonymous to network coding~\cite{pirandola:capacities}. 
As in classical networks, network coding can outperform routing when trying to maximize the flow of information through a quantum network - albeit at the expense
of requiring much more complex quantum operations.
On the one hand,
work in this area has an information-theoretic flavor in which limits to network flows are derived in terms of quantum capacities~\cite{aggregating,Zeng:capacities,lo:rateLoss,pirandola:capacities}. These works are of great
appeal to establish the ultimate limits of transmitting quantum information. We remark that of course achieving the capacities does in general require a full blown quantum computer,
and indeed it is known that even
arbitrarily large block sizes (and hence quantum computers and memories) are required to achieve some capacities of quantum channels~\cite{elkouss:blockLength}. We remark that in turn
we here precisely focus on building protocols on top of very simple elementary operations.

On the other hand, a number of works consider the creation of quantum states amongst network nodes using network coding methods~\cite{Bruss:codingRouter,Leung:networkCode,kobayashi:coding}. These works are highly relevant for 
distributing quantum states, and complementary to our more high level approach:
While their focus rests on the creation of states, we focus on how to marshal the resources
given by said states once they are created.
It is of course possible to generate highly entangled states shared by many network nodes, instead of,
for example, creating bipartite entangled states resulting in VQLs
or shared entanglement between a subset of the nodes such as graph states~\cite{bruss:graph}. 
We will discuss such possibilities further in Section~\ref{sec:discuss}.

Another meaning of the term quantum routing~\cite{strauch:ParallelStateTransfer,xue:stateTransfer,kay:stateRouting} 
occurs in the domain of quantum state transfer~\cite{bose:stateTransfer} (see~\cite{kay:stateTransfer} for a review). 
Here, one considers a network of spins or oscillators between which one can control the physical interactions.
Specifically, the goal is to design the Hamiltonian coupling the different spins leading to a time evolution
that will eventually transfer a quantum state from one place to another. 
In contrast, we here focus on high level approach of resource allocation given simple operations, also using
a combination of physical carriers for which couplings are often fixed. 
We remark that at the physical layer there is of course a notion of state transfer in all quantum networks, for example the transfer of a state from an optical cavity at one network node onto a photon to create entanglement between two
nodes~\cite{cirac:distantNodeStateTransfer}.

Finally, we recall
that all information transmission in a quantum network can in principle
be understood as transforming a network topology consisting of entangled links into another topology, as emphasized by e.g.~\cite{cirac:percol}. 
Concretely,~\cite{cirac:percol,siomau:percol,cirac:entDist,cuquet:percol} give concrete transformations of networks in which a fixed toplogy of noisy entangled links is transformed into another by a coordinated operation at each nodes
such that the resulting graph has a higher percolation threshold.

\subsubsection{Related work in classical networking}
Let us now consider the relation to routing methods known in classical networks.
First, let us remark that even given a procedure to create VQLs,
there are fundamental differences between classical and quantum networks:
a VQL is maximally dynamic in that it can in principle be established between any two nodes in the network,
but can be used only once.
Hence, while given a fixed static graph of VQLs one could apply a classical routing algorithm to the resulting topology,
we are here in a situation where we can specifically design the graph of VQLs to assist routing.
Moreover, in this design as well as in the resulting routing algorithm,
we want to pay attention to how VQLs can be reestablished.
Nevertheless, we can draw significant inspiration from classical routing schemes. Since we desire routing methods that make efficient local decisions, it is natural to consider hierarchical routing schemes in which the network is naturally divided into different ``levels'' which have long been suggested for large networks (see e.g.~\cite{kleinrock:routing}) and are also employed on the internet. Such schemes have seen use from routing on VLSI circuits~\cite{tsai:hybridRouting}, to wireless networks~\cite{karaki:routingSurvey,younis:routingSurvey}.
The scheme we consider here has most in common with compact routing schemes (see e.g.~\cite{Gavoille:Overview} for a survey,~\cite{Thorup:compact,eilama:compact,Eilam:CompactRouting} for more recent works and~\cite[Table 1]{abraham:lowDoubling} for an overview). Indeed, if we were to fix the VQLs permanently, then our routing scheme would lead to a rather
efficient compact routing scheme adapted to the particular topologies of VQLs on the sphere and ring. Finally, we remark that there do of course exist classical routing
algorithms (see e.g.~\cite{Dolev:Bubbles}) that are dynamic in the sense that they can deal with temporary failures of nodes and communication links. 
While 
this is a less dynamic situation than ours in which VQLs can be created between any nodes, it may be useful to consider their methods in combination with more complex
quantum links as discussed in Section~\ref{sec:discuss}. Finally, we remark that as already noted above, the notion of virtual connections is again a concept that is well established
in classical networks, already when ATM was being used~\cite{tanenbaum}. Again, while not immediately applicable we expect such ideas could give useful inspiration.

\subsection{Overview}\label{sec:OverviewIntro}
%\steph{The first paragraph is somewhat optional if we rephrase the ring, I am undecided whether to keep it}
%Note that throughout, we ignore the question of routing 
%classical communication, since already exist many solutions to this problems~\cite{moy1998ospf,rekhter2006bgp}. Since classical communication comes for free, we can also assume that the sender informs the receiver of the intention to set up
%a VQL. That is, the receiver knows the senders identity. 

Before turning to the specific case of the ring and sphere, let us briefly summarize the main ideas unifying the two cases.
Our first task is to distribute entanglement so that the path between each pair of nodes (according to the topology of the graph of VQLs) in the network is short $O(\log N)$.
At the same time we also need only $O(\log N)$ qubits of quantum storage at each network node.
This may not be so hard to achieve by itself, but we want to do this in a way
that will later enable an efficient routing algorithm and make replenishing consumed VQLs easy.
We remark that while future routing algorithms using the model of~\cref{fig:NetworkBox}
may draw great benefits from assigning VQLs depending on the load in the network,
we will work with a static graph to be replenished.
Key to our efficient routing algorithm is a smart hierarchical labelling of network nodes in the resulting network graph formed by the VQLs. 
Indeed, our algorithm forms an instance of a hierachical routing algorithm, where a hierarchical structure is imposed by the way we assign the VQLs between nodes.

The key idea in designing the network of VQLs is to start with a base structure, and successively approximate the desired network
through what is known as \emph{subdivision} of edges~\cite{loop1987smooth}.
Subdividing an edge is the process of replacing an edge with a node,
and connecting that node with the endpoints of the edge.
After subdivision, the new nodes are interconnected to reach the desired structure.
We iterate this process until the graph represents the physical network,
in both number of nodes and connections.
The previous graphs generated during the subdivision procedure can
now be used as a structure for distributing entanglement.
If we distribute entanglement not only along physical connections,
but also along edges of previous subdivision iterations
then these will form shortcuts through the network connecting far away nodes.

\begin{figure}
	\centering
	\begin{subfigure}[b]{0.32\textwidth}
		\centering
		\begin{tikzpicture}[scale=0.7]
			\node (0) at (0,0) {$0$};
			\node (2) at (4,0) {$2$};

			\draw[thick, blue, dotted] (0) -- (2);

			\node (1) at (2,2) {$1$};
			\node (3) at (2,-2) {$3$};
			\foreach \x in {0,1,2,3} {%
				\pgfmathparse{int(mod(\x - 1 + 4,4))};
				\draw (\x) -- (\pgfmathresult);
			}
		\end{tikzpicture}
		\caption{graph $G_2$}
	\end{subfigure}
	\begin{subfigure}[b]{0.33\textwidth}
		\centering
		\begin{tikzpicture}[scale=0.7]
			\node (0) at (0,0) {$0$};
			\node (4) at (4,0) {$4$};

			\draw[thick, blue, dotted] (0) -- (4);

			\node (2) at (2,2) {$2$};
			\node (6) at (2,-2) {$6$};
			\foreach \x in {0,2,4,6} {%
				\pgfmathparse{int(mod(\x - 2 + 8,8))};
				\draw[red, dashed] (\x) -- (\pgfmathresult);
			}

			\node (1) at (0,2) {$1$};
			\node (3) at (4,2) {$3$};
			\node (5) at (4,-2) {$5$};
			\node (7) at (0,-2) {$7$};
			\foreach \x in {0,1,...,7} {%
				\pgfmathparse{int(mod(\x - 1 + 8,8))};
				\draw (\x) -- (\pgfmathresult);
			}
		\end{tikzpicture}
		\caption{graph $G_3$}
	\end{subfigure}
	\begin{subfigure}[b]{0.33\textwidth}
		\centering
		\begin{tikzpicture}[scale=0.7]
			\node (0) at (0,0) {$0$};
			\node (8) at (4,0) {$8$};

			\draw[thick, blue, dotted] (0) -- (8);

			\node (4) at (2,2) {$4$};
			\node (12) at (2,-2) {$12$};
			\foreach \x in {0,4,8,12} {%
				\pgfmathparse{int(mod(\x - 4 + 16,16))};
				\draw[red, dashed] (\x) -- (\pgfmathresult);
			}

			\node (2) at (0,2) {$2$};
			\node (6) at (4,2) {$6$};
			\node (10) at (4,-2) {$10$};
			\node (14) at (0,-2) {$14$};
			\foreach \x in {0,2,...,14} {%
				\pgfmathparse{int(mod(\x - 2 + 16,16))};
				\draw[cyan,dashdotted] (\x) -- (\pgfmathresult);
			}

			\node (1) at (-1,1) {$1$};
			\node (3) at (1,3) {$3$};
			\node (5) at (3,3) {$5$};
			\node (7) at (5,1) {$7$};
			\node (9) at (5,-1) {$9$};
			\node (11) at (3,-3) {$11$};
			\node (13) at (1,-3) {$13$};
			\node (15) at (-1,-1) {$15$};

			\foreach \x in {0,...,15} {%
				\pgfmathparse{int(mod(\x - 1 + 16,16))};
				\draw (\x) -- (\pgfmathresult);
			}
		\end{tikzpicture}
		\caption{graph $G_4$}
	\end{subfigure}
	\caption{%
		VQLs on the ring. The outer edges (solid black) correspond to the physical links in the ring network,
		while the inner edges (blue dotted, red dashed, cyan dash-dotted)
		correspond to virtual quantum links enabled by entanglement distribution between distant network nodes.
		According to the recursive procedure we later refer to these graphs as 
		$G_2$, $G_3$ and $G_4$.
	}
	\label{ringRoutingTagged}
\end{figure}
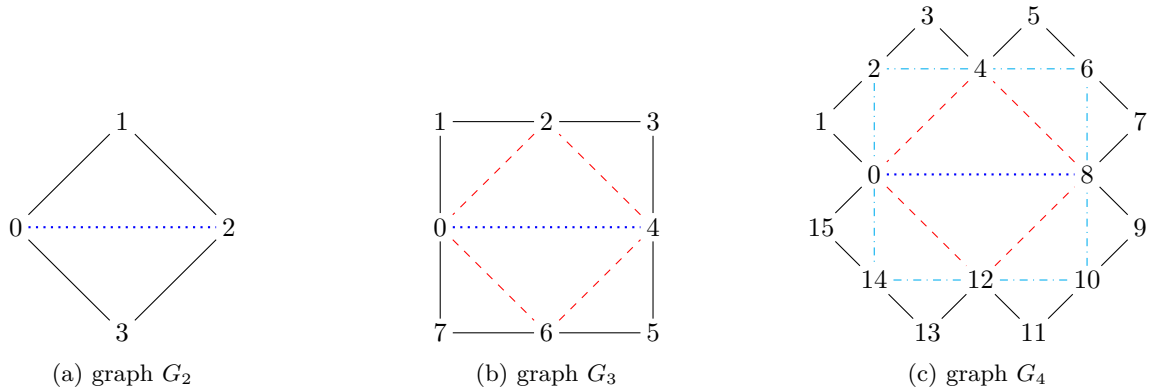

To illustrate the idea let us first consider the ring illustrated in
\cref{ringRoutingTagged}. We ``grow'' the network by successive subdivisions from the left to the right. Our procedure introduces a hierarchical structure since previous iterations can be understood as being lower in the hierarchy. 
Intuitively, we route by descending into a common level of the hierarchy at which we can use VQL shortcuts to bridge long distances.
We remark that for the ring we use a simplified labeling of the nodes. 
However, it would be possible to use a very similar labeling reflecting the recursive procedure with which the graph is created as in the case of the sphere. 
This more general labeling would also allow a situation in which the nodes are denser on one side of the ring than the other.

\begin{figure}
	\centering
	\begin{subfigure}{0.49\textwidth}
		\centering
		\includegraphics[width=0.47\textwidth]{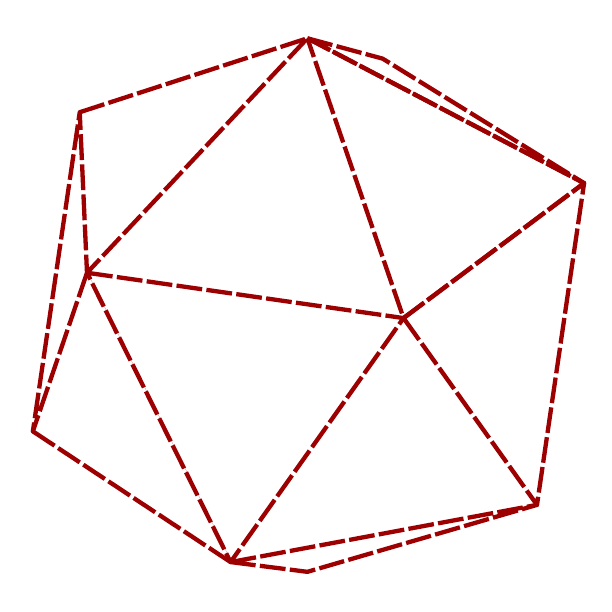}
		~
		\includegraphics[width=0.47\textwidth]{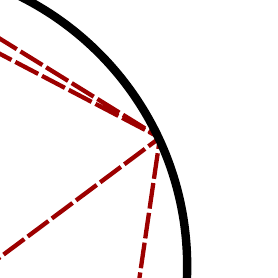}
		\caption{Base graph and close-up with sphere.}
		\label{fig:recursiveStructure1}
	\end{subfigure}
	~
	\begin{subfigure}{0.49\textwidth}
		\centering
		\includegraphics[width=0.47\textwidth]{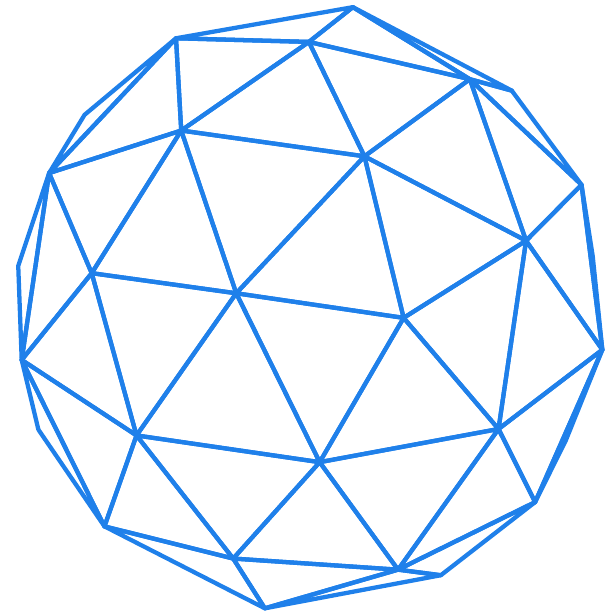}
		~
		\includegraphics[width=0.47\textwidth]{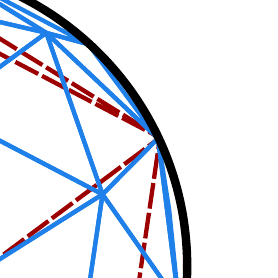}
		\caption{One recursive step and close-up with VQL shortcuts.}
		\label{fig:recursiveStructure2}
	\end{subfigure}
	\caption{
		VQLs on the sphere. We perform a recursive procedure to add more nodes to the network,
		and keep doing this until every node represent a physical node.
		In this figure the structure approximates the sphere more accurately the more
		recursive steps we perform (i.e. the more nodes there are in the network).
		The previous iterations can then function as structure for virtual quantum links (VQL) using
		entanglement.
		In \cref{fig:recursiveStructure2} we can see that the connections
		in previous iterations (red dashed, \cref{fig:recursiveStructure1})
		are added to the network (solid blue) as VQLs that effectively form 
		shortcuts that bridge nodes separated by longer distances.
		For example, shortest paths between vertices from \cref{fig:recursiveStructure1}
		take half as many hops in \cref{fig:recursiveStructure2} if we include the red dashed edges.
		Thus VQLs significantly reduce the distance between nodes in a network.
	}
	\label{fig:recursiveStructure}
\end{figure}

The case of the sphere, detailed in \cref{sec:sphere}, is in spirit very similar.
Again, all edges in the previous iteration
will become VQLs. Key to understanding this procedure
is that the resulting structure will indeed approximate the sphere.  The closest regular polyhedron, i.e.\ a platonic solid, which approximates a sphere is the icosahedron.
The icosahedron will be subdivided using a common algorithm in 3D modelling for subdividing triangular meshes, called Loop subdivision~\cite{loop1987smooth}.
Edges connecting the nodes in previous iterations (\cref{fig:recursiveStructure1}) will remain as VQLs in the next iteration (\cref{fig:recursiveStructure2}).

We emphasize that our method does not require the network to be the same everywhere on the sphere. If we want to consider a situation in which the nodes on one side of the sphere are denser than at another, we simply perform a further subdivision in that region. In essence, this means that nodes in sparser regions will not be present which does not affect our routing procedure.

We will analyse the properties of the two graphs generated in a such a way for the ring and the sphere,
where we will see that this approach is effective at reducing the diameter in a network
while still limiting the degree of nodes.
Given the recursive structure of the graph, it is also possible to define efficient routing algorithms. Finally, we consider how VQLs are replenished in Section~\ref{sec:replenishing}.

\subsection{Preliminaries: Graph Theory}\label{sec:GraphTheory}
A network can be modeled as a graph with the network nodes being represented by the vertices of the graph
and the links between the nodes being represented by the edges.
We denote a graph with $G=(V,E)$, where $V$ is its set of vertices in $G$
and $E\subseteq \left\{ \{\alpha,\beta\} \subseteq V : u\neq v \right\}$ is its set of edges.
We have restricted ourselves to simple graphs where the edges are undirected, unweighted,
and there is at most one edge between any two distinct vertices
and no edge to a vertex itself.

For a pair of vertices $\{\alpha,\beta\} \in E$, we say that $\alpha$ is adjacent to $\beta$,
which may also be denoted with $\alpha \edge \beta$.
For any vertex $\alpha \in V$ we refer to all the vertices adjacent to it as the neighbors of $\alpha$
and denote this set as
\begin{equation}
	N(\alpha) \coloneqq \left\{\beta : \alpha \edge \beta \right\}\,.\label{eq:neighbourhood}
\end{equation}
A walk of length $n$ from a vertex $\alpha$ to $\beta$ is a sequence of vertices $\alpha_0,\dotsc \alpha_n$
where $\alpha_i \edge \alpha_{i+1}$ for all $i\in\{0,\dotsc,n-1\}$,
$\alpha_0=\alpha$, and $\alpha_n=\beta$.
A path is walk with no repeated vertices. When studying networks, it is useful to
know how many network links we need to use to get from one node to another. Here, the relevant quantity
is the distance between the respective vertices in the graph.
Given vertices $\alpha,\beta\in V$, the distance between them,
\begin{equation}
	d(\alpha,\beta)\,,
\end{equation}
is defined as the length of the shortest walk from $\alpha$ to $\beta$.
In case the graph is not clear from the context, we specify it in the subscript, i.e., we write
$d_G(\alpha,\beta)$.
The diameter of a graph $G$, denoted $D(G)$, is defined as
\begin{equation}
	D(G) \coloneqq \max\set{d(\alpha,\beta): \alpha,\beta \in V}\}.
\end{equation}
The diameter of the graph corresponds to the number of network links that need to be used in the worst case to communicate between two nodes in the network.
Furthermore, we say that $H=(V',E')$ is a subgraph of $G=(V,E)$ if $V'\subseteq V$ and $E'\subseteq E$.
The subgraph $H$ is an induced subgraph when for any two vertices $\alpha,\beta \in V'$
it holds that $\{\alpha,\beta\} \in E' \iff \{\alpha,\beta\} \in E$.

\subsubsection{Graph Notation}\label{sec:graphNotation}
Let us define some of the notation we use to describe subdivided graphs.
The set of vertices in the subdivided graph $G$ is $V$, and let $E$ be the set of edges.
We will use the following notation:
\begin{align}
	\alpha,\beta,\gamma,\eta,\pi & \in V & \text{lowercase Greek letters for vertices,}\label{eq:allVertices}\\
	V_i &\subseteq V & \text{the vertices after $i$ subdivisions,} \label{eq:verticesKsubdivisions}\\
	G_i &  &\text{the subdivided graph after $i$ subdivisions.}
\end{align}
From now on, when we refer to $V$, $E$ or the graph $G$ we refer to the subdivided graph,
not general graphs.

An $(\alpha_1,\alpha_n)$-path is usually represented by $P_{\alpha_1,\alpha_n} = \alpha_1,\alpha_2, \ldots, \alpha_n$,
which is a list of vertices $\alpha_i \in V$, including the endpoints, with $i \in \mathbb N$.
A  useful function is the list concatenation operation, denoted by $\doubleplus$, that can also be used to concatenate paths,
which is defined as
\begin{equation}
	[\alpha_1,\ldots, \alpha_n] \doubleplus [\beta_1, \ldots, \beta_m] \coloneqq [\alpha_1,\ldots,\alpha_n,\beta_1,\ldots,\beta_m]\,.
	\label{eq:listConcat}
\end{equation}
% DO NOT REMOVE THIS 

\section{Ring Network}\label{sec:Ring}%
As a warmup, we consider a network of quantum nodes arranged in a ring topology,
where each node only has physical links to two neighbors.
We will show how to create VQLs using the physical links and entanglement swapping,
and present an efficient routing algorithm on the resulting network of VQLs.
The case of the ring contains many of the essential ideas,
but is more accessible due to many simplifications that arise.

We remark that there is two ways to use the routing algorithms below. The first follows the idea of classical routing, in which we send a bit - or in our case - qubit along 
a connection to the next node in the network. Using VQLs, this means we would teleport the qubit from one node to the next. This requires classical control information to be transmitted
to the receiving node. We will not be concerned with how the classical control information is relayed, and indeed it may folow a different network topology than the graph of VQLs.
The second way to look at routing is to use the routing algorithm not to send a qubit directly. Instead, our goal will be to form one big VQL between the sender and receiver using
entanglement swapping at each intermediary node. The sender may then relay the qubit to the receiver by teleportation along the new long distance VQL.
The second perspective makes it clear that routing algorithms on VQLs can indeed be understood as a way to manage the resource of entanglement in a quantum network.
We remark that indeed it is sufficient for each node to send all control information only to final receiver.

We will phrase our routing algorithms with this second perspective in mind. However, it is easy to instead replace the swapping operation by sending the qubit onwards at each node.

\subsection{Definition of the VQL graph}
The ring network is a graph on $N$ vertices, $C_N$, labelled by the elements of $\set{0, \dotsc, N-1}$.
We restrict $N$ to powers of $2$, such that $N=2^n$ for some natural $n$.
We assume there is a physical network $C_N$ so that each node $k$ has as neighbours the nodes $k-1, k+1 \pmod{N}$.
On top of the physical network we add VQLs to reduce the diameter $D(C_N)$ from $O(N)$ to $O(\log N)$.
At the same time we will also limit the size of the quantum memory required for the VQLs
by bounding the degree of all vertices to $O(\log N)$.

We construct a ring network with VQLs similar to \cref{ringRoutingTagged}.
For any $k\in\set{1,\dotsc, n-1}$, we add a cycle of length $2^{n-k}$ through nodes whose labels are divisible by $2^k$.
That is, the VQLs yield the following connections
\begin{equation}
	0 \sim 2^k \sim 2\cdot 2^k \sim 3\cdot 2^k \sim \dotso \sim
	2^{n-k-1}\cdot 2^k \sim 0
\end{equation}
which allow us to fast forward $2^k$ vertices when starting from a vertex whose label is divisible by $2^k$.
In graph theoretic terms, given a graph $C_N$ we construct a low-degree
graph $G_n = (V_n,E_n)$ with the same vertices as $C_N$.
We will assume that a VQL is prepared between any two nodes that correspond to adjacent vertices in the graph $G_n$.
We will show that $D(G_n) = O(\log N) = O(n)$.

\paragraph{Example}
To give an intuitive understanding of why $D(G_n) = O(n) = O(\log N)$ we suppose our network has $2^6$ nodes
and node $0$ wants to communicate with node $37 = 2^5+2^2+2^0$.
Then it could take the path $0, 2^5, 2^5+2^2, 2^5+2^2+2^0$, where the first hop uses a VQL that skips $2^5$ nodes,
the second hop uses a VQL that skips $2^2$ nodes and the last one follows the physical link between two adjacent nodes, which we can also see as a VQL if used to share entanglement.

Following the same idea we see that we can reach any node in at most six hops.
More generally in a network of $N=2^n$ nodes we will be able to find a path of length at most $2n$ between any two nodes by going via node 0.
\\\\
We now give a precise definition of the graph $G_n = (V_n,E_n)$ that we informally described above.
The vertex set is given by $V_n\coloneqq\set{0,\dotsc, N-1}$.
%Before defining the edge set, let us introduce a function $\gcdd : V_n \times V_n \to \mathbb{N}$ defined by
%\begin{align}
%  \gcdd(\alpha,\beta) \coloneqq \begin{cases}	
%			2^n & \text{if $a=b=0$}\\		
%			\max\set{2^k: \text{$k\in\mathbb{N}$ and $2^k$ divides $a$ and $b$}} & \text{otherwise}
%		  \end{cases}
%\end{align}
%As the notation suggests, $\gcdd(\alpha,\beta)$ is the largest power of two that divides both $a$ and $b$. Since, $\gcd(0,0) = 2^n$, strictly speaking, $\gcdd$ is also a function of $n$. However, this special case will not be of  importance to us, so for the sake of simplicity we have chose not to make this dependence explicit.
As stated before, we add a VQL that skips $2^k$ nodes at every node divisible by $2^k$ for some integer $k$.
Let us introduce a function $t: V_n \to \mathbb{N}$ that gives the largest such $k$ for a node:
\begin{align}
  t(\alpha) \coloneqq \begin{cases}
			n & \text{if $\alpha=0$}\\
			\max\set{k: \text{$2^k$ divides $\alpha$}} & \text{otherwise}
		  \end{cases}\,.
\end{align}
In other words, $t(\alpha)$ counts the number of twos in the prime factorization of $\alpha$. 
We also define $\gcdd: V_n \times V_n \to \mathbb{N}$ as
\begin{align}
  \label{eq:q}
  \gcdd(\alpha,\beta) \coloneqq 2^{\min\set{t(\alpha),t(\beta)}}.
\end{align}
As the notation suggests, $\gcdd(\alpha,\beta)$ is the largest power of two that divides both $a$ and $b$.
Since, $\gcd(0,0) = 2^n$, strictly speaking, $\gcdd$ is also a function of $n$. However, this special case will not be of  importance to us, so for the sake of simplicity we have chose not to make this dependence explicit.

With these tools at hand, we define the edge set $E_n$ of $G_n$.
For vertices $\alpha$ and $\beta$ of $G_n$, we have $\{\alpha,\beta\} \in E_n$ if and only if
\begin{equation}
  \abs{\alpha-\beta} \equiv \gcdd(\alpha,\beta) \pmod{2^n}\,.
\label{eq:routedge}
\end{equation}
This formalizes the idea that for any node with label $\alpha$ divisible by $2^k$
we have a VQL that allows us to skip $2^k$ nodes and go directly to $\alpha \pm 2^k \pmod{N}$.

It is not hard to see that $G_n$ contains the graph $C_N$ representing our ring network as a subgraph.
Indeed, for any pair of consecutive network nodes, $\alpha\in\set{0,\dotsc, N-1}$ and $\beta=\alpha+1 \pmod{N}$,
we have $\gcdd(\alpha,\beta)=1$ and thus $\alpha \edge \beta$ in $G_n$.
For an illustration, see a drawing of $G_4$ in \cref{ringRoutingTagged},
where the outer 16-cycle  corresponds to the physical links in the network and the remaining edges
correspond to VQL enabled by pre-shared entanglement.

\subsection{Properties of the Routing Graph}
%First, we observe that the routing graph $G_n$ contains the routing graphs $G_{k}$ for $k\le n$. This observation is useful for proving other properties of the routing graphs as well as for showing the correctness of the routing algorithm we propose in the next section. Second, we show in~\cref{lem:RingDiameter} that the diameter of $G_n$ is logarithmic in the number of vertices $\abs{V_n}$. This is important to us, since the diameter corresponds to the number of entanglement swaps two network nodes need to perform in the worst case in order to communicate to one another.
%Finally, in \cref{lem:nondecreasingmove} we establish that in order to find a shortest path between vertices $\alpha,\beta\in V_n$ one is never required to use edges from the graph $G_{n+k}$ for $k\ge1$. This fact will be useful when analysing the efficient routing protocol which we propose in the next section.

In this section we establish some properties of the graphs $G_n$. 
%Most notably we show that the diameter of $G_n$ is logarithmic in the number of nodes in the corresponding ring network (see~\cref{lem:RingDiameter}).
We start by observing that for all $k\le n$, the routing graph $G_k$ is isomorphic to an induced subgraph
of $G_n$. If we take a second look at \cref{ringRoutingTagged} we see that, up to re-labeling of the
vertices, the graph $G_4$ contains $G_3$ which in turn contains $G_2$. In addition, we can observe
that the re-labeling corresponds to dividing the vertex labels by two. Indeed, the subgraph of $G_4$
induced by its even vertices is isomorphic to $G_3$ via mapping vertex $2\alpha$ of $G_4$ to vertex $\alpha$
of $G_3$. More generally, we have the following:
%
%\todoi{Suggested text: We show that the even vertices of $V_n$ are isomorphic to the $V_{n-1}$,
%by providing a mapping and using \eqref{eq:routedge} to show adjacency.}

\begin{lemma}\label{lem:containment}
For all $n\ge 2$, the subgraph induced by the even vertices of $G_n$ is isomorphic to $G_{n-1}$
via mapping an even vertex $\alpha$ of $G_n$ to the vertex $\alpha/2$ of $G_{n-1}$.
\end{lemma}

\noindent\cref{lem:containment} allows us to bound the diameter of $G_n$ to $O(\log N)$.
The main idea for bounding the diameter is that any odd vertex
of $G_n$ is adjacent to an even vertex which by \cref{lem:containment} belongs to an induced subgraph
isomorphic to $G_{n-1}$.

%\noindent We can now to bound the diameter of $G_n$.
%This is important to us, since the diameter corresponds to the number of entanglement swaps
%two network nodes need to perform in the worst case in order to communicate to one another.
%\todoi{By using \cref{lem:containment} we can show the new vertices are directly connected
%to the isomorphic subgraph such that the overall graph diameter is of size $2n$.}

\begin{lemma}\label{lem:RingDiameter}
	For any $n\in\mathbb{N}$, we have that $D(G_n) \leq D(G_{n-1}) +2$ and $D(G_1)=1$, where $D(G)$ is the diameter of a graph $G$.
	In particular, we have $D(G_n) = O(\log N) = O(n)$.
\end{lemma}

\noindent
We also show that to find a shortest path between two vertices in $G_{n+k}$ that are also contained
in the subgraph isomorphic to $G_n$ we can restrict our attention to the edges in $G_n$. 
This will be useful for proving the optimality of the routing algorithm we propose in the next section.

\begin{lemma} \label{lem:gennobackedge}
For any $n, k \in \mathbb{N}$ and any two vertices $\alpha$ and $\beta$ of $G_n$ we have
$d_{G_n}(\alpha,\beta) = d_{G_{n+k}} (2^k \alpha,2^k \beta)$.
\end{lemma}

\noindent When looking for a shortest $(\alpha,\beta)$-path, \cref{lem:gennobackedge} can help us to narrow
down the choices for the vertex to proceed to after $\alpha$.
Specifically, we show that it suffices to restrict our attention to only two vertices:
\begin{corollary}
Let $\alpha$ and $\beta$ be two distinct vertices of some $G_n$.
If $t(\alpha) \le t(\beta)$ and $\alpha_{\pm} \coloneqq \alpha \pm 2^{t(\alpha)} \pmod{2^n}$, then
\begin{equation}
	d(\alpha_+,\beta)  = d(\alpha,\beta) - 1
	\quad \text{or} \quad
	d(\alpha_-,\beta)  = d(\alpha,\beta) - 1.
\end{equation}
In other words, either $\alpha$ is adjacent to $\beta$ or there exists a shortest $(\alpha,\beta)$-path of the form
\begin{equation}
  \alpha,\alpha_+,\dotsc, \beta \quad \text{or } \quad \alpha,\alpha_-,\dotsc, \beta.
\end{equation}
\label{cor:BestMove}
\end{corollary}

\subsection{Routing via a Shortest Path}
%\steph{There is a difference between finding a shortest path and routing. In routing we usually start only at one end. It would be better to rephrase this also as only 
%starting from one end, or else explain that the sender informs the receiver ahead of time that they want to communicate and establish a path. In routing it is strictly speaking
%not known by the receiver apriori that he'll be the receiver for some sender. It is perfectly natural in this path reservation that the sender first tells the receiver he wants
%to communicate though using classical communication.}
%\todoi{I clarified what we do with the path.}

In this section, we propose a routing algorithm that finds a shortest path between any two vertices of $G_n$ (see~\cref{alg:ring}). Again, we emphasize that this algorithm can also be understood as a procedure to marshall entanglement.
%We have chosen to specify our algorithm as a recursion but it can easily be rewritten as a \texttt{for} loop.
%First, without any additional communication cost, we assume that the sender $\alpha$ alerts the receiver $\beta$ with classical communication that she is trying to route some quantum information to her.
%\todoi{Why is this necessary?}
On a high level, the sender $\alpha$ finds the next node on a shortest path to $\beta$
and transmits the data and the identity of the receiver to this next node.
As before, we remark that transmitting data is equivalent to performing a corresponding entanglement swapping operation on the incoming and outgoing VQL.
This node becomes now the new sender and continues the routing algorithm until the data reaches the final receiver $\beta$. 

Any node can easily find the next node on the optimal path to a given node.
In fact, in our warmup example of the ring, due to the structure and the labeling, any node can find
the entire shortest path to a given node without having to store the network in her memory.
We provide below an algorithm to find the entire shortest path,
though we only use it (somewhat redundantly) to find the next node on a shortest path.
We present our routing algorithm this way in order to stress that it is local,
and it can easily be adapted to the case where nodes in the ring are denser in some regions.

The overall idea for finding a shortest path between the two vertices $\alpha$ and $\beta$ of $G_n$ is to proceed from both ends simultaneously by taking edges that lead to vertices in graphs $G_k$ with $k<n$
until a common vertex is reached. It is possible to calculate the optimal next nodes $\gamma_\pm$ given any node $\gamma$ by calculating $\gamma \pm 2^{t(\gamma)}$.
This is the only operation that is required in the algorithm.
Since the label of a receiver $\beta$ is known to any sender $\alpha$,
it can also locally calculate $\beta_\pm$ for any $\beta$ locally.
Thus, the routing can also be performed locally.

Every edge we take leads us to a vertex strictly lower in the hierarchy,
so we reach a vertex from $G_1$ in at most $n = O(\log N)$ steps which would yield a path of length at most $2n$.
In order to ensure that a shortest path is found, we augment the partially constructed path using the subroutine $\mathtt{bestMove}$ (see \cref{alg:BestMove}), where
we choose to take the edge that leads us to the vertex from the graph $G_k$ with the smallest possible $k$.
We provide a formal and complete argument in the proof of \cref{thm:RingOptimal}.

\begin{theorem}
For any $n\ge 1$ and vertices  $a$ and $b$ of $G_n$, the algorithm $\mathtt{path}(\alpha,\beta)$ returns a shortest $(\alpha,\beta)$-path.
\label{thm:RingOptimal}
\end{theorem}

\noindent A complete overview of routing data from a node $\alpha$ to node $\beta$ is given by the pseudocode \cref{alg:completeRingRouting}.
A shortest path is computed at every step by \cref{alg:ring} and the first element of this path is the next node in routing.
To calculate whether two nodes are close, we use a subroutine $\mathtt{path2}(\alpha,\beta)$
(see \cref{alg:path2}) that returns a path if $d(\alpha,\beta) \leq 2$.

\begin{algorithm}
	\LinesNumbered
	\DontPrintSemicolon
	\SetKwInOut{Data}{Data}
	\SetKwInOut{Input}{Input}\SetKwInOut{Output}{Output}
	\SetKwFunction{path}{path}
	\SetKwFunction{send}{send}
	\SetKwFunction{swap}{swap}
	\Data{%
		The label, $\alpha$, of the node itself;
		a \send{$\gamma, (\eta, \beta)$} procedure that sends a tuple of node identifiers $\eta, \beta\in V$ to the node $\gamma\in V$;
		and a procedure \swap{$\gamma,\eta$}
		that swaps entanglement shared with nodes $\gamma \in V$ and $\eta \in V$, and sends appropriate classical correction information to the receiver $\beta$.
	}
	\Input{$(\eta, \beta) \in V\times V$, where $\eta$ is the previous sender, and $\beta$ the final destination.}
	\SetKwProg{ringRoute}{Procedure}{ is}{end}
	\tcp{Run on reception of a request.}
	\ringRoute{$\mathrm{ringRoute((\eta,\beta))}$}{%
		\eIf{$\alpha = \beta$}{%
			\tcp{Destination reached.}
		}{%
			$P_{\alpha,\beta} \gets$ \path{$\alpha, \beta$} \tcp*{Call \cref{alg:ring}}
			$\gamma \gets P_{\alpha,\beta}[1]$ \tcp*{The next node is the second element of the path $P_{\alpha,\beta}$}
			\send{$\gamma, (\alpha,\beta)$} \tcp*{Forward the request to $\gamma$.}
			\tcp*{Unless we are the original sender, we now perform a swap.}
			\If{$\eta$ is not $\perp$}{% 
				\swap{$\eta, \gamma$} \tcp*{Perform an entanglement swapping operation(see \cref{fig:entanglementSwap}).}
			}
		}
	}
	\caption{%
		Routing from node $\alpha$ to node $\beta$ on the ring. The original sender $\alpha$ first executes the procedure \texttt{ringRoute} with inputs $(\perp,\beta)$, 
		and passes on the request to the next node. 
		All subsequent nodes $\alpha\in V$ execute the procedure \texttt{ringRoute} on reception of a request, which contains 
		the identifier of the requesting node $\eta$ and the final destination $\beta$. 
		If the destination of the request, $\beta \in V$, has not been reached,
		then $\alpha$ will calculate the next node on a shortest path to $\beta$, and perform a swapping operation.
	}
	\label{alg:completeRingRouting}
\end{algorithm}

\begin{algorithm}
	\LinesNumbered
	\DontPrintSemicolon
	\SetKwInOut{Input}{Input}\SetKwInOut{Output}{Output}
	\Input{vertices $\alpha,\beta\in \set{0,\dotsc,2^n-1}$ of $G_n$}
	\Output{A shortest path $[\alpha, \ldots, \beta]$ between vertices $\alpha$ and $\beta$}

	\SetKwProg{path}{Function}{ is}{end}
	\path{$\mathrm{path}(\alpha,\beta)$}{
		\uIf(\tcp*[f]{$\mathrm{path2}$ finds a path if $d(\alpha,\beta) \leq 2$ (see \cref{alg:path2}).}){%
			$\mathrm{path2}(\alpha,\beta) \neq \emptyset$
		}{%
			\Return $\mathrm{path2}(\alpha,\beta)$\;
		}
		\uElse(\tcp*[f]{Recursively call \emph{path} to find a shortest subpath from the best neighbour.}){%
			\eIf{$t(\alpha) \le t(\beta) $}{
				\Return $\alpha \doubleplus
						\mathrm{path}\big(\mathrm{bestMove(\alpha)},\beta\big)$\;
			}{%
				\Return $\mathrm{path}
					\big(\alpha, \mathrm{bestMove}(\beta)\big) \doubleplus \beta$\;
			}
        }
    }
\caption{Algorithm for finding a shortest path between two vertices of the graph $G_n$.}
\label{alg:ring}
\end{algorithm}

\begin{algorithm}
	\LinesNumbered
	\DontPrintSemicolon
	\SetKwInOut{Input}{Input}\SetKwInOut{Output}{Output}
	\Input{vertices $\alpha,\beta\in \set{0,\dotsc,2^n-1}$ of $G_n$}
	\Output{A shortest path $[\alpha, \ldots, \beta]$ of length at most 2 between vertices $\alpha$ and $\beta$.
		Otherwise, the empty result, $\bot$.}

	\SetKwProg{path}{Function}{ is}{end}
	\path{$\mathrm{path2}(\alpha,\beta)$}{
		\eIf(\tcp*[f]{Using \cref{eq:routedge}}){$\alpha \edge \beta$}{%
			\Return $[\alpha, \beta]$\;
		}{%
			\tcp{Use \cref{cor:BestMove} to check for 2\textsuperscript{nd} degree adjacency.}
			$\gamma \gets \argmin \set{t(\alpha), t(\beta)}$\;
			$\delta \in \set{\alpha,\beta} \setminus \set{\gamma}$\;
			$\gamma_\pm \gets \gamma \pm 2^{t(\gamma)}$\;
			\uIf{$\gamma_+ \edge \delta$}{%
				\Return $[\alpha, \gamma_+, \beta]$\;
			}
			\uElseIf{$\gamma_- \edge \delta$}{%
				\Return $[\alpha, \gamma_-, \beta]$\;
			}
			\Else{%
				\Return $\bot$ \tcp*{Return the empty result if $d(\alpha,\beta) > 2$.}
			}
		}
    }
\caption{%
	An $O(1)$ algorithm for finding a path of length at most $2$ between two vertices of the graph $G_n$.
	Note that the presence of an edge can be checked with \cref{eq:routedge}.
}
\label{alg:path2}
\end{algorithm}

\begin{algorithm}
	\LinesNumbered
	\DontPrintSemicolon
	\SetKwInOut{Input}{Input}\SetKwInOut{Output}{Output}
	\Input{vertex $\alpha\in\set{0,\dotsc,2^n-1}$ of $G_n$ with $t(\alpha)\le t(\beta)$ and $d(\alpha,\beta) > 2$.}
	\Output{%
		A neighbor $\gamma\in V$ of $\alpha$ that is guaranteed to be on a shortest path to any $\beta$.
	}

	\SetKwProg{bestMove}{Function}{ is}{end}
	\bestMove{$\mathrm{bestMove}(\alpha)$}{
		$\alpha_+ \gets v + 2^{t(\alpha)}$\;
		$\alpha_- \gets v - 2^{t(\alpha)}$\;
		\uIf{ $t(\alpha_+) > t(\alpha_-)$}{
			\Return $\alpha_+$}
		\uElseIf{$t(\alpha_+) < t(\alpha_-)$}{
				\Return $\alpha_-$\;
			}
		\Else{\Return $\mathrm{RandomChoice}\set{\alpha_+,\alpha_-}$}
    }
	\caption{An $O(1)$ algorithm for incrementing the path by one vertex.}\label{alg:BestMove}
\end{algorithm}

%\begin{algorithm}
%\SetAlgorithmName{Protocol}{protocol}{List of Protocols}
%	\LinesNumbered
%	\DontPrintSemicolon
%	\SetKwInOut{Input}{Input}\SetKwInOut{Output}{Output}
%	\Input{$n\ge 1; \alpha,\beta \in \set{0,\dotsc,2^n-1}$ }
%	\Output{A shortest path from $a$ to $b$}
%
%	\todoi{Old algorithm for comparison.}
%    $a_1 \gets a$\;
%	$b_1 \gets b$\;
%	$A,B \gets [\;]$\;
%	\For{ $i =1$ to $n$}{
%		$A$.append($a_i$)\;
%		$B$.prepend($b_i$)\;
%		\uIf{ $\set{a_i,b_i}\in E_n$}{
%			\Return $A$.append($B$)
%		}
%		\uElseIf{$a_i~\text{\upshape and}~b_i~\text{\upshape have a common neighbour}$}{
%			$A$.append(common-neighbour($a_i$,$b_i$))\;
%			\Return $A$.append($B$)\;
%		}
%		\Else{
%			\label{ln:ifa}\eIf {$(a_i\mod 2^i) = 0$} {
%				$a_{i+1} \gets a_i$
%			}{
%				$a_{i+1} \gets\argmin\set{p(x): x=a_i\pm 2^{i-1}}$
%			}
%
%			\eIf {$(b_i \mod 2^i) = 0$} {
%				$b_{i+1} \gets b_i$
%			}{
%				$b_{i+1} \gets\argmin\set{p(x): x=b_i\pm 2^{i-1}}$
%			}
%		}
%	}
%	return $A$.append($B$)
%\caption{\pname{RinglogRoute}}
%\label{ptr:logrout}
%\end{algorithm}

\section{Sphere Network}\label{sec:sphere}
Let us now consider the case of the sphere. As before we describe how to design the VQLs and an efficient routing algorithm on the graph of VQLs.
For now, we will simply assume VQLs have been created and there is a background process to replenish them.
%In \cref{sec:replenishing} we consider recreating VQLs.
As in the case of the ring, we also note that sending qubits is equivalent to performing entanglement swapping operations connecting the incoming and outgoing VQLs at a network node. 

We start by introducing the relevant graph of VQLs (\cref{sec:graphProperties})
that motivates our routing algorithm (\cref{sec:routing}).
We first give a routing algorithm with global information based on the graph properties (\cref{sec:globalRouting}).
Then, we devise a labelling scheme (\cref{sec:labelling})
and describe a local routing algorithm (\cref{sec:localRoutingAlgorithm}).
Finally, we perform an analysis of the space and time resources needed by our local routing algorithm (\cref{sec:complexityAnalysis}).
All statements in this section are later formally proved in \cref{sec:appendixSphere}.

\subsection{Definition of the VQL graph}\label{sec:networkModel}

%Furthermore, we give some theorems that are proved in the Appendix (\cref{sec:networkModelTechnical}

\subsubsection{Approximating a Sphere}
\begin{figure}[tb]
	\begin{subfigure}[b]{0.33\textwidth}
		\centering
		\includegraphics[width=0.5\textwidth]{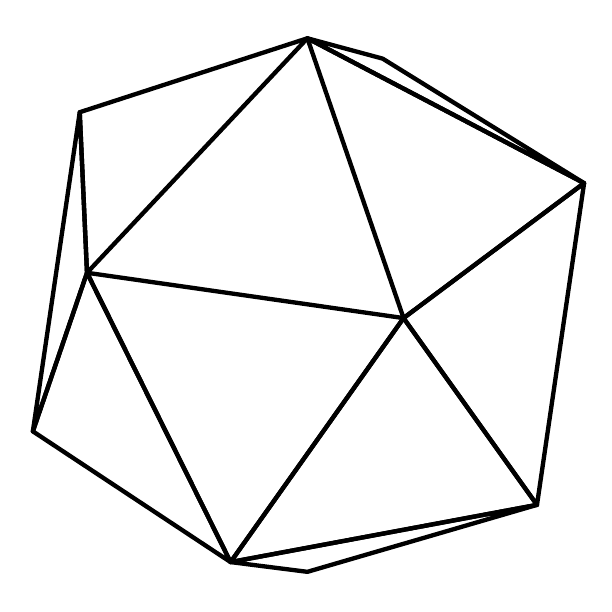}
		\caption{The base icosahedron}\label{fig:icoSub0}
	\end{subfigure}
	\begin{subfigure}[b]{0.33\textwidth}
		\centering
		\includegraphics[width=0.5\textwidth]{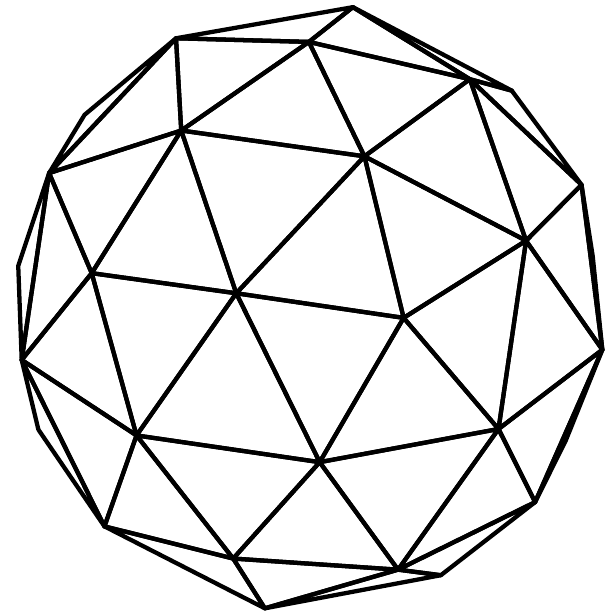}
		\caption{One subdivision}\label{fig:icoSub1}
	\end{subfigure}
	\begin{subfigure}[b]{0.33\textwidth}
		\centering
		\includegraphics[width=0.5\textwidth]{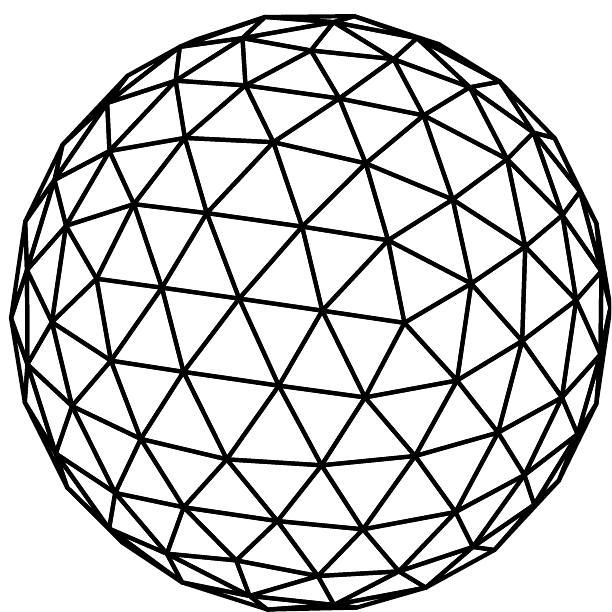}
		\caption{Two subdivisions}\label{fig:icoSub2}
	\end{subfigure}
	\caption{%
		Approximating the sphere by subdividing edges of the icosahedron.
		By iteratively performing this procedure, the resulting graph closer and closer approximates a sphere.
		Each edge represents a entanglement shared by vertices in the network.
		We use edges of previous subdivisions as VQL shortcuts in the network,
		e.g.\ \cref{fig:icoSub1} also contains the edges in \cref{fig:icoSub0}
		as long distance VQLs. 
		Entanglement is established through physical connections
		on the most fine-grained distribution of edges in the network.
	}
	\label{fig:icosahedronSubdivision}
\end{figure}

As a base graph we look at three-dimensional polyhedra with a uniform distribution, i.e. the five platonic solids.
We will add vertices to a platonic solid in such a way that the resulting graph better approximates a sphere.
Evidently, in our problem the nodes are a given, but it will be convenient to imagine that indeed we grow the graph by approximating the sphere
until all existing nodes can be accommodated.
For prior work we look at 3D modelling, where it is a common problem to approximate
smooth surfaces from few polygons.
A standard algorithm in 3D modelling for subdividing polyhedra is Loop subdivision~\cite{loop1987smooth}.
This algorithm requires triangular polygons, which leave only the
tetrahedron, octahedron and icosahedron to choose from.
Loop subdivision performs approximation iterations by what is called edge subdivision.
In each subdivision, a vertex is placed on the middle of edges and new nodes are connected to their nearby neighbours.
An example of Loop subdivision on the icosahedron is shown in \cref{fig:icosahedronSubdivision},
which illustrates how we can subdivide an icosahedron to approximate a sphere.
The platonic solid which when subdivided resembles a sphere the most, is the icosahedron.
We thus start with a graph representation of the icosahedron,
which is scaled so that all $12$ vertices are placed on the sphere.

Loop subdivision provides us with a structure for assigning long-distance entangled links.
Let us designate the icosahedron as the initial graph $G'_0$,
and subdividing $G'_i$ results in the graph $G'_{i+1}$.
If the final graph is denoted by $G'_k = (V_k, E_k)$,
then $|V_k|$ coincides with the number of network nodes, which are uniformly distributed over the sphere.
Furthermore, $E_k$ will coincide with the physical links in the network.
We will add all the edges of the previous subdivision graphs $G'_i$ for $i <k$ to function as VQLs
in the network and define our VQL network as
\begin{equation}
	G_k = (V_k, \bigcup_{i=0}^k E_i)\,.
\end{equation}
Note that this is in contrast with the ring network, where $E_i$ includes all $E_h$ for $h \in [i]$.
These VQLs connect nodes that do not have a direct physical link and can be created with entanglement swapping.
The pseudocode for the subdivision algorithm is given in \cref{alg:subdivisionIcosahedron}.

\begin{algorithm}[tb]
	\LinesNumbered
	\DontPrintSemicolon
	\SetKwInOut{Data}{Data}
	\SetKwInOut{Input}{Input}
	\SetKwInOut{Output}{Output}
	\Data{$G_0 = (V_0, E_0)$ the icosahedron.}
	\Input{$k \in \mathbb N$, the number of subdivisions.}
	\Output{$G = (V,E)$, the subdivided graph.}
	\SetKwProg{Fn}{Function}{ is}{end}
	\Fn{$\mathrm{subdivide}(k)$}{
		\For{$i \gets 0$ \KwTo $k-1$}{
			\tcp{Generate the nodes}
			$L_{i+1} \gets \emptyset$\;
			%$p \gets$ empty Map $: V \to E$ \tcp*{A mapping of a vertex to its edge}
			\For{$e \in E_i$}{\label{alg:subdivisionMidpoint}
				$\alpha \gets$ midpoint of $e$ on sphere.\;
				$L_{i+1} \gets L_{i+1} \cup \{\alpha\}$\;
				%$p(\alpha) \gets e$ \tcp*{Remember which edge created $\alpha$}
			}
			\tcp{Generate the edges}
			$E_{i+1} \gets \emptyset$\;
			\For{$\alpha \in L_{i+1}$}{
				%$(e= \{\beta_1,\beta_2\}) \gets p(\alpha)$\;
				$\{\beta_1, \beta_2\} \gets p(\alpha)$ \tcp*{$p(\alpha)$ is later defined in \cref{eq:directParent}}
				$E_{i+1} \gets E_{i+1} \cup \{ \{\alpha,\beta_1\}, \{\alpha,\beta_2\}\}$ \tcp*{Create edges to vertices in $V_i$}\label{alg:subdivisionParents}
				\;
				\tcp{The vertices $\beta_3,\beta_4$ that form triangles with $\beta_1$ and $\beta_2$}
				$\{\beta_3, \beta_4\} = \{ \beta \in N(\beta_1) \cap N(\beta_2) : \ly(\beta_1) = \ly(\beta_2) \lor \ly(\beta) = i\}$\;
				%$\mathbb E \gets$ edges in the triangles $\{\beta_1,\beta_2,\beta_3\}, \{\beta_1,\beta_2,\beta_4\}$\;
				$\mathbb E \gets \{ \{\beta_1, \beta_3\}, \{\beta_1, \beta_4\}, \{\beta_2, \beta_3\}, \{\beta_2, \beta_4\} \}$\;
				$\mathbb V \gets \{\gamma \in L_{i+1} : p(\gamma) \in \mathbb E\}$ \tcp*{The adjacent vertices in $L_{i+1}$}
				$E_{i+1} \gets E_{i+1} \cup
				\left\{ \{\alpha,\gamma\} : \gamma \in \mathbb V \right\}$
				\tcp*{Create edges to vertices in $L_{i+1}$}\label{alg:subdivisionSiblings}
			}
			$V_{i+1} \gets V_i \cup L_{i+1}$
		}
		\Return $G_k = (V_k, \bigcup_{i=0}^k E_i)$\;
	}
	\caption{Subdivision algorithm of the icosahedron.
		A vertex is placed on each edge in $E_i$ and then connected to nearby vertices.
		The new vertices are grouped in $L_{i+1}$, and all edges in $E_{i+1}$.
		The cumulative set of vertices created after $i$ iterations is $V_i$,
		so that the graph after $i$ subdivisions is $G_i = (V_i, E_i)$.
	}
	\label{alg:subdivisionIcosahedron}
\end{algorithm}

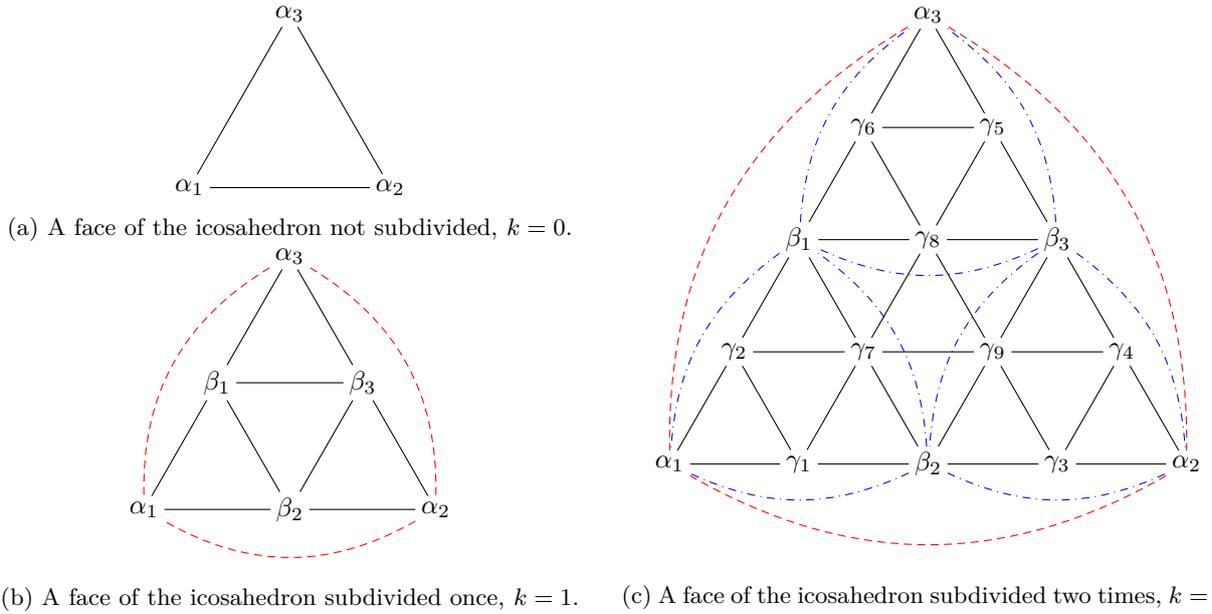
\begin{figure}[tb]
	\centering
	\begin{tabular}{c}
	\begin{subfigure}{0.48\textwidth}
		\centering
		\begin{tikzpicture}[scale=0.33]
			% Alpha
			\node (a1) at (0,0) {$\alpha_1$};
			\node (a2) at (8,0) {$\alpha_2$}
				edge (a1);
			\node (a3) at (4,7) {$\alpha_3$}
				edge (a1)
				edge (a2);
		\end{tikzpicture}
		\caption{A face of the icosahedron not subdivided, $k=0$.}
	\end{subfigure}
	\\
	\begin{subfigure}{0.48\textwidth}
		\centering
		\begin{tikzpicture}[scale=0.48]
			% Alpha
			\node (a1) at (0,0) {$\alpha_1$};
			\node (a2) at (8,0) {$\alpha_2$}
				edge [bend left, red, densely dashed] (a1);
			\node (a3) at (4,7) {$\alpha_3$}
				edge [bend right, red, densely dashed] (a1)
				edge [bend left, red, densely dashed] (a2);

			% Beta
			\node (b1) at (2,3.5)	{$\beta_1$}
				edge (a1)
				edge (a3);
			\node (b2) at (4,0)		{$\beta_2$}
				edge (a1)
				edge (a2)
				edge (b1);
			\node (b3) at (6,3.5)	{$\beta_3$}
				edge (a2)
				edge (a3)
				edge (b2)
				edge (b1);
		\end{tikzpicture}
		\caption{A face of the icosahedron subdivided once, $k=1$.}
		\label{fig:subdividedOnce}
	\end{subfigure}
	\end{tabular}
	\begin{subfigure}{0.48\textwidth}
		\centering
		\begin{tikzpicture}[scale=0.85]
			% Alpha
			\node (a1) at (0,0) {$\alpha_1$};
			\node (a2) at (8,0) {$\alpha_2$}
				edge [bend left, red, densely dashed] (a1);
			\node (a3) at (4,7) {$\alpha_3$}
				edge [bend right, red, densely dashed] (a1)
				edge [bend left, red, densely dashed] (a2);

			% Beta
			\node (b1) at (2,3.5)	{$\beta_1$}
				edge [bend right=25, blue, dashdotted] (a1)
				edge [bend left=25, blue, dashdotted] (a3);
			\node (b2) at (4,0)		{$\beta_2$}
				edge [bend left=25, blue, dashdotted] (a1)
				edge [bend right=25, blue, dashdotted] (a2)
				edge [bend right=25, blue, dashdotted] (b1);
			\node (b3) at (6,3.5)	{$\beta_3$}
				edge [bend left=25, blue, dashdotted] (a2)
				edge [bend right=25, blue, dashdotted] (a3)
				edge [bend right=25, blue, dashdotted] (b2)
				edge [bend left=25, blue,dashdotted] (b1);

			% Gamma
			\node (g1) at (2,0)		{$\gamma_1$}
				edge (a1)
				edge (b2);
			\node (g2) at (1,1.75)	{$\gamma_2$}
				edge (a1)
				edge (b1)
				edge (g1);
			\node (g3) at (6,0)		{$\gamma_3$}
				edge (a2)
				edge (b2);
			\node (g4) at (7,1.75)	{$\gamma_4$}
				edge (a2)
				edge (b3)
				edge (g3);
			\node (g5) at (5,5.25)	{$\gamma_5$}
				edge (a3)
				edge (b3);
			\node (g6) at (3,5.25)	{$\gamma_6$}
				edge (a3)
				edge (b1)
				edge (g5);
			\node (g7) at (3,1.75)	{$\gamma_7$}
				edge (b1)
				edge (b2)
				edge (g1)
				edge (g2);
			\node (g8) at (4,3.5)	{$\gamma_8$}
				edge (b1)
				edge (b3)
				edge (g5)
				edge (g6)
				edge (g7);
			\node (g9) at (5,1.75)	{$\gamma_9$}
				edge (b2)
				edge (b3)
				edge (g3)
				edge (g4)
				edge (g7)
				edge (g8);
		\end{tikzpicture}
		\caption{A face of the icosahedron subdivided two times, $k=2$.}
		\label{fig:subdividedTwice}
	\end{subfigure}

	\caption{Example steps in the subdivision algorithm on one face of the icosahedron,
		for given subdivide$(k)$.
		In the first subdivision, $\beta_i$ are placed on edges $\{\alpha_i, \alpha_j\} \in E_0$,
		and connected to other nearby $\beta_i$.
		The edges between $\alpha_i$ are kept in the graph (red dashed).
		The second iteration better shows which edges are added on a subdivision.
		For example, $\gamma_8$ is connected to all $4$ nearby $\gamma_i$
		and the vertices $\beta_1$ and $\beta_3$ on the edge that was subdivided into $\gamma_8$
		(black solid).
		Again, the edges of the previous graph are kept (red, dash \& blue, dash dot) in the graph.
		The colored and dashed and/or dotted edges are long-distance connections that reduce the graph diameter.
		The physical connections are those edges that are on the last layer.
		So if $k=2$ is the final subdivision, then the black solid edges in \cref{fig:subdividedTwice} are the physical connections.
	}
	\label{fig:faceIcosahedron}
\end{figure}

\paragraph{Example}
We have illustrated a face of the graph $G_k$ for $k \in \{0,1,2\}$ in \cref{fig:faceIcosahedron}. 
The physical links between nodes are contained in the edges that were generated last,
the solid black edges.
Entanglement can be established between two adjacent nodes using the physical link.
Moreover, all other dotted edges represent a VQL, i.e. an entangled pair between the endpoints,
where we have created the long distance entanglement by using entanglement swapping.
For example, to create an edge $\{\alpha_1,\alpha_3\} \in E_1$ in \cref{fig:subdividedOnce}
we perform an entanglement swap in $\beta_1 \in V_1$ on
$\{\alpha_1, \beta_1\}, \{\beta_1, \alpha_3\} \in E_1$ that were created using physical links.
%If purification of the entanglement is successful, perfect entanglement is established between $\alpha_1$ and $\alpha_3$.
%If it is not, the process is restarted to try again.
%Once successful, the nodes $\alpha_1$ and $\alpha_3$ can communicate directly using this entangled pair,
%as if they were adjacent (disregarding classical communication time).
Performing entanglement swapping uses up the entanglement $\alpha_1 \edge \beta_1$ and $\beta_1 \edge \alpha_3$ though,
so edges between these nodes will have to be replenished.
A similar procedure is also followed by $\beta_2$ and $\beta_3$ in $V_1$ to create $\{\alpha_1, \alpha_2\}$ and $\{\alpha_2, \alpha_3\}$ in $E_1$ respectively.
For more subdivisions, such as in \cref{fig:subdividedTwice},
we perform the above procedure 3 times to create the edge $\{\alpha_1, \alpha_3\} \in E_2$.
First by $\gamma_2$ and $\gamma_6$ in $V_2$ to create
$\{\alpha_1, \beta_1\}, \{\beta_1, \alpha_3\} \in E_2$,
then by $\beta_1 \in V_2$ to create $\{\alpha_1, \alpha_3\} \in E_2$.

In addition to the notation defined in
\cref{eq:allVertices,eq:verticesKsubdivisions}
we will use the following notation:
\begin{align}
	L_i &= V_i \setminus V_{i-1} & \text{the vertices generated in the $i$-th subdivision,} \label{eq:layerVertices}\\
	E_i &\subseteq E & \text{only the edges generated in the $i$-th subdivision,} \label{eq:edgesKsubdivisions}\\
	G_i &= (V_i, \bigcup_{i=0}^k E_i) & \text{the subdivided sphere graph after $i$ subdivisions.}
\end{align}
A useful term is that of a \emph{layer}, which may be compared to the layers in an onion.
For each subdivision another layer is added to the graph.
Layer $i$ refers to all vertices and edges in $L_i$ and $E_i$.
When we refer to going up layers, that means that we are increasing $i$ for $L_i$ and $E_i$.
Conversely, going down the layers means decreasing $i$.

\subsubsection{Properties of the Routing Graph}\label{sec:graphProperties}
The subdivided icosahedron must meet two requirements.
First, the graph must have a small diameter,
and second, it must have a small degree for each vertex.
A small diameter is desired to reduce the number of entanglement swaps necessary to establish a communication.
If two vertices are not adjacent then entanglement swaps have to be performed until they are,
only then can they communicate securely.
We will show that the graph diameter scales logarithmically in the number of nodes,
i.e.\ an exponential improvement if we were to use only the physical links.
\begin{proposition}[Graph Diameter]\label{prop:problemDiameter}
	The diameter of the subdivided graph $D(G_k)$, where $k \in \mathbb N_0$, is
	\[
		D(G_k) \leq 2k+3 =  \log_2\left(\frac{N-2}{10}\right) + 3\,.
	\]
\end{proposition}
Furthermore, we upper bound the degree of each node.
For every edge connected to a node, some quantum bit must be stored in a quantum memory.
These quantum memories are currently limited in size,
and for practical applications must be kept small.
We show that the degree of every node scales logarithmically in the total number of nodes.
\begin{proposition}[Vertex Degree]\label{prop:3vertexDegree}
	The degree of every vertex $v\in V$ in the subdivided graph as a function of $N$ is upper bounded by
	\[
		\deg(v) \leq 3\log_2\left(\frac{N-2}{10}\right) + 5.
	\]
\end{proposition}

\subsubsection{Shortest Path Structure}\label{sec:shortestPathStructure}
To find a routing algorithm we take a look at the structure that is present in the network.
An important property of every vertex, except those in the base icosahedron,
is that it has been generated from an edge in the subdivision algorithm.
This edge has two endpoints, which in turn are also generated from edges, etc.
The relation structure that this assumes is tree-like, but may contain splitting and joining branches.
Because two vertices generate one vertex on subdivision, we call them \emph{parents} of the \emph{child} vertex.
The parents of the parents are then, of course, grandparents.

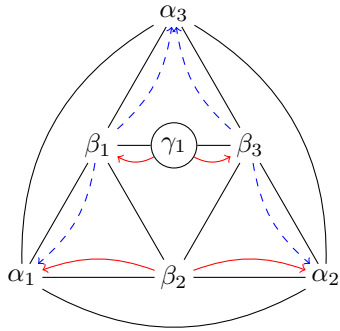
\begin{figure}[tb]
	\centering
	\begin{tikzpicture}[scale=0.5]
		% Alpha
		\node (a1) at (0,0) {$\alpha_1$};
		\node (a2) at (8,0) {$\alpha_2$}
			edge[bend left] (a1);
		\node (a3) at (4,7) {$\alpha_3$}
			edge[bend right] (a1)
			edge[bend left] (a2);

		% Beta
		\node (b1) at (2,3.5)	{$\beta_1$}
			edge (a1)
			edge (a3)
			edge[->,  bend right=20, blue, dashed] (a3)
			edge[->, bend left=20, blue, dashed] (a1);
		\node (b2) at (4,0)		{$\beta_2$}
			edge (a1)
			edge (a2)
			edge (b1)
			edge[->, bend right=20, red] (a1)
			edge[->,  bend left=20, red] (a2);
		\node (b3) at (6,3.5)	{$\beta_3$}
			edge (a2)
			edge (a3)
			edge (b2)
			edge[->,  bend left=20, blue, dashed] (a3)
			edge[->, bend right=20, blue, dashed] (a2);

		% Gamma
		\node[draw, circle] (g1) at ($(b1)!0.5!(b3)$) {$\gamma_1$}
			edge (b1)
			edge (b3)
			edge[->, bend left, red] (b1)
			edge[->, bend right, red] (b3);
	\end{tikzpicture}
	\caption{
		The notions of \emph{children} and \emph{parents} allows us to easily express
		relations between vertices over multiple iterations of the subdivision algorithm.
		A child vertex is a vertex generated on an edge between two vertices (parents)
		in the subdivision algorithm.
		The parents of $\beta_2$ are $p(\beta_2) = \{\alpha_1, \alpha_2\}$,
		and for $\gamma_1$ are $p(\gamma_1) = \{\beta_1, \beta_2 \}$ (red, solid arrow).
		The grandparents of $\gamma_1$ are $\{\alpha_1,\alpha_2, \alpha_3\}$ (blue, dashed arrow).
		There are no grandparents for $\beta_2$ because its parent $\alpha_1 \in V_0$.
	}
	\label{fig:childParent}
\end{figure}

We precisely define the terminology used with the subdivided graph.
The layer function maps vertices to their layer number
\begin{gather}
	\ly \colon V \to \mathbb N\\
	\ly(\alpha) = k : \alpha \in L_k\,. \label{eq:layer}
\end{gather}
And the parent function gives the vertices that are the endpoints of the edge that generated $\alpha$
if $\ly(\alpha) > 0$, otherwise it is undefined:
\begin{gather}
	p \colon L_i \text{ with } i>0 \to V \times V\\
	p(\alpha) = \{ \beta : \beta \in N(\alpha), \ly(\beta) < \ly(\alpha) \}\,.  \label{eq:directParent}
\end{gather}

\paragraph{Example}
To illustrate what the parents are, we give an example using \cref{fig:childParent}.
The parents of $\beta_2$ are
\begin{equation*}
	p(\beta_2) = \{\alpha_1, \alpha_2\}\,,
\end{equation*}
and of $\gamma_1$ they are
\begin{equation*}
	p(\gamma_1) = \{\beta_1, \beta_3\}.
\end{equation*}
Since all $\alpha_i$ are on layer zero, they do not have any parents,
so $\beta_2$ does not have grandparents, but $\gamma_1$ does have grandparents
\begin{equation*}
	\bigcup_{\pi \in p(\gamma_1)} p(\pi) = \left\{\alpha_1, \alpha_2, \alpha_3\right\}\,.
\end{equation*}

It turns out that vertices on the same layer that are adjacent must share a parent,
which we call the \emph{common parent}.
We use the notation $\pi_{\alpha,\beta}$ to signify a common parent of any $\alpha,\beta \in V$ that are connected.
The name $\pi$ is intended as a mnemonic for \textbf{p}arent.
\begin{proposition}[Common Parent]
	Consider two vertices $\alpha_1, \alpha_2 \in L_k$ so that $\alpha_1 \edge \alpha_2$, for $k \in \mathbb N_1$.
	Then $\alpha_1$ and $\alpha_2$ have a unique common parent $\pi_{\alpha_1,\alpha_2}$.
\end{proposition}
The sphere graph also has the property that when routing between two vertices,
it is never shorter to route through edges on a higher layer than either vertex.
This means that when routing between two vertices on a layer lower than $k$,
we may ignore any edges $E_h$ and vertices $L_h$ with $h>k$.
This greatly reduces the number of options when routing between vertices
if the layer of the destination is known.
\begin{theorem}[No Higher Edge]
	If $m$ is the distance between vertices $\alpha_0,\alpha_m \in V$,
	and $k = \max(\ly(\alpha_0),\ly(\alpha_m))$,
	then for all shortest paths between $\alpha_0$ and $\alpha_m$ of length $m$ it holds that they do not use an edge in $E_h$, where $h > k$.
\end{theorem}
Now that we know higher layers are not useful for routing,
we can start analysing what happens if two vertices are far apart.
We show that when a vertex $\alpha_0$ routes to a vertex on a lower layer, $\alpha_m$,
then there always exists a shortest path through a parent of $\alpha_0$.
\begin{lemma}[Lower Layer Path]\label{lem:3LowerLayerPath}
	Consider vertices
	\begin{equation}	
		\alpha_0, \alpha_m \in V : \ly(\alpha_1) < \ly(\alpha_m)\,.
	\end{equation}
	with $d(\alpha_0,\alpha_m) = m$.
	Then there exists a shortest $(\alpha_0,\alpha_m)$-path also of length $m$
	that contains only vertices $\beta \in V_h, h < \ly(\alpha_0)$
	between $\alpha_0$ and $\alpha_m$, except $\alpha_0$.
\end{lemma}
Additionally, when two vertices on the same layer $\alpha_0,\alpha_m \in L_k$
are at least 3 distance apart $d(\alpha_0,\alpha_m) \geq 3$,
then there is a shortest path between them that uses only lower layer vertices.
So when we can guarantee that two vertices are some distance apart,
we may assume that there are some parents $\pi_0 \in p(\alpha_0)$ and $\pi_m \in p(\alpha_m)$,
that have $d(\pi_0,\pi_m) = d(\alpha,\beta) -2$.
\begin{theorem}[Three Hops]
	\label{thm:3threeHops}
	Consider a shortest path $P_{\alpha_0,\alpha_{m}}$ of length $m \geq 3$
	between two vertices on the same layer $\alpha_0, \alpha_{m} \in L_k, k>0$.
	Then there exists a path also of length $m$ that contains only vertices in
	$V_h, h < k$ between $\alpha_0$ and $\alpha_{m}$,
	except for $\alpha_0$ and $\alpha_{m}$.
\end{theorem}
This hints at the presence of a routing algorithm which routes through the parents
when it knows that $\alpha_0$ and $\alpha_m$ are not on the same layer,
or when they are at least at a distance of three.
However, vertices always have two parents,
so it remains to see which parent should be chosen.

\paragraph{Example}
To illustrate how the common parent works, we give a small example using~\cref{fig:childParent} once more.
From the figure it is possible to see that the common parent of $\beta_1$ and $\beta_2$ is
\[
	\pi_{\beta_1, \beta_2} = \alpha_1\,,
\]
because both share that same parent.
Another example is $\pi_{\beta_2, \beta_3} = \alpha_3$.
However, $\alpha_1$ and $\alpha_2$ do not share a common parent even though they are connected,
since they are on the base layer.
And $\gamma_1$ and $\beta_2$ also do not share a common parent,
because they are not connected.
While $\gamma_1$ and $\beta_2$ do share a common ancestor ($\alpha_1$ or $\alpha_2$),
the common parent is only applicable to direct parents.

\subsection{Routing Algorithm}\label{sec:routing}
Using the structure of the icosahedron and its subdivisions, we can devise an efficient routing algorithm. 
We will give a high-level overview of our approach and reasoning for a routing algorithm here and
in \cref{sec:technicalDetails} we will go into the technical details of the results and the proofs.

\subsubsection{Global Routing Algorithm}\label{sec:globalRouting}
We first look at designing a routing algorithm for finding a shortest path that uses global information.
Our end goal is an algorithm that uses only local information,
where each node can route using only information about nearby nodes as we will see in \cref{sec:localRoutingAlgorithm}.
Still, it is helpful to approach this from a global perspective since it is easier to understand the
idea of the algorithm which can then be recursively expressed in a clean manner.
On the way there, we design a labelling scheme that can later be used for local routing as well.

The general intuition of the routing algorithm is that when two vertices are not close to each other,
then it is possible to route towards a lower layer as indicated in the
\nameref{thm:3threeHops} Theorem (\cref{thm:3threeHops}).
Routing may then proceed from these lower layer vertices.
If nodes are close to each other, then we may find a path using a standard routing algorithm,
such as Dijkstra's algorithm~\cite{kleinberg2006algorithm}.

We give two rules with which the algorithm will always be able to pick the best parent
if the endpoints are sufficiently far apart.
Consider vertices $\alpha, \beta \in V$.
When $\alpha$ and $\beta$ are far away, we can route through lower layer vertices.
Since it may be necessary to go down another layer,
it is intuitively better to choose the parent that is on the lowest layer
so that we reach faster the lowest layer in a path between $\alpha$ and $\beta$.
From this intuition we have formalised two rules
that we can apply if $d(\alpha, \beta) > 6$.
The first rule checks whether the node $\alpha$ has a parent which is on a lower layer.
\begin{definition}[Parent Rule]\label{def:parentRule}
	Consider vertices $\alpha \in L_i$ with $i \geq 1$ and $\beta \in V$, where $d(\alpha,\beta) >6$.
	Let $\{\pi_1, \pi_2\} = p(\alpha)$.
	If $\ly(\pi_1) < \ly(\pi_2)$, then pick $\pi_1$ over $\pi_2$ when routing from $\alpha$ to $\beta$.
\end{definition}
This means that if the routing algorithm has to choose between $\pi_1$ and $\pi_2$,
it should always choose $\pi_1$.
If instead both parents are on the same layer, then it is interesting to see if they
contain the possibility to reach a lower layer faster by looking at the grandparents.
The Grandparent Rule looks one step ahead and checks
whether there exists a grandparent which is on a lower layer than the other grandparents.
\begin{definition}[Grandparent Rule]\label{def:grandparentRule}
	Consider vertices $\alpha \in L_i$, with $i\geq 2$, and $\beta\in V$, where $d(\alpha,\beta) >6$
	and where $\forall\gamma \in p(\alpha)$ there exists $p(\gamma)$.
	Let $\{\pi_1, \pi_2\} = p(\alpha)$ and $\ly(\pi_1) = \ly(\pi_2)$ so that \cref{def:parentRule} does not apply.
	Furthermore, let there be a grandparents $\{\gamma_1, \gamma_2\} = p(\pi_1)$ and $\{\gamma_2, \gamma_3\} = p(\pi_2)$.
	Here $\gamma_2$ is the common parent of $\pi_1$ and $\pi_2$ according to \cref{prop:commonParent}.
	If $\ly(\gamma_1) < \ly(\gamma_2)$ and $\ly(\gamma_1) < \ly(\gamma_3)$, then pick $\pi_1$ over
	$\pi_2$ when routing from $\alpha$ to $\beta$.
\end{definition}
Note that both rules can be applied symmetrically, since a shortest path from $\alpha$ to $\beta$
on a simple graph is also a shortest path from $\beta$ to $\alpha$. In case none of the two rules apply we let the routing algorithm pick a random parent. 
We will formally state the optimality of the routing algorithm using these rules after the example.

\begin{figure}[tb]
	\centering
	\begin{tikzpicture}
		% Alpha
		\node (a1) at (0,0) {$\alpha_1$};

		% Beta
		\node (b1) at (2,3.5)	{$\beta_1$};
		\node (b2) at (4,0)		{$\beta_2$};

		\draw[mediumlightgray] (b1) -- (2.75, 4.8125);
		\draw[mediumlightgray] (b2) -- (5.5, 0);

		% Gamma
		\node (g1) at (2,0)		{$\gamma_1$}
			edge (a1)
			edge (b2);
		\node[circle,draw,blue] (g2) at (1,1.75)	{$\gamma_2$}
			edge (a1)
			edge (b1)
			edge (g1)
			edge[->, bend left, blue, densely dashed] (b1)
			edge[->, bend right, blue, densely dashed] node[black, left] {\checkmark} (a1);
		\node (g7) at (3,1.75)	{$\gamma_7$}
			edge (b1)
			edge (b2)
			edge (g1);

		\node[circle, draw, red](e1) at (2, 1.75) {$\eta_1$}
			edge(g2)
			edge(g7)
			edge[->, bend left, red, dashdotted] node[black,below] {\checkmark} (g2)
			edge[->, bend right, red, dashdotted] (g7);

		\node (g8) at (4,3.5) {}
			edge[mediumlightgray] (b1)
			edge[mediumlightgray] (g7);

		\node (g9) at (5,1.75)	{}
			edge[mediumlightgray] (g7)
			edge[mediumlightgray] (b2);

		\node (i1) at (1.25,4.8125) {}
			edge[mediumlightgray] (b1);

		\node (i2) at (0.25,3.5) {}
			edge [mediumlightgray] (b1)
			edge [mediumlightgray] (g2);

		\node (i3) at (-0.75, 1.75) {}
			edge[mediumlightgray] (g2)
			edge[mediumlightgray] (a1);

		\node (i4) at (-2, 0) {}
			edge[mediumlightgray] (a1);

		\node (destination) at (9, 2.5) {destinations $\rightarrow$};
		\node (destination2) at (-4, 2.5) {$\leftarrow$ destinations};
		%\draw[help lines] (0,0) grid (8,7);
	\end{tikzpicture}
	\caption{
		Pictured is a small fraction of the subdivided graph,
		where $\ly(\alpha_i) < \ly(\beta_j) < \ly(\gamma_k) < \ly(\eta_\ell)$.
		When $\gamma_2$ has to route to a far away vertex (specified in the text) on a lower or equal layer,
		it needs to choose between $\alpha_1$ and $\beta_1$.
		Because $\ly(\alpha_1) < \ly(\beta_1)$ it is better to choose $\alpha_1$ according to \cref{def:parentRule},
		since a lower layer vertex is better suited to route over long distances with longer edges.
		When $\eta_1$ has to route far away, it needs to choose between $\gamma_2$ and $\gamma_7$.
		Initially, $\ly(\gamma_2) = \ly(\gamma_7)$ so they seem equally good.
		However, $\gamma_2$ is one hop away from $\alpha_1$ which can route far away faster than either parent of $\gamma_7$ ($\beta_1$ and $\beta_2$).
		\cref{def:grandparentRule} then states it is better to route through $\gamma_2$ than $\gamma_7$.
	}
	\label{fig:algRules}
\end{figure}
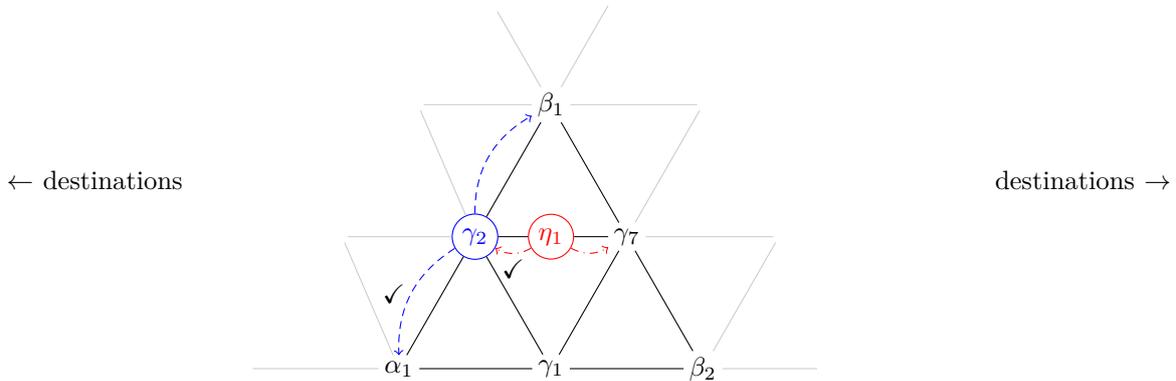

\paragraph{Example}
We give an example of why the lower layer parent is better than the higher layer parent using \cref{fig:algRules}.
Depicted is a fraction of the subdivided graph where $\gamma_2$ wants to route to a destination more than 6 hops away,
where it is assumed that the layer of the destination is lower or equal to that of $\gamma_2$.
Since the destination is far away, it can to route to a parent (\cref{lem:3LowerLayerPath} or \cref{thm:3threeHops}).
Because $\ly(\alpha_1) < \ly(\beta_1)$ the parent rule (\cref{def:parentRule}) applies,
and thus $\alpha_1$ is chosen over $\beta_1$, since it has better edges to route to far away destinations.
We do this, because if $\ly(\beta_1)$ is still higher than the destination,
then we would need to route to a parent of $\beta_1$.
However, $\alpha_1 \in p(\beta_1)$ so we could have skipped $\beta_1$ and went directly to $\alpha_1$.
We later formally state this in \cref{lem:4optimalChoice}.

The grandparent rule (\cref{def:grandparentRule}) applies to $\eta_2$ in \cref{fig:algRules},
where we also assume that the destination is more than 6 hops away and on a lower or equal layer,
so a choice has to be made between $\gamma_2$ and $\gamma_7$.
Initially, both parents look equally good to route through since $\ly(\gamma_2) = \ly(\gamma_7)$
and the Parent Rule does not apply.
However, $\gamma_2$ is adjacent to $\alpha_1$ which is on a lower layer than either parent of $\gamma_7$.
Thus routing through $\gamma_2$ is better when routing to far away destinations,
since it is possible to reach a lower layer faster through $\gamma_2$.

\paragraph{Routing Optimality}
Given the choices that the algorithm makes, and the Parent and Grandparent Rules, we can prove
that the algorithm takes an optimal choice given that $\alpha$ is far away from $\beta$.
A sufficient condition for applying the Parent and Grandparent Rules in the routing algorithm
is that $d(\alpha, \beta) > 6$.
To guarantee these conditions we perform Dijkstra's algorithm~\cite{kleinberg2006algorithm}
limited to the 6th-order neighbourhood of $\alpha$:
\begin{equation}
	N^\alpha = \{\gamma \in V : d(\alpha, \gamma) \leq 6\}\,.
\end{equation}
This results in the global routing algorithm given in \cref{alg:spherePath}.
\cref{alg:globalSpherePath} gives a step-by-step implementation of the global routing algorithm for routing and entanglement swapping.
The routing algorithm uses \cref{alg:bestParent} to apply the Parent and Grandparent Rules in selecting a parent if $d(\alpha,\beta) >6$. 
If $\ly(\alpha) \leq \ly(\beta)$ then the Parent and Grandparent Rules are applied to $\alpha$, because it is on a higher layer and can go to a parent.
However, if $\ly(\beta) < \ly(\alpha)$ the Rules are applied to $\beta$,
because that vertex must then go to a parent as implied by \cref{lem:3LowerLayerPath}.
We will later see that the neighbourhood is limited in such a way
that the complexity of this algorithm is still acceptable (\cref{sec:complexityAnalysis}).
First, we show that selecting a parent according to the Parent and Grandparent Rules
is an optimal choice when $d(\alpha,\beta) >6$.
\begin{lemma}[Optimal Choice]\label{lem:4optimalChoice}
	Consider vertices $\alpha,\beta \in V$ where $d(\alpha,\beta) > 6$ and $\ly(\alpha) \geq \ly(\beta)$.
	Then the routing algorithm (\cref{alg:spherePath}) finds the next node in a shortest path, namely it chooses a parent $\pi_1 \in p(\alpha)$ so that $d(\pi_1, \beta) = d(\alpha,\beta)-1$.
\end{lemma}
Combining \cref{lem:4optimalChoice}, Dijkstra's algorithm for $d(\alpha,\beta) \leq 6$,
and the symmetry of a shortest path between $\alpha$ and $\beta$
allow us to prove the optimality of the routing algorithm.
\begin{theorem}[Sphere Routing Optimality]
	Consider $\alpha,\beta \in V$, where $\alpha$ is the sender and $\beta$ the receiver.
	Then the routing algorithm (\cref{alg:spherePath}) finds a next node in a shortest path, namely it chooses a vertex $\pi_1 \in N(\alpha)$ so that $d(\pi_1, \beta) = d(\alpha,\beta) - 1$,
	or a vertex $\pi_2 \in N(\beta)$ so that $d(\alpha, \pi_1) = d(\alpha, \beta) -1$.
\end{theorem}
To guarantee termination of the routing algorithm,
the base case must always find a path if given
two vertices on the base layer $\alpha, \beta \in V_0$,
i.e.\ two vertices for which a step to the parents is not possible.
The diameter of the icosahedron is 3,
which means that $d(\alpha, \beta) \leq 3$,
so $\beta \in N^\alpha$ and Dijkstra is able to find a shortest path on the base icosahedron.

\begin{algorithm}
	\LinesNumbered
	\DontPrintSemicolon
	\SetKwInOut{Data}{Data}
	\SetKwInOut{Input}{Input}\SetKwInOut{Output}{Output}
	\SetKwFunction{path}{path}
	\SetKwFunction{send}{send}

	\Data{%
		The label, $\alpha$, of the node itself;
		the data required for calling \path{} (see \cref{alg:spherePath});
		a procedure \send{$\gamma, (\eta, P_{\alpha,\beta}, \beta)$} that sends a tuple of $\eta, \beta\in V$ and a path $P_{\alpha,\beta} \in [V]$ to $\gamma\in V$;
		and a procedure \swap{$\eta,\gamma$}
		that swaps entanglement shared with $\eta$ and $\gamma$, and sends appropriate classical correction information to the final destination $\beta$.
	}
	\Input{%
		A tuple $(\eta, P_{\alpha,\beta}, \beta)$,
		where $\eta \in V$ is the previous sender $\eta$ from which the request is received, $\beta \in V$ is the final destination,
		and $P_{\alpha,\beta}$ is the path from $\alpha$ to $\beta$. 
	}
	
	\SetKwProg{globalSphereRoute}{Procedure}{ is}{end}
	\tcp{Run on reception of a request.}
	\globalSphereRoute{$\mathrm{globalSphereRoute}((\eta, P_{\alpha,\beta}, \beta))$}{%
		\uIf{$\alpha = \beta$}{%
			\tcp{Destination reached.}
		}
		\uElse{%
			\tcp{If we are the original sender, we compute the path.}
			\If{$P_{\alpha,\beta}$ is $\perp$}{%
				$P_{\alpha,\beta} \gets$ \path{$\alpha, \beta$} \tcp*{Call \cref{alg:spherePath} to calculate a complete path to $\beta$.}
			}
			\tcp{Forward the request to the second element in $P_{\alpha,\beta}$}
			$\gamma \gets P_{\alpha,\beta}[1]$\;
			\send{$\gamma, (\alpha, P_{\alpha,\beta}[1,\ldots,\abs{P_{\alpha,\beta}}], \beta)$}\;

			\tcp{Unless we are the original sender, perform a swapping operation}
			\If{$\eta$ is not $\perp$}{%
				\swap{$\eta, \gamma$}\;
			}
		}
	}
	\caption{%
		A procedure similar to \cref{alg:completeRingRouting} for routing incoming requests
		on the sphere network using global information from $\alpha$ to $\beta$. The initial sender $\alpha$ first runs {\tt globalSphereRoute} with inputs
		$(\perp, \perp, \beta)$.
		We insert a complete shortest path to the receiver $\beta$ given
		by \texttt{path}$(\alpha,\beta)$ into the request forwarded to the next node,
		so that following nodes only have to take the next element from the path.
	}
	\label{alg:globalSpherePath}
\end{algorithm}

\begin{algorithm}[tb]
	\DontPrintSemicolon
	\SetKwInOut{Data}{Data}
	\SetKwInOut{Input}{Input}
	\SetKwInOut{Output}{Output}
	\Data{$N^\alpha$ the 6th-order neighbourhood of $\alpha$}
	\Input{$\alpha,\beta \in V$, sender and receiver respectively.}
	\Output{$P_{\alpha,\beta}$ the path from $\alpha$ to $\beta$}
	\SetKwProg{Fn}{Function}{ is}{end}
	\Fn{$\mathrm{path}(\alpha, \beta)$}{%
		\eIf{$\beta \in N^\alpha$}{%
			\Return dijkstra($\alpha, \beta, N^\alpha$) 
			\tcp*{Dijkstra on the subgraph $N^\alpha$.}
		}{%
			\tcp{Increment the path to a parent in the label of $\alpha$ or $\beta$.}
			\eIf{$\ly(\beta) > \ly(\alpha)$}{%
				\Return $\text{path}(\alpha, \text{bestParent}(\beta)) \doubleplus \text{bestParent}(\beta)$\;
			}(\tcp*[f]{Else $\ly(\alpha) \geq \ly(\beta)$}){%
				\Return $\text{bestParent}(\alpha) \doubleplus \text{path}(\text{bestParent}(\alpha),\beta)$
			}
		}
	}
	\caption{%
		Routing Algorithm with global information, which includes both $\alpha$ and $\beta$.
		Once $d(\alpha, \beta) \leq 6$,
		Dijkstra's algorithm limited to the 6th-order neighbourhood of $\alpha$ will find a shortest path to $\beta$.
		If $d(\alpha,\beta) > 6$, then the vertex on the highest layer must take a step to a parent.
		To choose a parent on a shortest path
		we apply the Parent and Grandparent Rules (\cref{def:parentRule,def:grandparentRule}),
		as given in the \texttt{bestParent} function (\cref{alg:bestParent}).
	}
	\label{alg:spherePath}
\end{algorithm}

\begin{algorithm}[tb]
	\DontPrintSemicolon
	\SetKwInOut{Data}{Data}
	\SetKwInOut{Input}{Input}
	\SetKwInOut{Output}{Output}
	\Data{The parents $p(\alpha)$ and grandparents $\bigcup_{\pi \in p(\alpha)} p(\pi)$}
	\Input{$\alpha \in V$, the vertex to choose the right parent for.}
	\Output{$\pi \in p(\alpha)$, the parent to route through.}
	\SetKwProg{Fn}{Function}{ is}{end}
	\Fn{$\mathrm{bestParent}(\alpha)$}{
		\uIf(\tcp*[f]{Check the parent rule first.}){\textup{\cref{def:parentRule} applies}}{%
			\Return $\pi_1$ from \cref{def:parentRule}\;
		}
		\uElseIf(\tcp*[f]{Then check the grandparent rule.}){\textup{\cref{def:grandparentRule} applies}}{%
			\Return $\pi_1$ from \cref{def:grandparentRule}\;
		}
		\Else{%
			\Return $\text{randomElement}(p(\alpha))$ \tcp*{Pick a random parent otherwise.}
		}
	}
	\caption{%
		Routing step in case $d(\alpha,\beta) > 6$.
		The vertex on the highest layer ($\alpha$) performs a step to a parent on a shortest path
		that is chosen according to the Parent and Grandparent Rules (\cref{def:parentRule,def:grandparentRule}).
	}
	\label{alg:bestParent}
\end{algorithm}

\subsubsection{Labelling}\label{sec:labelling}
We assign to every vertex/node a label that uniquely identifies the vertex and allows
for a routing algorithm, where each vertex needs only information about its local neighborhood to run it.
As seen in the previous section, we also need to know which layer the vertex is on and how far away it is from the receiver, in order to decide for the endpoints $\alpha,\beta\in V$ whether $\ly(\alpha) \geq \ly(\beta)$ and whether $d(\alpha,\beta) > 6$. On the other hand, the label must be small in size, as it functions as a data header included in every transmission over the network, indicating where the data must go.
If we construct a labelling scheme with the above properties, we will be able to construct an efficient local routing algorithm.

What's more, the labelling scheme can be adapted to work for graph structures similar to the sphere,
as long as there is a child-parent relationship.
Additionally, it is possible to randomize the selection of the parent in case of multiple equivalent options,
and in doing so decrease the load of an edge, which is the number of entanglement swaps it must perform per time unit.
We further discuss the possibilities of the labelling scheme in \cref{sec:discuss}.

Every node is labelled in a hierarchical routing scheme similar to Internet Protocol (IP)
which is widely used for routing in current computer networks~\cite{postel1981ip}.
A node is assigned a unique ID, which is simply a unique integer.
The label contains this unique ID, and the node's ancestry tree which includes parents, then grandparents, etc.
It is constructed as a list of sets of vertex IDs, i.e. $[\{a\}, \{b,\dots\},\{\dots\},\dots]$,
where $a,b\in \left[1, \dots, N\right]$ are the unique IDs.
Since the ID is unique we can equate a vertex $\alpha \in V$ to its ID $a$, so that we can use them interchangeably.
Each set in the label denotes a group of equally good parents for routing.

The label indicates which vertices in the ancestry tree are the fastest way to reach
lower layers.
The inverse is also true,
it is possible to construct a shortest path to the labelled vertex from any vertex in the label.
As we have seen in the Parent Rule (\cref{def:parentRule}) and the Grandparent Rule (\cref{def:grandparentRule})
some vertices are better suited for this purpose.
In the labelling we will embed these rules, so that the local algorithm is easily able to
find the parents that satisfy these rules.

First up is the Parent Rule.
Any vertices that are unfavored by the Parent Rule are not included in the label.
We define the function $p^{\text{good}}$ which returns the set of nodes that
are not unfavored by the Parent Rule:
\begin{equation}
	p^{\text{good}}(\mathbb A) = \{\beta \in p(\mathbb A): \forall\gamma \in p(\mathbb A), \ly(\beta) \leq \ly(\gamma)\}\,.
\end{equation}
We can implement the Grandparent Rule by referencing the Parent Rule, since a parent fulfils the Grandparent Rule if it has parents that are also good vertices according to $p^{\text{good}}$.
We can show this if we consider the parents $\{\pi_1, \pi_2\} = p(\alpha)$, where $\ly(\pi_1) = \ly(\pi_2)$ because the Parent rule is inconclusive.
Then we can assume that $\{\gamma_1, \gamma_2\} = p(\pi_1)$ and $\{\gamma_2, \gamma_3\} = p(\pi_2)$,
where $\gamma_2$ is the common parent of $\pi_1$ and $\pi_2$, since $\pi_1 \edge \pi_2$.
If $\pi_1$ is preferred by the Grandparent Rule,
that implies $\ly(\gamma_1) < \ly(\gamma_2)$ and $\ly(\gamma_1) < \ly(\gamma_3)$.
So $p^{\text{good}}(\{\pi_1, \pi_2\}) = \{\gamma_1\}$, since both $\gamma_2$ and $\gamma_3$ are on a higher layer than $\gamma_1$.
Thus if
\begin{equation}
	p(\pi_2) \cap p^{\text{good}}(\{\pi_1, \pi_2\}) = \emptyset\,,\label{eq:grandparentRule}
\end{equation}
then $\pi_2$ is unfavored by the Grandparent Rule.
A similar argument holds for when $\pi_2$ is favored by the Grandparent rule.
This leads us to the filter $f(\mathbb A)$ which filters out nodes in the set
$\mathbb A$ that are unfavored by the Parent and Grandparent Rules.
Let
\begin{equation}
	\mathbb B = p^{\text{good}}(\mathbb A)\,,
\end{equation}
then we have to check if the grandparents of $\alpha$ exist, i.e.\ $p(\beta) : \forall\beta \in \mathbb B$.
We can do this by taking any $\beta \in \mathbb B$ and checking whether $\beta \in V_0$.
We only have to do this once, since all $\beta$ must be on the same layer according to the Parent Rule.
If the grandparents do exist, then we apply the Grandparent Rule as we saw in \cref{eq:grandparentRule}
\begin{equation}
	f(\mathbb A) = \begin{cases}
		\mathbb B & \text{if for any $\beta \in \mathbb B$ holds that $\beta \in V_0$,}\\
		\{\beta \in \mathbb B: p(\beta) \cap p^{\text{good}}(\mathbb B) \ne \emptyset \} & \text{otherwise}.
	\end{cases}
	\label{eq:labelFilter}
\end{equation}
Using the definition of the filter $f$, it is possible to define the label which satisfies the Parent Rule at every entry.
At the first entry of the label is the vertex ID itself,
then come the parents of the vertex that satisfy the Parent Rule and the Grandparent Rule.
We repeat this process for the grandparents, great-grandparents, etc. until the base layer has been reached.
As such we build the label which satisfies the Parent and Grandparent Rules,
so that every vertex in the label is either preferred by the Parent or Grandparent Rule or the rules do not apply because no parent is preferred.
\begin{gather}
	\la(\mathbb A) = \begin{cases}
		[\mathbb A] & \text{ if } \exists\alpha\in \mathbb A : \alpha \in V_0\,,\\
		[\mathbb A]  \doubleplus \la(f(\mathbb A)) & \text { otherwise.}
	\end{cases}
\end{gather}
From this recursive definition we can define the label of a vertex $\alpha \in V$ as
\begin{gather}
	\la(\alpha) = \la(\{\alpha\})\,. \label{eq:label}
\end{gather}
Pseudocode for the construction of the label is given in \cref{alg:labelling}.

\begin{algorithm}[tb]
	\DontPrintSemicolon
	\SetKwInOut{Data}{Data}
	\SetKwInOut{Input}{Input}
	\SetKwInOut{Output}{Output}
	\Data{$G = (V,E)$, the complete graph.}
	\Input{$\alpha \in V$}
	\Output{$\la(\alpha)$, the label of $\alpha$}
	\SetKwProg{Fn}{Function}{ is}{end}
	\Fn{$\mathrm{label}(\alpha)$}{
		$\la(\alpha)_1 = \{\alpha\}$\;
		$i \gets 1$\;
		\While{$\forall\alpha \in \la(\alpha)_i: \alpha \not\in V_0$}{%
			$\la(\alpha)_{i+1} \gets f(\la(\alpha)_i)$\;
			$i \gets i+1$\;
		}
		\Return $[\la(\alpha)_1, \la(\alpha)_2, \ldots, \la(\alpha)_i]$\;
	}
	\caption{%
		Labelling of vertices.
		When the graph is constructed, each vertex is given a label that will allow efficient routing.
		Function $f(\cdot)$ is defined in \cref{eq:labelFilter}.
		The labelling is similar to IP addressing, because a hierarchical structure is used.
	}
	\label{alg:labelling}
\end{algorithm}

It is important to note that the size of the labelling is in fact small. Let $[\ldots]_k$ denote the $k$-th element of a list.
We can prove that the size of the labelling is bounded by $\abs{\la(\alpha)_i} \leq 3$ for any $i$.
This shows that the size of the label is upper bounded by $\abs{\la(\alpha) } = O\left(\log(V)\right)$.
Furthermore, every two vertices for an entry in the label are adjacent and on the same layer.
\begin{lemma}[Label Bound]\label{lem:5labelBound}
	The label size is bounded as $\abs{ \la(\alpha)_\ell } \in \{1,2,3\}$,
	for any $\alpha \in V$ and $\ell \in \mathbb N$.
	Additionally, for any $\beta, \gamma \in \la(\alpha)_\ell : \beta \ne \gamma$
	it holds that $\beta \edge \gamma$ and $\ly(\beta) = \ly(\gamma)$.
\end{lemma}

\paragraph{Example}
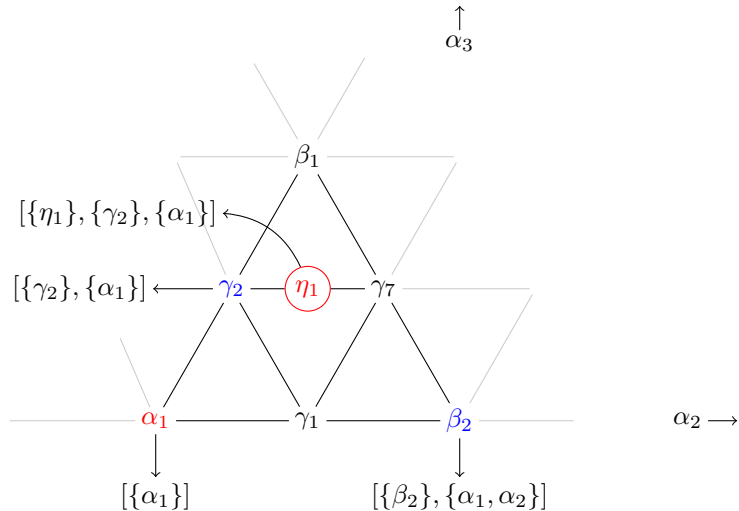
\begin{figure}[tb]
	\centering
	\begin{tikzpicture}
		% Alpha
		\node[red] (a1) at (0,0) {$\alpha_1$};
		\node (a2) at (7,0) {$\alpha_2$};
		\node (a3) at (4,5) {$\alpha_3$};
		
		\draw[->] (a2) -- +(0.65,0);
		\draw[->] (a3) -- +(0,0.5);

		% Beta
		\node (b1) at (2,3.5)	{$\beta_1$};
		\node[blue] (b2) at (4,0)		{$\beta_2$};

		\draw[mediumlightgray] (b1) -- (2.75, 4.8125);
		\draw[mediumlightgray] (b2) -- (5.5, 0);

		% Gamma
		\node (g1) at (2,0)		{$\gamma_1$}
			edge (a1)
			edge (b2);
		\node[blue] (g2) at (1,1.75)	{$\gamma_2$}
			edge (a1)
			edge (b1)
			edge (g1);
		\node (g7) at (3,1.75)	{$\gamma_7$}
			edge (b1)
			edge (b2)
			edge (g1);

		\node[draw, circle, red] (e1) at (2, 1.75) {$\eta_1$}
			edge(g2)
			edge(g7);

		\node (g8) at (4,3.5) {}
			edge[mediumlightgray] (b1)
			edge[mediumlightgray] (g7);

		\node (g9) at (5,1.75)	{}
			edge[mediumlightgray] (g7)
			edge[mediumlightgray] (b2);

		\node (i1) at (1.25,4.8125) {}
			edge[mediumlightgray] (b1);

		\node (i2) at (0.25,3.5) {}
			edge [mediumlightgray] (b1)
			edge [mediumlightgray] (g2);

		%\node (i3) at (-0.75, 1.75) {}
		\node (i3) at (-0.5, 1.17) {}
			%edge[mediumlightgray] (g2)
			edge[mediumlightgray] (a1);

		\node (i4) at (-2, 0) {}
			edge[mediumlightgray] (a1);

		\node (labela1) at ($(a1) + (0,-1)$) {$[\{\alpha_1\}]$}
			edge[<-] (a1);
		\node (labelb2) at ($(b2) + (0,-1)$) {$[\{\beta_2\}, \{\alpha_1, \alpha_2\}]$}
			edge[<-] (b2);
		\node (labelg2) at ($(g2) + (-2,0)$) {$[\{\gamma_2\}, \{\alpha_1\}]$}
			edge[<-] (g2);
		\node (labele1) at ($(g2) + (-1.5, 1)$) {$[\{\eta_1\}, \{\gamma_2\}, \{\alpha_1\}]$};
		\draw[<-] (labele1.east) to[bend left] (e1);

		%\draw[help lines] (0,0) grid (8,7);
	\end{tikzpicture}
	\caption{
		Pictured are the vertex labels of a fraction of the subdivided graph where $\ly(\alpha_i) < \ly(\beta_j) < \ly(\gamma_k) < \ly(\eta_\ell)$.
		The labels of the nodes $\alpha_1, \beta_2, \gamma_2$ and $\eta_1$ are printed.
		The label (\cref{eq:label}) applies the Parent Rule (\cref{def:parentRule}) and the Grandparent Rule (\cref{def:grandparentRule})
		to include only vertices which are suitable for routing far away.
	}
	\label{fig:labelExample}
\end{figure}
In \cref{fig:labelExample} we give an example on how the label is constructed.
For $\alpha_1$, the ID is $\alpha_1$, and it is already on layer 0, thus
\[
	\la(\alpha_1) = [\{\alpha_1\}]\,.
\]
A node on the first layer is $\beta_2$, which has as parents $\alpha_1$ and $\alpha_2$.
Since $\ly(\alpha_1) = \ly(\alpha_2)$ both parents pass the Parent Rule (\cref{def:parentRule}),
and since $\alpha_1, \alpha_2 \in V_0$ they also pass the Grandparent Rule.
Furthermore, $\alpha_1 \in V_0$ so the label stops here, at
\[
	\la(\beta_2) = [\{\beta_2\}, \{\alpha_1, \alpha_2\}]\,.
\]
Located on the second layer is $\gamma_2$, with parents $\{\alpha_1, \beta_1\}$.
In this case $\ly(\alpha_1) < \ly(\beta_1)$,
so $\beta_1$ is not included in the label according to the Parent Rule.
Because $\alpha_1 \in V_0$, the label ends here, becoming
\[
	\la(\gamma_2) = [\{\gamma_2\}, \{\alpha_1\}]\,.
\]
Lastly, $\eta_1$ is located on the third layer, with parents $\{\gamma_2, \gamma_7\}$.
Both parents pass the Parent Rule, because $\ly(\gamma_2) = \ly(\gamma_7)$.
However, $\gamma_2$ has access to a better parent than $\gamma_7$,
namely $\alpha_1$.
According to the Grandparent Rule only $\gamma_2$ is selected.
Then comes the next iteration, where the parents of $\gamma_2$ are investigated,
which are $\{\alpha_1, \beta_1\}$.
As we have seen before, only $\alpha_1$ is included because of the Parent Rule,
so that
\[
	\la(\eta_1) = [\{\eta_1\}, \{\gamma_2\}, \{\alpha_1\}]\,.
\]

\subsubsection{Local Routing Algorithm}\label{sec:localRoutingAlgorithm}
We will adapt the global routing algorithm given in \cref{alg:spherePath} to an algorithm
that does not require global information to route.
This is important for any routing algorithm that is used in real world networks.
If a network node has to calculate a shortest route with a global routing algorithm,
then each vertex must store the entire network structure.
That would mean the locally stored data grows linearly with the network size in every vertex and edge.
However, we would prefer an algorithm that does not require much memory
even when the number of nodes grows large, ideally with memory logarithmic in the size of the network.
Thus we try not to store the entire network structure,
and restrict the data to what every vertex knows about itself and its local neighborhood.
Furthermore, we argue that the local algorithm is equivalent to its global counterpart,
so that the proof of optimality for the global algorithm will also guarantee
the optimality of the local routing algorithm.

In a local algorithm we assume the perspective of a single network node $\alpha$ who wishes to send a qubit - or equivalently - establish a long distance VQL with $\beta$ using {\tt localSphereRoute}.
The main idea of converting the global routing algorithm to a local one is that basically the label of the destination $\beta$ contains all the necessary information to locally route to it without needing to have knowledge of where $\beta$ is in the network and what is its neighborhood.
%For example, we cannot calculate \texttt{bestParent}($\beta$) in \cref{alg:spherePath}
%since we do not know the neighbourhood of $\beta$.
%During the global algorithm $\beta$ moves to the \texttt{bestParent}($\beta$) if for the layer
%of $\alpha$ and $\beta$ holds that $\ly(\alpha) > \ly(\beta)$,
%and from the available information on $\beta$ (the label) we cannot predict which layer $\beta$ is on.
%We only have the IDs that are stored in $\la(\beta)$.

\newcommand{\betarecv}{\beta^{\text{recv}}}

The $\beta$ in recursive calls of \texttt{path}$(\alpha,\beta)$
is often replaced by \texttt{bestParent}($\beta$).
To avoid confusion, we fix the receiver as $\betarecv$.
With $\beta$ we then refer to current $\beta$ in a recursive call of \texttt{path}$(\alpha,\beta)$.
We know two things which allow is to route locally:
\begin{enumerate}
	\item In the global algorithm, the destination variable $\beta$ can only be a vertex that appears in $\la(\betarecv)$, since in the recursion of the global routing algorithm $\beta$
		can only be replaced with \texttt{bestParent}($\beta$).
	\item All possible \texttt{bestParent}($\beta$) nodes during routing of the global
		routing algorithm are contained in $\la(\betarecv)$.
\end{enumerate}
Because the global routing algorithm has been shown to be correct,
we can conclude that as long as there is no $\beta' \in \la(\betarecv)$ for which $d(\alpha, \beta') \leq 6$
then we can route to \texttt{bestParent}($\alpha$).
Once there is such a $\beta'$, we have reached the part of the global routing algorithm
where $d(\alpha,\beta) \leq 6$ and we route to $\beta'$.
Afterwards, we find a shortest $(\beta', \betarecv)$-path.

Routing should go up the layers now, instead of down.
For some $i$ it holds that $\beta' \in \la(\betarecv)_i$.
Let us now consider which node $\gamma$ is on a shortest path towards $\betarecv$.
Since we know that the global routing algorithm is restricted to nodes in $\la(\betarecv)$
we know that
\[
	\gamma \in \la(\betarecv)_{i-1}\,.
\]
It does not matter which $\gamma \in \la(\betarecv)_{i-1}$ is chosen as the next step in the path.
The global algorithm also chooses $\gamma$ at random when routing from $\la(\betarecv)_{i-2}$,
since all $\gamma$ meet both the Parent and Grandparent rules (\cref{def:parentRule,def:grandparentRule}).
Thus any $\gamma \in \la(\betarecv)_{i-1} \cap N(\beta')$ will be reachable from $\beta'$,
and is still on a shortest path towards $\betarecv$.
The existence of $\gamma$ follows from the definition of the label (\cref{eq:label}),
because there must be a child of $\beta'$ that is added to the label.
Thus there must be a $\gamma$ in the label for which
\[
	\gamma \in \la(\betarecv)_{i-1} : \beta' \in p(\gamma)\,.
\]
We now repeat this selection of $\gamma$ with a new $\beta'$ until we reach the receiver $\betarecv$.
All in all, this results in the local routing algorithm given in \cref{alg:localRoute}.
A step-by-step description of how to perform routing and entanglement swapping using the local routing algorithm is given in \cref{alg:sphereStepByStep}.
For simplicity we ignore the complete path returned by \texttt{localRoute}.
It can be used to speed up routing for the following $5$ nodes by taking the head of the list and routing there.

\begin{algorithm}
	\LinesNumbered
	\DontPrintSemicolon
	\SetKwInOut{Data}{Data}
	\SetKwInOut{Input}{Input}\SetKwInOut{Output}{Output}
	\SetKwFunction{localRoute}{localRoute}
	\SetKwFunction{send}{send}
	\Data{%
		The label, $\alpha$, of the node itself;
		a procedure \send{$\gamma, (\eta, \beta)$} that sends a tuple of $\eta, \beta\in V$ to $\gamma\in V$;
		and a procedure \swap{$\gamma,\eta$}
		that swaps the entanglement shared with $\gamma$ and $\eta$, and sends appropriate classical correction information
		to the final destination $\beta$.
	}
	\Input{A tuple $(\eta, \beta) \in V \times V$, where $\eta$ is the previous sender and $\beta$ is the destination.}
	\SetKwProg{localSphereRoute}{Procedure}{ is}{end}
	\tcp{Run on reception of a request.}
	\localSphereRoute{$\mathrm{localSphereRoute((\eta, \beta))}$}{%
		\eIf{$\alpha = \beta$}{%
			\tcp{Destination reached.}
		}{%
			\tcp{Forward the request the second node ($\gamma$) on a shortest path ($P_{\alpha,\beta}$).}
			$\gamma \gets$ \localRoute{$\alpha, \beta$}$[1]$ \tcp*{Call \cref{alg:localRoute}}
			\send{$\gamma, (\alpha, \beta)$}\;

			\tcp{Unless we are the original sender, perform a swapping operation.}
			\If{$\eta$ is not $\perp$}{%
				\swap{$\eta, \gamma$}\;
			}
		}
	}
	\caption{%
		A procedure for routing incoming packets on the sphere network from $\alpha$ to $\beta$, or equivalently, for creating a long distance VQL between $\alpha$ and $\beta$.
		The original sender $\alpha$ first runs {\tt localSphereRoute} with inputs $(\perp,\beta)$ where $\beta$ is the final 
		destination. 
		Every subsequent node $\alpha \in V$ runs the procedure on reception of a request.
		Compared to the global routing algorithm (\cref{alg:globalSpherePath})
		it is not necessary to calculate and forward a complete $P_{\alpha,\beta}$ path.
	}
	\label{alg:sphereStepByStep}
\end{algorithm}

\begin{algorithm}
	\DontPrintSemicolon
	\SetKwInOut{Data}{Data}
	\SetKwInOut{Input}{Input}
	\SetKwInOut{Output}{Output}
	\Data{$\alpha \in V$, the current node.\\
		$N^\alpha \subseteq V$, the IDs of the 6th-order neighbourhood of $\alpha$.}
	\Input{$\beta \in V$, the destination.}
		%$d \in \{0,1\}$, $d$ stands for `down', signalling if the path should travel down $\la(\beta)$ to $\beta$.}
	\Output{$[V]$, a path of the next nodes to route to.}
	\SetKwProg{Fn}{Function}{ is}{end}
	\Fn{$\mathrm{localRoute}(\beta)$}{%
		\uIf{$\alpha = \beta$}{%
			Destination reached
		}
		\uElseIf{$\alpha \in \la(\beta)_i : i \in \left\{ 1,\dots,\abs{\la(\beta)} \right\}$}{%
			\tcp{Go to a child of $\alpha$ that is in the label of $\beta$ and is on a lower layer.}
			%$i : \alpha \in \la(\beta)_i$\;
			\Return randomElement$(\la(\beta)_{i-1} \cap N(\alpha))$ \tcp*{Choose randomly on multiple options}
		}
		\Else{%
			$L_\beta \gets \cup_j \la(\beta)_j$\;
			\eIf{$L_\beta \cap N^\alpha \ne \emptyset$}{%
				\tcp{Use Dijkstra's algorithm limited to $N^\alpha$ to find a shortest path.}
				\Return $\argmin_{\gamma \in N^\alpha \cap L_\beta} \left\{|P_{\alpha,\gamma}| : P_{\alpha,\gamma} = \text{dijkstra}(\alpha, \gamma, N^\alpha) \right\}$\;
			}{%
				\tcp{Go to a parent in the label of $\alpha$ that is on a lower layer.}
				\Return randomElement$(\la(\alpha)_2)$ \tcp*{Choose randomly on multiple options}
			}
		}
	}
	\caption{Used by {\tt localRouteSphere} to find the next node on the path to the final destination $\beta$.}
	\label{alg:localRoute}
\end{algorithm}

\subsubsection{Analysis of the Local Routing algorithm}\label{sec:complexityAnalysis}
We show that the local routing algorithm, as well as the routing when the path is at most of length 6, are efficient and use little memory. Since the algorithm is local, we look at the necessary resources per node.
We first prove that the classical memory of every node scales logarithmically with the number of nodes $N$.
\begin{theorem}[Memory size]
	The classical memory size per node for the local routing algorithm (\cref{alg:localRoute}) is $O(\log^6 N)$.
\end{theorem}
\noindent Furthermore, we show that the running time of the routing algorithm per node also
scales logarithmically in the number of nodes $N$.
\begin{theorem}[Local Running Time]
	The running time per node of the local routing algorithm (\cref{alg:localRoute}) is $O(\log N)$.
\end{theorem}
\noindent Since the diameter of the graph is $D(G) = 4\log_2\left( \frac{n-2}{10} \right) + 3 = O(\log N)$
(\cref{prop:problemDiameter}) and every node uses $O(\log N)$ time,
the total running time of the local routing algorithm is at most $O(\log^2 N)$.

\section{Replenishing Entanglement}\label{sec:replenishing}

So far we have assumed that a background process creates entanglement as specified.
However, a key part is the feasibility of replenishing the entanglement in the network
and the background process as a whole.
In this section we will introduce a simple model for the cost of replenishment.

We distinguish two operations: Creation of an entangled pair on a physical link,
and an entanglement swap plus purification.
Furthermore, we assume that there are no failures when purifying entanglement.
Depending on the implementation, the bottleneck may be the entanglement creation operation or the entanglement swapping.  
As a simplification, we assume that both operations take $1$ unit of time.
We restrict the number of operations per edge to one per time step.
As a result, entanglement swapping prohibits any other operation on either of the involved edges in the same time step;
conversely, it implies that the creation operation can be performed at all physical connection concurrently.
A further restriction we can impose is that every node may only perform one entanglement swapping
in every time step.
(See also \cref{fig:NetworkBox} for a summary of our assumptions.)
With these assumptions and restrictions we calculate the number of time steps needed to initiate the graph structure
for the ring and the sphere graphs.

\begin{theorem}[Entangling Time]\label{thm:4entanglingTime}
	Consider the subdivided graph $G_k = (V_k, \cup_{i \in \{0,\dots,k\}} E_i)$
	with no entanglement distributed.
	Then the number of time steps $T(E)$ required to establish entanglement along $E$ is
	\begin{equation*}
		T(E) = 2k + 1 = O(\log N)\,.
	\end{equation*}
\end{theorem}

\noindent Now that we know how much time it takes to initiate the graph,
what about recreating entanglement after it has been used for routing or when it has decohered?
We will show that the number of time steps required after routing scales
linearly with the number of edges that have to be replaced.
\begin{theorem}[Edge Entangling]\label{thm:edgeEntangling}
	Consider a ring or sphere graph $G_k = (V, E)$ where entanglement has been established along
	all edges $E$.
	If at any point in time the entanglement along the edges $S \subseteq E$ has been consumed,
	then, it takes at most $2 \abs{S}$ time steps to establish once again entanglement along all edges $E$.
\end{theorem}
Note that by combining the above two theorems, we have that replenishing the entanglement in the entire graph
can take time at most 
\begin{equation}
	\min(2|S|, 2k+1) = O(\log N)\,.
\end{equation}
%Since $D(G_k) \leq 2k + 1$ for the ring (\cref{lem:RingDiameter}) and $D(G_k) \leq 2k + 3$ for the sphere (\cref{prop:problemDiameter}),
%we can conclude that $2\abs{S} \leq 4k + 6$ for any path on these subdivided graphs.
%For an arbitrary set of nodes $S$ we know that they can be replenished in less than linear time.
%In any case, we know that recreating entanglement over the entire network takes $2k + 1$ time steps,
%so the maximum entanglement refresh time is upper bounded by $2k+1 = O(\log N)$.

\section{Robustness of the Network}\label{simulations}
\begin{figure}[tb]
	\centering
	\includegraphics[width=\textwidth]{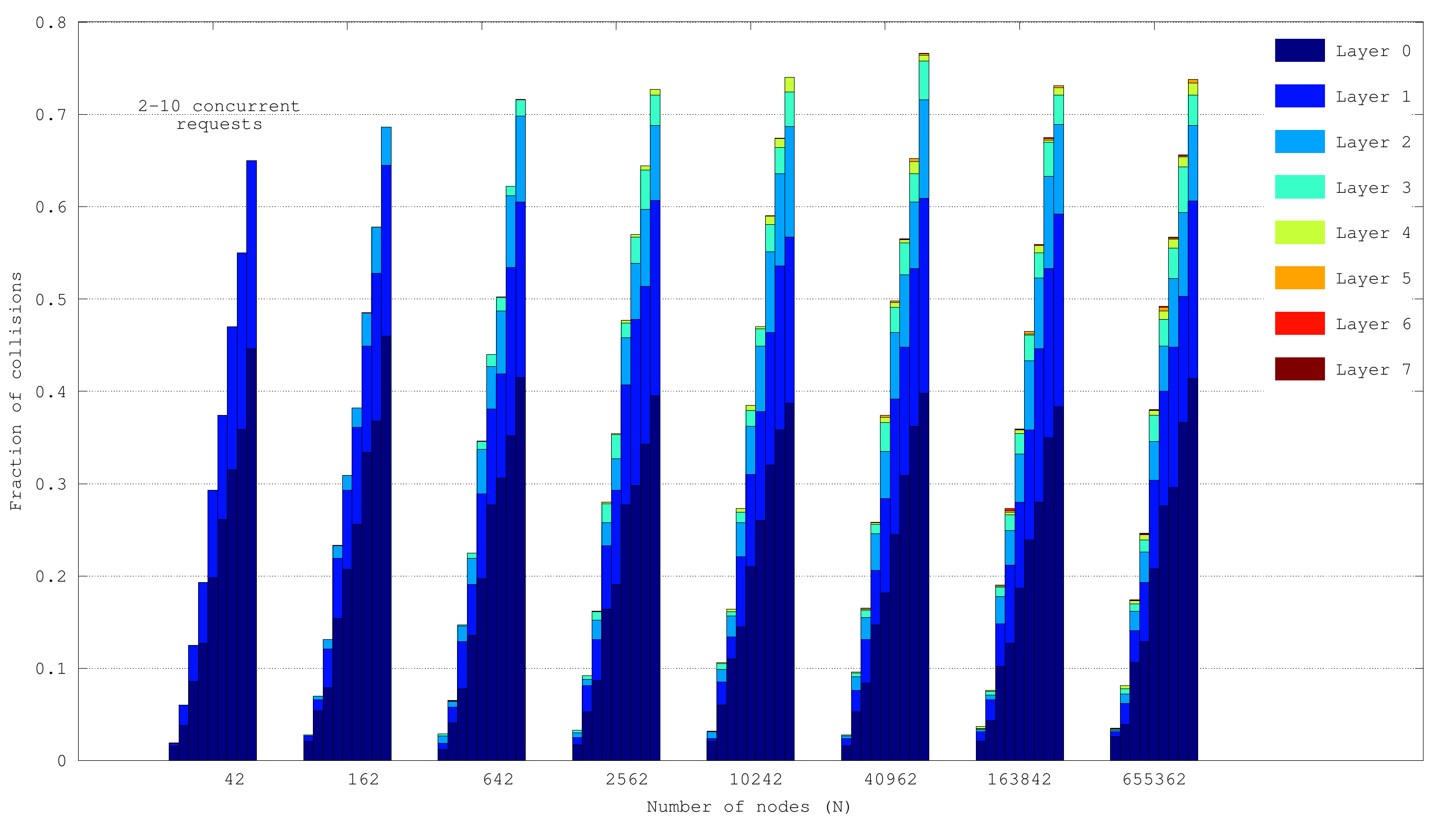}
	\caption{%
		The single-use nature of Virtual Quantum Links may prevent multiple users from using the same path.
		Using simulations we have checked how robust our networks are for multiple concurrent requests.
		We draw $2-10$ pairs from all nodes randomly and record the collisions on the lowest (and most costly) layer.
		$1,000$ samples were recorded, explaining the small variations in the peaks.
		The results show that 7 concurrent requests in the sphere
		already results in a collision in more than $0.5$ of the samples for all $N$.
		There also is no large dependence on $N$.
		From the segmented bar heights we see that if a collisions occurs
		there is likely to be a collision in the lower layers.
	}\label{fig:simCollisions}
\end{figure}
In classical networks requests can be queued when there are multiple requests contending over the same connection.
But for VQLs the connection may only be used once until it is replenished. Waiting for links to be relenished occurs a significant cost, since in practise the lifetime of qubits is short.
In practise, we thus expect that extension of our algorithms will divert traffic along alternative VQLs instead of waiting.

Nevertheless, it would be interesting to see how well our routing algorithms can in fact deal with multiple requests by routing them simultaneously. In other words, we would like to gain insight in how often multiple requests result to a contention of an edge, which we call a collision. Furthermore, we want to know where such collisions occur often,
because we expect that long-distance VQLs, those with a low layer number, have a higher load.

To this end we have performed simulations of the number of collisions under load,
i.e.\ when there are multiple simultaneous requests to use the network and two requests attempt to use the same VQL.
We have implemented a simulation program to generate the ring and sphere graphs from this paper
and perform routing on them~\cite{github}.
In the simulation we draw $2,4, \ldots, 20$ nodes uniformly from the set of all nodes $V$ without replacement.
We then group them into pairs, where each node is used once, and generate paths between them
using the routing algorithms from this paper.
If there are any two paths that contain the same edge then this is counted as a collision.
When multiple collisions occur only the lowest layer collision is counted,
because these are the most expensive VQLs to replace.
Drawing nodes uniformly results in longer distances between nodes than a real-world case
where communications are usually more localised.
We expect that this results in more collisions because of higher loads on lower layers.

Due to the recursive structure of our network and the symmetry in the network topologies considered, we expect tohave a significant amount
of collisions at lower layers. Indeed, drawing pairs of nodes at random on ring would yield nodes on opposite ends of the network with large probability, requiring the use
of the lowest layer. 
Indeed, we estimate that the odds of drawing two nodes such that
a path between them uses the middle VQL approaches $\frac{1}{2}$ for increasing $N$.
This intuition is indeed confirmed in our simulation.
The fraction of requests with a collision for the ring approaches $1$
for increasing number of nodes $N$ and number of concurrent requests.

The sphere simulation results are given in \cref{fig:simCollisions}.
First off, it is clear that path collisions is also a significant problem for the sphere,
but less so than the ring.
Even for few concurrent requests we encounter collisions frequently.
An increasing number of nodes $N$ also does not mitigate many of the collisions.
For larger $N$ it does seem the case that lower layer collisions decrease somewhat,
but this is compensated by more collisions on higher layers.
We conclude that when working with quantum networks it is necessary to address collisions.
Furthermore, on our graphs there are often collisions for concurrent requests on lower layer VQLs.
This indicates there is high load on these links.
If we consider the VQLs on layer 0 and 1, these account for over $50\%$ of collisions,
even more so on the sphere.
Thus the load on lower layer edges should be addressed.

Two types of solutions can be considered, the first being an algorithmic one:
Specifically, one could introduce fallbacks in the routing algorithms to recover from collisions.
It is possible to route around an unavailable VQL in our networks
because each link is made by entanglement swapping two shorter VQLs.
Alternatively, we could handle load by randomizing the routing algorithms
where we slightly increase the (expected) path length
by consuming very expensive and coveted long distance VQLs only with some probability.
Or the graph itself could be modified in some to way to reduce the load on the lower layer edges.
It would also be interesting to see analytical results of the behaviour of our procedure under load.
This would open the way for a qualitative analysis of network robustness.

A second type of solution is to use more sophisticated quantum nodes in the ``core'' of the network, which can store slightly more qubits, meaning that the VQLs at lower levels would in fact correspond to multiple VQLs, that is, multiple entangled pairs, between the same network nodes.

\section{Discussion and open questions}\label{sec:discuss}
We have taken a first step towards formalizing a quantum network and providing efficient routing algorithms for simple network topologies. Much work, however, remains to be done.
The main idea is to exploit the resource of short physical entangled links to create VQLs
that reduce the diameter of the network to a logarithmic size
and keep the degree of the nodes logarithmic. 
Indeed, we emphasize that our inability to send qubits over long physical distances demands tricks such as creating VQLs, and we have already discussed the advantages of doing so ahead of time.

We proceeded to devise efficient routing algorithms on the corresponding graph of VQLs. Equivalently, we can understand such algorithms as a way to marshall the resource of entanglement in the network.
We have exhibited a procedure to create VQLs for the ring and for the sphere network
that minimizes the number of qubits stored at each node
(i.e.\ the number of VQLs using the node as endpoint),
restricts the necessary operations on the nodes only to very few entanglement swapping operations,
and allows for a labelling that leads to a very efficient routing algorithm making only local decisions.
It would be interesting to see how this approach extends to other network structures such as the grid.
We note that the central idea of the routing procedure is to use the labels assigned by hierarchical subdivisions.
The same idea of constructing labels (and the resulting routing algorithm)
could be applied whenever we perform similar subdivisions
- even if we iteratively perform two different forms of subdivisions.
For example for the grid, we essentially want to cover a square.
We start by an empty square and subdivide the outer edges in two.
We then proceed to introduce a node in the middle of the square.
This process results in the overall square being subdivided into four squares.
We can recursively apply the same procedure to subdivide the four new squares to approximate a grid.

Evidently, what we have done here leaves much to be desired since we have deliberately abstracted away many of the underlying issues in creating quantum networks;
however, we believe that by introducing a cost to the VQLs and introducing lifetimes
(i.e., the classical communication delay is captured)
our more general model can form a useful abstraction
that allows classical ideas to be used in the quantum domain in order to manage entanglement resources.
It is clear that such routing protocols do not maximize the overall flow of quantum information in the network,
but rather focus on using only very simple quantum operations and state creation mechanisms.
Given the significant difficulty in experimental implementations such an approach may be more realistic.

In this work we have only worked with bipartite entangled links.
In the quantum domain, entanglement can, however, be shared amongst more than two network nodes. 
For the classical reader, we remark that such multi-node entanglement does not allow the teleportation
of a qubit between any pair of nodes.
However, one can use such multi-node entanglement to create a bipartite entangled link
between one pair of nodes if the other nodes measure their qubits and classically communicate their measurement results.
This operation can quite often be done in parallel to the teleportation or entanglement swapping operation.
In principle, it would thus be possible to create an entangled state amongst all network nodes
in which each node holds precisely one qubit,
and nevertheless we can
-- by performing suitable measurements --
create a VQL between any pair of nodes,
making routing in some sense trivial\footnote{%
	For the classical reader, we remark that the quantum problem is not analogous to broadcasting information, since qubits cannot be copied.
}.
The number of qubits in memory would then merely facilitate a number of simultaneous requests
to create a VQL between the sender and receiver.
In practice, however, multi-node entanglement is difficult to create and it is extremely fragile:
noise such as dephasing at just one of the nodes involved would destroy the entire entangled state in the network.
Evidently, quantum error-correction can help protect such entangled states, but in order to do
so we would need a large amount of redundancy,
i.e.\ a significant amount of extra qubits at each network node.
Even bipartite entanglement is subject to noise,
but in this regime entanglement becomes easier to replenish.
This is useful not just to recreate links,
but also to trade-off error-correction against simply retrying,
i.e., after a certain time we always replenish the entanglement from scratch instead of trying to correct errors. 

We expect that a combination of the more classical techniques of resources allocation we have examined here
and genuine quantum techniques involving more complicated forms of entanglement will be very fruitful.
In particular, one could use quantum methods to create multi-node entangled states shared amongst a smaller subset of nodes~\cite{bruss:graph}.
This would allow a trade-off between partially trivializing routing by using multi-node states amongst the subset,
and the difficulty to create and maintain such entanglement.
We remark that the problem of routing,
i.e.\ path allocation and also managing such entangled resources is a significant departure
from classical protocols in which links are always established between two
nodes.
If we were to work with quantum states shared between three nodes for example,
we would route on a hypergraph in which VQL edges connect three vertices.
Note that in this regime many of the peculiarities we considered here are retained,
e.g.\ entangled hyperlinks can only be used once.
Furthermore, we require additional classical communication to utilize the multi-node entanglement. 

Our general model can also capture the fact that creating long VQLs can be more demanding than short VQLs.
Adding weights to the links of the network and then finding shortest paths on weighted graphs
can reduce the load on long VQLs.
We can also try to deal with collisions by dynamically changing the weights of the links of the network.
For example, by assigning an $\infty$ weight to a link was just used by a routing request until it is replenished.

\subsection*{Acknowledgments}
ES, TI, and SW were supported by STW Netherlands, NWO VIDI and an ERC Starting Grant. 
LM was supported by UK EPSRC under grant EP/L021005/1.
IK was supported by an ERC Starting Grant QCC.
%\printbibliography
% \bibliographystyle{alpha}
\bibliography{network}
\appendix

\subsection{Technical Details for the Ring Network}
\subsubsection{Properties of the VQL Graph}
%--------------------------------%

%First, we observe that the routing graph $G_n$ contains the routing graphs $G_{k}$ for $k\le n$. This observation is useful for proving other properties of the routing graphs as well as for showing the correctness of the routing algorithm we propose in the next section. Second, we show in~\cref{lem:appRingDiameter} that the diameter of $G_n$ is logarithmic in the number of vertices $\abs{V_n}$. This is important to us, since the diameter corresponds to the number of entanglement swaps two network nodes need to perform in the worst case in order to communicate to one another.
%Finally, in \cref{lem:nondecreasingmove} we establish that in order to find a shortest path between vertices $\alpha,\beta\in V_n$ one is never required to use edges from the graph $G_{n+k}$ for $k\ge1$. This fact will be useful when analysing the efficient routing protocol which we propose in the next section.

\noindent In this section we establish some properties of the graphs $G_n$. Most notably we show that the diameter of $G_n$ is logarithmic in the number of nodes in the corresponding ring network (see~\cref{lem:appRingDiameter}).

We start by observing that for all $k\le n$, the routing graph $G_k$ is isomorphic to a subgraph of $G_n$.
This observation will be useful for bounding the diameter of $G_n$ as well as for showing
the correctness of the algorithm for finding the shortest path in \cref{sec:appShortestPath}.

\begin{lemma}\label{lem:appContainment}
For all $n\ge 2$, the subgraph induced by the even vertices of $G_n$ is isomorphic to $G_{n-1}$ via
mapping an even vertex $\alpha$ of $G_n$ to the vertex $\alpha/2$ of $G_{n-1}$.
\end{lemma}

\begin{fancyproof}
Let $H_{n-1} = (\tilde{V}_{n-1},\tilde{E}_{n-1})$ be the subgraph induced by the even vertices of
$G_n$. To establish the lemma statement we show that the map $f: \tilde{V}_{n-1} \to V_{n-1}$ defined by
\begin{align}
  f(\alpha) = \alpha/2
\end{align}
preserves both adjacency and nonadjacency. 
To accomplish this we need to establish that for all $\alpha,\beta\in\tilde{V}_{n-1}$
we have $\set{\alpha, \beta} \in \tilde{E}_{n-1}$ if and only if $\set{f(\alpha),f(\beta)} \in E_{n-1}$.

First, according to the definition of an induced subgraph (see \cref{sec:GraphTheory}), for
all $\alpha,\beta\in\tilde{V}_n$ we have $\{\alpha,\beta\} \in \tilde{E}_{n-1} $ if and only if
$\{\alpha,\beta\} \in E_n$.
Next, from~\eqref{eq:routedge} we get that $\set{\alpha, \beta} \in E_{n}$ if and only if
\begin{align}
& \abs{\alpha-\beta} \equiv \gcdd(\alpha,\beta) \pmod{2^n} \\
	\Leftrightarrow \;
	& \abs[\Big]{\frac{\alpha}{2}-\frac{\beta}{2}} \equiv 
	\frac{\gcdd(\alpha,\beta)}{2} \pmod{2^{n-1}} \\
	\Leftrightarrow \;
& \abs{f(\alpha)-f(\beta)} \equiv q \big( f(\alpha),f(\beta) \big) \pmod{2^{n-1}} 
\label{eq:isomonly}
%\Leftrightarrow \;&\frac{f(\alpha)}{q(f(\alpha),f(\beta))} \equiv \frac{f(\beta)}{q(f(\alpha),f(\beta))}+1 \pmod {\frac{2^{n-1}}{q(f(\alpha),f(\beta))}},
%
%&\frac{a}{q(\alpha,\beta)} \equiv \frac{b}{q(\alpha,\beta)}+1 \pmod {\frac{2^n}{q(\alpha,\beta)}},\\
%\Leftrightarrow \;&\frac{a/2}{q(\alpha,\beta)/2} \equiv \frac{b/2}{q(\alpha,\beta)/2}+1 \pmod {\frac{2^{n-1}}{q(\alpha,\beta)/2}}\,,\\
%\Leftrightarrow \;&\frac{f(\alpha)}{f(q(\alpha,\beta))} \equiv \frac{f(\beta)}{f(q(\alpha,\beta))}+1  \pmod {\frac{2^{n-1}}{f(q(\alpha,\beta))}}\,,\\
%\label{eq:isomonly}
%\Leftrightarrow \;&\frac{f(\alpha)}{q(f(\alpha),f(\beta))} \equiv \frac{f(\beta)}{q(f(\alpha),f(\beta))}+1 \pmod {\frac{2^{n-1}}{q(f(\alpha),f(\beta))}},
\end{align}
where the last equivalence follows from the fact that for distinct even vertices $\alpha$ and $\beta$ we have
\begin{align}
 \frac{\gcdd(\alpha,\beta)}{2} = 
 \gcdd\left(\frac{\alpha}{2},\frac{\beta}{2}\right) = 
% 2^{\min\set{t(\alpha),t(\beta)} - 1} = 
% 2^{\min\set{t(a/2),t(\beta/2)}} =
 \gcdd(f(\alpha),f(\beta)).
\end{align}
To complete the proof it remains to note that \cref{eq:isomonly} holds if any only if $\set{f(\alpha),f(\beta)} \in E_{n-1}$.
\end{fancyproof}

We are now ready to bound the diameter of $G_n$. This is important to us, since the diameter corresponds to the number of entanglement swaps two network nodes need to perform in the worst case in order to communicate to one another.

\begin{lemma}
	For any $n\in\mathbb{N}$, we have that $D(G_n) \leq D(G_{n-1}) +2$ and $D(G_1)=1$, where $D(G)$ is the diameter of a graph $G$.
	In particular, we have $D(G_n) = O(\log N) = O(n)$.
	\label{lem:appRingDiameter}
\end{lemma}

\begin{fancyproof}
From \cref{lem:appContainment} we know that that the subgraph $H_{n-1}$ induced by the even vertices of $G_n$ is isomorphic to $G_{n-1}$ for all $n\ge 2$. Therefore, $D(H_{n-1}) = D(G_{n-1})$.

For any vertex $\gamma$ of $G_n$, we consider an even vertex $\gamma'$.
We let $\gamma'\coloneqq\gamma$ if $\gamma$ is even and we let $\gamma\coloneqq \gamma+1 \pmod{2^n}$ if $\gamma$ is odd.
In the latter case, we have that $\gcdd(\gamma,\gamma')=1$ and hence it follows from~\cref{eq:routedge}
that $\{\gamma,\gamma'\} \in E_{n}$.
Thus, for any vertex $\gamma$ of $G_n$, we have $d(\gamma,\gamma')\le 1$ and $\gamma'\in H_{n-1}$.
Using this simple observation we now have that for any two vertices $\alpha,\beta\in V_n$:
\begin{align}
	d_{G_n}(\alpha,\beta) &\le d(\alpha,\alpha') + d(\alpha',\beta')+ d(\beta,\beta'),\\
				 &\leq d(\alpha',\beta') + 2 \\
		 		 &\leq D(H_{n-1}) + 2\\
		 		 &= D(G_{n-1}) + 2.
\end{align}
Since for any two vertices $\alpha$ and $\beta$ of $G_n$ we have $d_{G_n}(\alpha,\beta) \le D(G_{n-1})+2$,
it follows directly from the definition of diameter that $D(G_n) \le D(G_{n-1})+2$.
\end{fancyproof}

\noindent We now show that if two vertices of $G_{n+k}$ are also contained in the subgraph isomorphic to $G_n$ then
in order to find a shortest path between them we can restrict our attention to the edges in this subgraph.
This will be useful for proving the correctness of the algorithm for finding the shortest path
we propose in \cref{sec:appShortestPath}.

\begin{lemma} \label{lem:appGenNoBackEdge}
For any $n, k \in \mathbb{N}$ and any two vertices $\alpha$ and $\beta$ of $G_n$ we have, $d_{G_n}(\alpha,\beta) = d_{G_{n+k}} (2^k \alpha,2^k \beta)$.
\end{lemma}

\begin{fancyproof}
Let us first establish the $k=1$ case.
It will be useful for us to consider the graph $H_{n} = (\tilde{V}_n,\tilde{E}_n)$ induced by the even
vertices of $G_{n+1}$.
From \cref{lem:appContainment} we know that $G_{n}$ is isomorphic to $H_{n}$ via mapping a vertex
$\alpha$ of $G_n$ to the vertex $2\alpha$ of $H_n$.
Therefore, $d_{G_n}(\alpha,\beta) = d_{H_{n}}(2\alpha,2\beta)$ for all $\alpha,\beta\in V_n$ and to
get that $d_{G_n}(\alpha,\beta) = d_{G_{n+1}}(2a,2b)$, %reach the desired conclusion
it suffices to show that $d_{H_n}(2\alpha,2\beta) = d_{G_{n+1}}(2\alpha,2\beta)$.
We accomplish this using a proof by contradiction. 
To this end, assume that $d_{H_n}(\alpha,\beta) > d_{G_{n+1}}(\alpha,\beta)$ for some $\alpha,\beta\in \tilde{V}_n$. 
Since $H_n$ is an induced subgraph of $G_{n+1}$, our assumption implies that any shortest $(\alpha,\beta)$-path $P_{ab}$ in $G_{n+1}$ must contain vertices from $V_{n+1}\setminus \tilde{V}_n$. Let us consider the subpaths $P_{\alpha'\beta'}$ of $P_{ab}$ which start and end with vertices from $H_n$ but have all other vertices belonging to $V_{n+1}\setminus \tilde{V}_n$. 
For at least one of these subpaths $P_{\alpha'\beta'}$ it must be that $d_{H_n}(\alpha',\beta') > d_{G_{n+1}} (\alpha',\beta')$
as otherwise we could replace all of them with paths of the same length contained entirely in $H_n$.
Let us now focus on some such $P_{\alpha'\beta'}$.
%Since $H_n$ is an induced subgraph of $G_{n+1}$, we can find $\alpha',\beta' \in \tilde{V}_n$ such that $d_{H_n}(\alpha',\beta') > d_{G_{n+1}} (\alpha',\beta')$ and all the inner vertices of any shortest $(\alpha',\beta')$-path $P_{\alpha'\beta'}$ belong to the set $V_{n+1} \setminus \tilde{V}_n$.
%
Since any $k \in {V_{n+1} \setminus \tilde{V}_n}$ is odd and no two odd vertices are adjacent,
we conclude that the path $P_{\alpha'\beta'}$ has length $2$,
i.e., $P_{\alpha'\beta'} = (\alpha',\gamma,\beta')$ for some odd $\gamma$.
This implies that $d_{G_{n+1}}(\alpha',\beta') = 2$.
Furthermore, since any odd vertex $\gamma$ is adjacent to only $\gamma+1$ and $\gamma-1 \pmod {2^{n+1}}$,
we obtain $\abs{\alpha'- \beta'} \equiv 2 \pmod {2^{n+1}}$.
Combining this with the fact that $\alpha'$ and $\beta'$ are even we see that $\alpha'$ is adjacent to $\beta'$
in $H_n$.
So $d_{H_n}(\alpha',\beta') = 1 < 2$, a contradiction.
Therefore, $d_{G_{n+1}}(2\alpha,2\beta) = d_{H_n}(2\alpha,2\beta) = d_{G_n}(\alpha,\beta)$ for any
$\alpha,\beta\in V_n$.

%Now that we have established the $k=1$ case, observe that for any $k$ we have $d_{G_n}(\alpha,\beta) = d_{G_{n+1}}(\alpha,\beta) = \dotso = d_{G_{n+k}}(\alpha,\beta)$, which completes the proof.
Now that we have established the $k=1$ case, observe that for any $k$ we have 
\begin{equation}
  d_{G_{n}}(\alpha,\beta) = d_{G_{n+1}}(2\alpha,2\beta) = d_{G_{n+2}}(2^2\alpha,2^2\beta) =
   \dotso = d_{G_{n+k}}(2^k \alpha, 2^k \beta),
\end{equation}
which completes the proof.
\end{fancyproof}

\noindent When looking for a shortest $(\alpha,\beta)$-path, \cref{lem:appGenNoBackEdge} can help us
to narrow down the choices for the vertex to proceed to after $\alpha$.
\begin{corollary}
Let $\alpha$ and $\beta$ be two distinct vertices of some $G_n$.
If $t(\alpha) \le t(\beta)$ and \mbox{$\alpha_{\pm} \coloneqq a \pm 2^{t(\alpha)} \pmod{2^n}$,} then
\begin{equation}
	d(\alpha_+,\beta)  = d(\alpha,\beta) - 1
	\quad \text{or} \quad
	d(\alpha_-,\beta)  = d(\alpha,\beta) - 1.
\end{equation}
In other words, either $\alpha$ is adjacent to $\beta$ or there exists a shortest $(\alpha,\beta)$-path of the form
\begin{equation}
  \alpha,\alpha_+,\dotsc, \beta \quad \text{or } \quad \alpha,\alpha_-,\dotsc, \beta.
\end{equation}
\end{corollary}

\begin{fancyproof}
Recall \cref{eq:q} and let $k\in\mathbb{N}$ be such than $\gcdd(\alpha,\beta) = 2^k$. Since $t(\alpha)\le t(\beta)$, we have $a =  2^k \alpha'$ and $b = 2^k \beta'$ for some odd  $\alpha'$ and some natural $\beta'$.  By \cref{lem:appGenNoBackEdge} we have that $d_{G_{n}}(\alpha,\beta) = d_{G_{n-k}}(\alpha',\beta')$. Since $\alpha'$ is odd its only neighbors in $G_{n-k}$ are $\alpha'_+ \coloneqq \alpha'+1$ and $\alpha'_- \coloneqq \alpha'-1$, where the arithmetic here and later on is performed modulo $2^n$. So the shortest $(\alpha',\beta')$-path must go through one of these two vertices and $d_{G_{n-k}}(\alpha',\beta') - 1 = d_{G_{n-k}}(\alpha'+1,\beta')$ or $d_{G_{n-k}}(\alpha',\beta') - 1 = d_{G_{n-k}}(\alpha'-1,\beta')$.
Applying \cref{lem:appGenNoBackEdge} again gives 
\begin{gather}
  d_{G_{n-k}}(\alpha'+1,\beta') = d_{G_{n}}(2^k(\alpha'+1),\beta)  = d_{G_{n}}(\alpha_+,\beta)\,,\\
  %\intertext{and}
  d_{G_{n-k}}(\alpha'-1,\beta') = d_{G_{n}}(2^k(\alpha'-1),\beta)  = d_{G_{n}}(\alpha_-,\beta)\,.
\end{gather}
Therefore, $d_{G_{n}}(\alpha,\beta)-1 = d_{G_{n}}(\alpha_+,\beta)$ or $d_{G_{n}}(\alpha,\beta)-1 = d_{G_{n}}(\alpha_-,\beta)$ and the proof is complete.
\end{fancyproof}

\subsubsection{Finding the Shortest Path}\label{sec:appShortestPath}

\begin{figure}
	\centering
	\begin{tikzpicture}
		\node (0) at (0,0) {$0$};
		\node (8) at (4,0) {$8$};

		\draw (0) -- (8);

		\node (4) at (2,2) {$4$};
		\node (12) at (2,-2) {$12$};
		\foreach \x in {0,4,8,12} {%
			\pgfmathparse{int(mod(\x - 4 + 16,16))};
			\ifthenelse{\x=8}{}{\draw (\x) -- (\pgfmathresult);}
		}

		\node (2) at (0,2) {$2$};
		\node (6) at (4,2) {$6$};
		\node (10) at (4,-2) {$10$};
		\node (14) at (0,-2) {$14$};
		\foreach \x in {0,2,...,14} {%
			\pgfmathparse{int(mod(\x - 2 + 16,16))};
			\ifthenelse{\x=8}{}{\draw (\x) -- (\pgfmathresult);}
		}

		\node (1) at (-1,1) {$1$};
		\node (3) at (1,3) {$3$};
		\node (5) at (3,3) {$5$};
		\node (7) at (5,1) {$7$};
		\node (9) at (5,-1) {$9$};
		\node (11) at (3,-3) {$11$};
		\node (13) at (1,-3) {$13$};
		\node (15) at (-1,-1) {$15$};

		\foreach \x in {0,...,15} {%
			\pgfmathparse{int(mod(\x - 1 + 16,16))};
			\ifthenelse{\x=5 \OR \x=6}{}{\draw (\x) -- (\pgfmathresult);}
		}

		\node[above right of=5, draw, circle, minimum size=1.75em] {$\alpha$}
			edge[bend right, ->] (5);
		\node[above right of=6, draw, circle, minimum size=1.75em] {$\alpha_2$}
			edge[bend right, ->] (6);
		\node[above of=4, draw, circle, minimum size=1.75em] {$\alpha_1$}
			edge[->] (4);
		\node[right of=8, draw, circle, minimum size=1.75em] {$\alpha_2^+$}
			edge[->] (8);
\node[right of=11, draw, circle, minimum size=1.75em] {$\beta$}
			edge[ ->] (11);

		\draw[->, red, dashed] (5) -- (6);
		\draw[->, red, dashed] (6) -- (8);
		\draw[->, blue, dashdotted] (5) -- (4);
		\draw[->, blue, dashdotted] (4) -- (8);
	\end{tikzpicture}
	\caption{Example illustration for the proof of \cref{thm:appRingOptimal}.
		If we call $\mathtt{path}(\alpha,\beta)$ with $\alpha=5$ and $\beta=11$ then it invokes
		$\mathtt{bestMove(\alpha)}$ which returns $\alpha_1 = 4$ suggesting to proceed via
		this vertex.
		We argue that if, in fact,  a shortest $(\alpha,\beta)$-path $P$ proceeded via $\alpha_2
		= 6$, then the algorithm can catch up with $P$ in the next step.
		According to \cref{cor:BestMove} we can assume that $P$ proceeds to either $\alpha_2^-
		\coloneqq \alpha_2 - 2 = 4$ or to $\alpha_2^+ \coloneqq \alpha_2 + 2 = 8$.
		If $P$ proceeded along the red dashed edges to $\alpha_2^+ = 8$ then we can use the
		blue dotted edges to reach vertex 8 in two steps by going via the vertex $\alpha_1$
		chosen by the algorithm.
		Finally, note that $P$ cannot proceed to $\alpha_2^-$ from $\alpha_2$ as it would
		contradict the optimality assumptions.
		Therefore, the algorithm's choice to proceed to vertex $\alpha_1 = 4$ and invoke $\mathtt{path}(4,11)$ was indeed optimal.
	}
\label{fig:ringRoutingOptimality}
\end{figure}
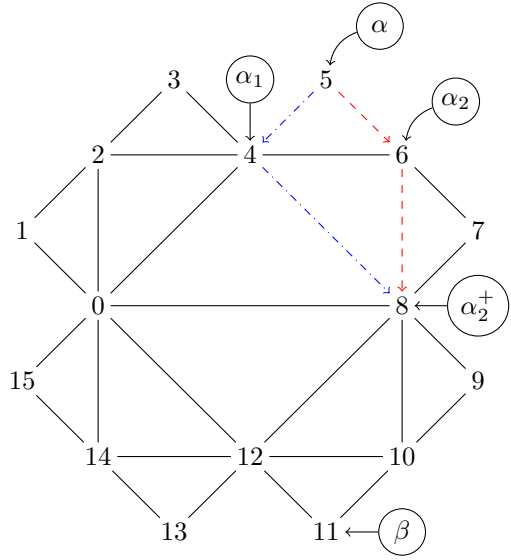

\begin{theorem}\label{thm:appRingOptimal}
	For any $n\ge 1$ and vertices  $\alpha$ and $\beta$ of $G_n$,
	the algorithm $\mathtt{path}(\alpha,\beta)$ (\cref{alg:ring}) returns a shortest $(\alpha,\beta)$-path.
\end{theorem}

\begin{fancyproof}
To find a shortest $(\alpha,\beta)$-path the algorithm chooses a vertex $\gamma\in\set{\alpha,\beta}$ and then
calls $\mathtt{bestMove}(\gamma)$ to find a neighbor $\eta$ of $\gamma$ that must lie on a shortest
$(\alpha,\beta)$-path.
Depending on the choice of $\gamma$, the algorithm then proceeds to a recursive call of $\mathtt{path}(\eta,a)$
or $\mathtt{path}(\eta,\beta)$.
So the algorithm returns a path of the form
\begin{equation}
  \alpha,\eta, \dotsc, \beta
  \quad \text{or } \quad
  \alpha,\dotsc, \eta, \beta\,.
\end{equation}
where $\mathtt{path}(\eta,\beta)$ or $\mathtt{path}(\alpha,\eta)$ has been used to fill in the dots.
Therefore, to show that the algorithm indeed returns a shortest $(\alpha,\beta)$-path it suffices to
argue that $\mathtt{bestMove}(\gamma)$ returns a vertex $\eta$ that lies on a shortest $(\alpha,\beta)$-path.
It will be essential that the algorithm chooses to call $\mathtt{bestMove}(\gamma)$ for a vertex
$v\in\set{\alpha,\beta}$ which satisfies $t(\gamma) = \min\set{t(\alpha),t(\beta)}$. As in earlier
proofs we will perform the arithmetic involving vertex labels modulo $2^n$ without explicit mention.

To simplify notation let us assume that the algorithm chooses $\gamma=\alpha$ (see
\cref{fig:ringRoutingOptimality} for an example).
Note that the choice of $\gamma$ is based on the values $t(\alpha)$ and $t(\beta)$, and $\alpha$ can be chosen
only if $t(\alpha) \le t(\beta)$.
From \cref{cor:BestMove} we know that either $\alpha_+: = a + 2^{t(\alpha)}$ or $\alpha_- \coloneqq \alpha -
2^{t(\alpha)}$ is a good move, i.e., $d(\alpha_+,\beta) = d(\alpha,\beta)-1$ or $d(\alpha_-,\beta)
= d(\alpha,\beta)-1$.
We see that $\mathtt{bestMove}(\alpha)$ will return one of the vertices $\alpha_+$ or $\alpha_-$
depending on the values of $t(\alpha_+)$ and $t(\alpha_-)$.
More precisely, $\mathtt{bestMove}$ will choose $\alpha_1\in\set{\alpha_+,\alpha_-}$ such that $t(\alpha_1)
= \max\set{t(\alpha_+),t(\alpha_-)}$.
Let us denote the vertex not chosen by our algorithm with $\alpha_2$ so that $\set{\alpha_1,\alpha_2}
= \set{\alpha_+,\alpha_-}$.
Since we know that $d(\alpha_1,\beta) = d(\alpha,\beta)-1$ or $d(\alpha_2,\beta) = d(\alpha,\beta)-1$
our goal is to show that whenever $d(\alpha_2,\beta) = d(\alpha,\beta)-1$ it must be that also
$d(\alpha_1,\beta) = d(\alpha,\beta)-1$.
Informally speaking, we want to show that going through $\alpha_2$ is never better than the choice
made by our algorithm which proceeds through $\alpha_1$.
So let us assume that $d(\alpha_2,\beta) = d(\alpha,\beta)-1$ and aim to show that instead of going
through $\alpha_2$ we might as well go through $\alpha_1$.

Let $r$ be the odd number such that $a = r2^{t(\alpha)}$. Then we have
\begin{equation}
  \alpha_+ = (r+1) 2^{t(\alpha)}
  \quad \text{and} \quad
  \alpha_- = (r-1) 2^{t(\alpha)}.
\end{equation}
Since $r$ is odd, both $r+1$ and $r-1$ are divisible by two but precisely one of them is divisible by four.
Since $\mathtt{bestMove}(\alpha)$ chooses the vertex that yields a larger value for the function $t$,
so we know that $\alpha_1$ is divisible by $2^{t(\alpha)+2}$ but $\alpha_2$ is not.
Therefore, we have that $t(\alpha_2) = t(\alpha)+1$ and $t(\alpha_1) \ge t(\alpha)+2$.
Also, for some odd $r_2$ we have
\begin{equation}
	\alpha_2 = r_2 2^{t(\alpha_2)}
	\quad \text{and} \quad
	\alpha_1 \in\set[\big]{(r_2+1)2^{t(\alpha_2)},(r_2-1)2^{t(\alpha_2)}}.
\label{eq:a12}
\end{equation}
We now separately analyze the case when $t(\alpha)<t(\beta)$ and when $t(\alpha)=t(\beta)$. 

First, assume that $t(\alpha) < t(\beta)$.
In this case $t(\alpha_2) \le t(\beta)$.
Since $d(\alpha,\beta)>2$ because $\mathtt{path2}(\alpha,\beta) = \emptyset$,
we know that $\alpha_2\neq \beta$.
Therefore, we can apply \cref{cor:BestMove} to vertices $\alpha_2$ and $\beta$ to conclude that there is a shortest $(\alpha,\beta)$-path of the form
\begin{equation} 
	P^+ \coloneqq \alpha,\alpha_2,\alpha_2^+,\dotsc,\beta
	\quad \text{or} \quad
	P^-\coloneqq \alpha,\alpha_2,\alpha_2^-,\dotsc,\beta\,,
\label{eq:Paths}
\end{equation}
where $\alpha_2^{\pm} \coloneqq \alpha_2 \pm 2^{t(\alpha_2)} = (r_2 \pm 1) 2^{t(\alpha_2)}$.
Recalling \cref{eq:a12} we see that $\alpha_1 = \alpha_2^+$ or $\alpha_1 = \alpha_2^-$. Since the difference between $\alpha_2^+$ and $\alpha_2^-$ is $2^{t(\alpha_2)+1}$, we conclude that
\begin{equation}
	\set{\alpha_2^{+},\alpha_2^-} \subset 
	\set{\alpha_1, \alpha_1 \pm 2^{t(\alpha_2)+1}}\,.
\label{eq:Sets}
\end{equation}
Since $t(\alpha_1) \ge t(\alpha) + 2 = t(\alpha_2)+1$, we know that $\alpha_1$ is divisible by $2^{t(\alpha_2)+1}$.
%Therefore, $\alpha_1$ is equal or adjacent to all the vertices in the right-hand side set of~\eqref{eq:Sets}.
It now follows that $\alpha_1$ is adjacent or equal to both  $\alpha_2^+$ and $\alpha_2^-$.
This observation lets us redirect the paths $P^+$ and $P^-$ from~\eqref{eq:Paths} through $\alpha_1$
instead of $\alpha_2$ without increasing their length.
Since at least one of those paths had length $d(\alpha,\beta)$, we have shown that $d(\alpha_1,\beta)
= d(\alpha,\beta)-1$ as desired.

We now turn to the case when $t(\alpha)=t(\beta)$.
In this case, we have $t(\beta)<t(\alpha_2)$. 
Therefore, from \cref{cor:BestMove} it follows that we can choose $\beta_1 \in \set{ \beta +
2^{t(\beta)},\beta - 2^{t(\beta)}}$ so that $d(\alpha_2,\beta_1)=d(\alpha_2,\beta)-1$.
Since $t(\beta+2^{t(\beta)}) > t(\beta)$ and $t(\beta-2^{t(\beta)})>t(\beta)$ we conclude that $t(\beta_1) \ge t(\beta)+1$.
Recalling that $t(\alpha_2) = t(\alpha) +1 = t(\beta)+1$, we obtain $t(\beta_1) \ge t(\alpha_2)$.
Since $d(\alpha,\beta)>2$, we also know that $\alpha_2 \neq \beta_1$ which lets us apply \cref{cor:BestMove}
to vertices $\alpha_2$ and $\beta_1$.
%Therefore, we can apply \cref{cor:BestMove} to $\alpha_2$ and $b_1$ to obtain that 
This tells us that
$d(\alpha_2^+,\beta_1) = d(\alpha_2,\beta_1)-1$ or $d(\alpha_2^-,\beta_1) = d(\alpha_2,\beta_1)-1$
where $\alpha_2^{\pm} \coloneqq \alpha_2 \pm 2^{t(\alpha_2)}$.
In combination with the fact that both $\alpha_2$ and $\beta_1$ lie on some shortest $(\alpha,\beta)$-path 
(i.e., $d(\alpha_2,\beta_1) = d(\alpha,\beta)-2$) we get that there exists a shortest $(\alpha,\beta)$-path of the form  
\begin{equation} 
	\alpha,\alpha_2,\alpha_2^+,\dotsc,\beta
	\quad \text{or} \quad
	\alpha,\alpha_2,\alpha_2^-,\dotsc,\beta\,.
\end{equation}
%where $\alpha_2^{\pm} \coloneqq \alpha_2 \pm 2^{t(\alpha_2)}$.
Now we note that we are in the same situation as in the previous case.
Therefore, we can finish off our argument using the same steps as taken after~\eqref{eq:Paths}.
\end{fancyproof}
\subsection{Technical Details for the Sphere Network}\label{sec:appendixSphere}
The proofs are structured into two sections, following the organisation of the main part of the paper on the sphere network.
Each subsection handles the corresponding theorems claimed in the main part of the paper.
We start by a short outline of the results proved in each section.

\paragraph{\nameref{sec:appendixGraphProperties}}
We first prove that the network meets the requirements set for the diameter (\cref{prop:diameter})
and the degree of vertices (\cref{prop:vertexDegree}).
This also leads to relations between the number of subdivisions $k$,
the number of vertices $\abs{V_k} = N$, and the number of edges $\abs{E_k}$.

\paragraph{\nameref{sec:appShortestPathStructure}}
The parents of a node have a few simple properties that are formalized in propositions:
the common parent, the connectedness of parents and the layer of a vertex.
We use these properties in proofs to show that there exist certain parents that we can route through.

These are followed by two key theorems for routing, \cref{thm:noHigherEdge} (\nameref{thm:noHigherEdge})
and \cref{thm:threeHops} (\nameref{thm:threeHops}).
\nameref{thm:noHigherEdge} shows that any shortest path between vertices does not contain edges on higher layers than the endpoints of the path.
This greatly reduces the number of possibilities when routing, since we can ignore nodes
that are on a higher layer.
We also show that when a shortest path runs from a high layer to a lower layer
there is always a shortest path that immediately goes to a lower layer in \cref{lem:lowerLayerPath}.
To simplify further proofs we have defined a \emph{triangle} (\cref{def:triangle}),
because some triangles have restrictions on their configurations (\cref{prop:triangleShape,cor:triangleIntoTriangle}).
The \nameref{thm:threeHops} Theorem shows that if two vertices are on the same layer and at least 3 hops apart,
then there must be a shortest path that goes through the parents of both endpoints.
These Theorems already hint towards a routing algorithm that makes use of the parents of
nodes to route, as there almost always seems to be some parent that is on a shortest path,
except when they are already nearby.

\paragraph{\nameref{sec:proofOptimality}}
We will first show that there exists some structure in the graph
if the for the endpoints $\alpha,\beta \in V$ it holds that $d(\alpha,\beta)>6$ in \cref{lem:optimalChoicePath}.
Then we show that the global routing algorithm (\cref{alg:spherePath}) produces a shortest path.
We use Dijkstra's algorithm limited to the 6th-degree neighbourhood
to make sure that $d(\alpha,\beta) >6$.
That allows us to use the \nameref{thm:threeHops} Theorem to guarantee that some parent is in a shortest path.
To choose the best parent, we have defined the Parent and Grandparent Rules (\cref{def:parentRule,def:grandparentRule}).
Using the graph structure from \cref{lem:optimalChoicePath}
we will show that following the Parent and Grandparent Rules
results in an optimal algorithm in \cref{lem:optimalChoice}.

\paragraph{\nameref{sec:appendixLabelling}}
In order to turn our global routing algorithm, which finds a shortest path using information from both the sender and the receiver, into to a local routing algorithm, where the sender only uses local information, we introduce a hierarchical labelling (\cref{eq:label}) scheme which takes the Parent and Grandparent Rules into account, in order to store in the label the nodes which satisfy these rules.
This way the sender knows how to find the next node in a shortest path by looking only at the label of the receiver. 
Needless to say, we need to show that the size of the labelling remains small. A priori, since we are storing the parents, grandparents, etc. it seems that the label grows exponentially.
Nevertheless, we show in \cref{lem:labelBound} that the size of the label is limited to just 3 entries per layer, which makes its total size only logarithmic. 

\paragraph{\nameref{sec:complexity}}
For correctness, we argue that the local algorithm is equivalent
to the global algorithm in \cref{sec:localRoutingAlgorithm}.
Furthermore, it is important that the local routing algorithm is fast and uses little memory.
Since the algorithm is local, we look at the running time and memory size per vertex.
We show that the memory size and the running time both scale logarithmically with the number of vertices $N$ (\cref{thm:spaceComplexityRouting} and \cref{thm:timeComplexityRouting}).

\subsubsection{Definition of the VQL Graph}\label{sec:networkModelTechnical}
In this section we will prove some properties of the subdivided graph
that are important for the physical implementation as well as the usefulness of the network model.
We prove how the number of subdivisions $k$ relates to the number of nodes $|V|$ and edges $|E|$,
and what the maximum degree is of any node in the graph.
The degree is related to the amount of quantum memory required to implement such a node.
Furthermore, we look at the diameter of the graph.
With a small diameter we know that few entanglement swaps are necessary to connect sender and receiver.

\paragraph{Properties of the Routing Graph}\label{sec:appendixGraphProperties}
First we look at the size of $|V| = N$ and $|E|$ depending on the number of subdivisions $k$.
Let $i$ be the current iteration of the subdivision algorithm (\cref{alg:subdivisionIcosahedron}).
Per subdivision, a vertex is placed on each edge $e \in E_i$.
A new vertex is connected to 2 vertices in $V_{i}$,
and 4 neighboring new vertices in $L_{i}$.
The edges to the new vertices should not be counted double,
thus every new vertex adds $(2+4/2) = 4$ edges.
The icosahedron starts with $30$ edges, so that
\begin{align}
	|E_i| &= \begin{cases}
			30 & \text{ if $i=0$} \,,\\
			4|E_{i-1}| & \text{ otherwise}
	\end{cases}\\
	&= 4^i\cdot 30\,.
	\label{eq:numberEdges}
\end{align}
The total number of edges is
\begin{align}
	|E| = |\bigcup_{i=0}^k E_i| &= \sum_{i=0}^k 4^i \cdot 30 \\
		&= \frac{30(1-4^{k+1})}{1-4} \\
		&= 10\cdot 4^{k+1} - 10\,,
\end{align}
where we have used the direct formula for a geometric series.
A vertex is placed on the midpoint of each edge in $E_i$,
so that the number of vertices $V_i$ \eqref{eq:verticesKsubdivisions} is
\begin{align}
|V_i| &=
	\begin{cases}
			12 & \text{ if $i=0$}\\
			|V_{i-1}| + |E_{i-1}| & \text{ otherwise}
	\end{cases} \\
	&= |V_{i-2}| + |E_{i-2}| + |E_{i-1}| \\
	&= |V_0| + |E_0| + \ldots + |E_{i-1}|\\
	&= 12 + \sum_{\ell=0}^{i-1} 4^\ell 30 \\
	&= 12 + \frac{30(1-4^i)}{1-4} \\
	&= 10\cdot 4^i + 2\,.
	\label{eq:numberVertices}
\end{align}
The total number of nodes $|V| = |V_k|$.

Next we analyse the diameter and degree of the graph.
A small degree will allow the algorithm to run with smaller quantum memories,
while a smaller diameter is desired to have less entanglement swaps per communication.
Less swaps results in a lower error rate, since each swap introduces noise into the state.

\begin{proposition}[Graph Diameter]\label{prop:diameter}
	The diameter of the graph $D(G_k)$, where $k \in \mathbb N_0$ is the number of subdivision iterations, is
	\[
		D(G_k) \leq 2k+3 =  \log_2\left(\frac{N-2}{10}\right) + 3\,.
	\]
\end{proposition}

\begin{fancyproof}
	We give a proof using a recurrence relation over $k$.
	Let $D(G)$ be the diameter function on a graph $G$.
	Let $G_k = (V_k, \bigcup_{l=0}^k E_k)$ be the graph at iteration $k$.
	We also need an exact formula for $|V_k|$ which is found in \cref{eq:numberVertices}
	\begin{equation}
		|V_k| = 10\cdot4^k + 2 \iff k 
			= \log_4\left(\frac{|V_k|-2}{10}\right)
			= \frac{1}{2}\log_2\left(\frac{N-2}{r0}\right)\,. \label{eq:nRelatedToK}
	\end{equation}

	\basis $D(G_0) \leq 2\cdot 0 + 3 = 3$, which holds because the icosahedron has a diameter of 3.

	\ih Assume that $D(G_{k-1}) = 2(k-1) + 3$.

	\is
		For any vertex $\alpha \in L_k$ it is possible to reach a vertex in layer $L_{k-1}$ in one step.
		Thus
		\begin{align}
			D(G_k) &\leq D(G_{k-1}) + 2 \\
				&\stackrel{\mathrm{IH}}{=} 2(k-1) + 3 + 2 \\
				&= 2k + 3 \\
				&\stackrel{\eqref{eq:nRelatedToK}}{=} \log_2\left(\frac{N-2}{10}\right) + 3\,.
		\end{align}
		As such the diameter upper bound is proven.
\end{fancyproof}
\begin{proposition}[Vertex Degree]\label{prop:vertexDegree}
	The degree of every vertex $v\in V$ in the subdivided graph as a function of $N$ is upper bounded by
	\[
		\deg(v) \leq 3\log_2\left(\frac{N-2}{10}\right) + 5.
	\]
\end{proposition}
\begin{fancyproof}
	We show that the degree is upper bounded by calculating the number of vertices
	any vertex is connected to, when there have been $k$ subdivisions.
	A vertex on the base icosahedron $\alpha \in V_0$ is connected to 5 vertices $\beta \in V_0$.
	However, any other vertex $\alpha \in L_i$, for $i \ne 0$, is connected to 6 vertices $\beta \in V_i$, 
	because of how the algorithm connects new vertices to their parents and their neighbours.
	In every subdivision iteration $j: j>i$ a new midpoint vertex $\gamma$ is added to $L_{j+1}$,
	and an edge $\{\alpha, \gamma\} \in E_{j+1}$ as well.
	This edge is in turn again subdivided on the next iteration ($j+1$).
	Because the degree of the vertex $\alpha$ is the same as its number of edges,
	each vertex will add as many nodes as its degree in each subdivision iteration.
	Thus the degree of a vertex $\alpha \in V$ is upper bounded by
	\begin{align}
		\deg(\alpha) &\leq \begin{cases}
			5(k+1) & \text{if $\alpha \in V_0$}\\
			6k &\text{otherwise,}
		\end{cases}\\
		&\leq 6k + 5\\
		&\stackrel{\eqref{eq:nRelatedToK}}{=} 3\log_2 \left( \frac{N-2}{10} \right) + 5\,.
	\end{align}
	Thus proving the upper bound on the degree of every vertex.
\end{fancyproof}
\noindent Thus both the degree and the diameter are logarithmic in the number of vertices $N$.

\paragraph{Shortest Path Structure}\label{sec:appShortestPathStructure}
To find an efficient routing algorithm it is useful to know
what kind of structure shortest paths have.
From this structure, we then formulate an efficient routing algorithm.
First we look at the layers of the vertices and their parents.
Later, we look at restrictions on shortest paths in the graph, and how the distance and layer of nodes dictate what nodes the path uses.

A path is a sequence of vertices that are adjacent.
We show that adjacent vertices have a \emph{common parent},
which is of interest when we want to reroute a path through parent nodes.
A common parent of vertices $\alpha,\beta \in V$ is defined as
\begin{align}
	\pi_{\alpha,\beta} &\in \Pi_{\alpha,\beta} = \begin{cases}
		\{\alpha\} & \text{ if } \alpha=\beta\,,\\
		(p(\alpha) \cup \{\alpha\}) \cap (p(\beta) \cup \{\beta\}) & \text {otherwise,}
	\end{cases}
	\label{eq:commonParent}
\end{align}
so that the common parent of a node and itself, is itself.
The common parent of two different nodes is either a common parent in $p(\alpha) \cap p(\beta)$,
or if $\alpha \in p(\beta)$ then it is $\alpha \in \Pi_{\alpha,\beta}$ and vice versa.
We show that two vertices on the same layer $\alpha \edge \beta$
always share a unique common parent.
Furthermore, we show that the two parents of a node are always connected to each other.
\begin{proposition}[Common Parent]\label{prop:commonParent}
	Consider two vertices $\alpha_1, \alpha_2 \in L_k$ so that $\alpha_1 \edge \alpha_2$, for $k \in \mathbb N_1$.
	Then $\alpha_1$ and $\alpha_2$ have a unique common parent $\pi_{\alpha_1,\alpha_2}$.
\end{proposition}
\begin{fancyproof}
	If $\alpha_1 = \alpha_2$ then $\pi_{\alpha_1,\alpha_2} = \alpha_1$.
	We know that $\ly(\alpha_1) = \ly(\alpha_2)$ and $\alpha_1 \edge \alpha_2$.
	On \cref{alg:subdivisionSiblings} of the subdivision algorithm (\cref{alg:subdivisionIcosahedron})
	the vertex $\alpha_1$ is connected to its neighbours $\Gamma = (N(\alpha) \cap L_k)$ of layer $k$.
	In this step two triangles are defined to connect $\alpha_1$ to its neighbours,
	where each triangle only contains edges with at least one endpoint being in $p(\alpha_1)$.
	The edges of these triangles generate the set $\Gamma$ and $\alpha_1$.
	Then by construction it holds that all neighbours $\Gamma$ have a parent in common
	\[p(\alpha_1) \cap p(\gamma) \ne \emptyset \text{ for all } \gamma \in \Gamma.\]
	Since $\ly(\alpha_1) > 0$ the parent $p(\alpha_1)$ exists, idem for $p(\alpha_2)$
	Furthermore, $\alpha_1 \ne \alpha_2$, so
	$p(\alpha_1) \ne p(\alpha_2)$,
	because the edge $p(\alpha_1) \in E$ only generates one vertex.
	Thus there is a unique common parent
	\[
		\pi_{\alpha_1,\alpha_2} \in p(\alpha_1) \cap p(\alpha_2)\,.
	\]
\end{fancyproof}
\begin{proposition}[Parents Connected]\label{prop:parentsConnected}
	Consider some vertex $\alpha \in V_k, k > 0$.
	Then, for its parents $\{\beta_1, \beta_2\} = p(\alpha)$, it holds that there is an edge $\{\beta_1,\beta_2\} \in E$.
\end{proposition}
\begin{fancyproof}
	By construction of the graph, specifically on \cref{alg:subdivisionParents} of the graph generation algorithm (\cref{alg:subdivisionIcosahedron}),
	$\alpha$ is connected to its parents.
	These parents are chosen such that they are connected on \cref{alg:subdivisionMidpoint} of the graph generation algorithm,
	because $\alpha$ is placed as a midpoint on the edge.
	Thus proving that there is an edge $\{\beta_1, \beta_2\} \in E$.
\end{fancyproof}
And finally, we also show that the layer of a vertex depends on the maximum layer of its parents.
\begin{proposition}[Vertex Layer]\label{prop:vertexLayer}
	Consider a vertex $\alpha \in L_k : k > 0$, then 
	\begin{equation}
		\ly(\alpha) = \max_{\beta \in p(\alpha)} \{\ly(\beta)\}+1\,.
	\end{equation}
\end{proposition}
\begin{fancyproof}
	By construction of the graph (\cref{alg:subdivisionMidpoint} of \cref{alg:subdivisionIcosahedron}),
	there is an edge $e=\{\beta_1,\beta_2\} \in E_{k-1}$ of which $\alpha$ is the midpoint.
	The endpoints of this edge are the parents of $\alpha$, $\{\beta_1,\beta_2\} = p(\alpha)$.
	Furthermore, we know that $\ly(\beta_1) \leq k-1$ and $\ly(\beta_2) \leq k-1$,
	because they are parents of $\alpha$.
	Since $e \in E_{k-1}$, we know that $\ly(\beta_1) = k-1$
	or $\ly(\beta_2) = k-1$,
	so $k = \max(\ly(\beta_1), \ly(\beta_2)) + 1$
	because one of them must be on layer $k-1$.
\end{fancyproof}

An important property of the sphere graph, is that when routing between vertices
it is never useful to use edges in $E_k$ for higher $k$ than either vertex.
Let $\alpha,\beta \in V$ be endpoints of a routing request.
Then we can use this proposition to exclude children on a higher layer from being on a shortest path.
Since we will often assume $\ly(\beta) \leq \ly(\alpha)$ this allows us to ignore the children of $\alpha$.
\begin{theorem}[No Higher Edge]
	\label{thm:noHigherEdge}
	If $m$ is the distance between vertices $\alpha_0,\alpha_m \in V$,
	and $k = \max(\ly(\alpha_0),\ly(\alpha_m))$.
	Then, for all shortest paths between $\alpha_0$ and $\alpha_m$ of length $m$, it holds
	that they do not use an edge in $E_h$, where $h > k$.
\end{theorem}

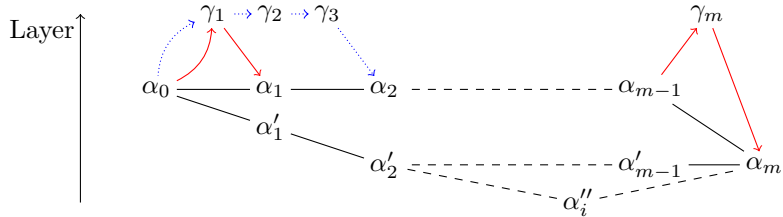
\begin{figure}
	\centering
	\begin{tikzpicture}
		\node (a0) at (0,0) {$\alpha_0$};
		\node (am) at (8,-1) {$\alpha_m$};

		% normal path 1
		\node (a1) at (1.5,-0.5) {$\alpha'_1$}
			edge (a0);
		\node (a2) at (3,-1) {$\alpha'_2$}
			edge (a1);
		\node (am1) at (6.5,-1) {$\alpha'_{m-1}$}
			edge (am)
			edge [dashed] (a2);

		% Common ancestor
		\node (ai) at (5.55, -1.5) {$\alpha''_i$}
			edge [dashed] (a2)
			edge [dashed] (am);

		% normal path 2
		\node (a12) at (1.5,0) {$\alpha_1$}
			edge (a0);
		\node (a22) at (3,0) {$\alpha_2$}
			edge (a12);
		\node (am12) at (6.5,0) {$\alpha_{m-1}$}
			edge (am)
			edge [dashed] (a22);

		% Higher layer path
		\node (g1) at (0.75,1) {$\gamma_1$}
			edge [<-, red, bend left] (a0)
			edge [<-, blue, bend right, densely dotted] (a0)
			edge [->, red] (a12);
		\node (g2) at (1.5, 1) {$\gamma_2$}
			edge [<-, blue, densely dotted] (g1);
		\node (g3) at (2.25, 1) {$\gamma_3$}
			edge [<-, blue, densely dotted] (g2)
			edge [->, blue, densely dotted] (a22);
		\node (gm1) at (7.25, 1) {$\gamma_{m}$}
			edge [<-, red] (am12)
			edge [->, red] (am);

		\draw [->] (-1,-1.5) -- (-1,1) node[anchor=north east] {Layer};
	\end{tikzpicture}
	\caption{An illustration of the proof of \cref{thm:noHigherEdge}.
		In black are given three possible shortest paths.
		It is not possible to construct a shortest path by replacing a hop with going through a higher layer (red, solid arrow).
		Additionally, any number of hops through a higher layer, may be replaced by going through a lower layer instead (blue, dotted arrow).
	}
	\label{fig:noHigherEdge}
\end{figure}

\begin{fancyproof}
	We give a proof by induction over $m$, where $d(\alpha_0, \alpha_m) = m$.
	An illustration of the proof is given in \cref{fig:noHigherEdge}.

	\basis $m=1$.
		By construction of the graph, we know that all edges in $E_i$ are subdivided to obtain $E_{i+1}$.
		This continues recursively until the highest layer is reached.
		This implies that
		\begin{align}
			\forall \alpha_0 \forall \alpha_1 \in V_k\,, \quad \{\alpha_0,\alpha_1\} \notin E_h \label{eq:lowerNodesHigherEdge}\,.
		\end{align}
		Thus there is no shortest path of length 1 that uses an edge in $E_h$.

	\ih
		If there are two vertices $\alpha_0, \alpha_m \in V$ such that the distance between them is of length $m = \ell-1$
		and $k = \max(\ly(\alpha_0), \ly(\alpha_m))$,
		then there exists no path of length $\ell-1$ that uses an edge in $E_h, h>k$.

	\is $m=\ell$.
		First consider all shortest paths $P_{\alpha_0,\beta_{m-1}}$ of length $m-1$ from $\alpha_0$ to a vertex $\beta_{m-1}$ that satisfy
		\begin{equation*}
			\beta_{m-1} \in (V_k \cap N(\alpha_m))\,.
		\end{equation*}
		By the Induction Hypothesis (IH) all $P_{\alpha_0,\beta_{m-1}}$ cannot contain an edge in $E_h$,
		or they would be longer than $m-1$.
		This means that for any edge in $E_k$, two or more edges have to be traversed in $E_h$.
		When $P_{\alpha_0,\beta_{m-1}}$ is extended using an edge in $E_h$,
		then the destination cannot be reached in one hop.
		Thus the path using an edge in $E_h$ is longer than $m$ hops.

		\begin{subproof}[Subproof No Higher Layer Vertex]
			Let us prove by contradiction that there are also no paths of length $m-1$ to some vertex on a higher layer
			$\gamma_{m-1} \in L_i : i > k$
			\begin{equation*}
				P_{\alpha_0,\gamma_{m-1}} = \alpha_0, \gamma_1, \ldots, \gamma_{m-2}, \gamma_{m-1}
			\end{equation*}
			that satisfy $\gamma_{m-1} \in N(\alpha_m)$,
			such that $\alpha_m$ can be reached in $m$ hops through $\gamma_{m-1}$.
			For a contradiction assume there is some path $P_{\alpha_0,\gamma_{m-1}}$.
			Then we know that $\gamma_{m-2} \not\in p(\gamma_{m-1})$,
			or else $\{\gamma_{m-2}, \alpha_m\} \in E$ according to \cref{prop:parentsConnected}
			because also $\alpha_m \in p(\gamma_{m-1})$ as $\ly(\alpha_m) < \ly(\gamma_{m-1})$.
			This would contradict that the distance between $\alpha_0$ and $\alpha_m$ is $m$.
			Nor is $\gamma_{m-1} \in p(\gamma_{m-2})$ because that would contradict the IH
			for the path $P_{\alpha_0, \gamma_{m-1}}$,
			as it would use an edge to a higher layer than $\ly(\gamma_{m-1})$.
			So
			\begin{equation}
				\ly(\gamma_{m-2}) = \ly(\gamma_{m-1})\,.
			\end{equation}
			\cref{prop:commonParent} implies there is some unique and distinct common parent $\pi_{\gamma_{m-2},\gamma_{m-1}}$.
			Let us call this parent $\pi_{\gamma_{m-2}, \gamma_{m-1}} = \pi_{2,1}$ for simplicity.
			The path
			\begin{align*}
				P_{\alpha_0,\gamma_{m-2}} = \alpha_0,\gamma_1,\ldots,\gamma_{m-2}
			\end{align*}
			has length $m-2$,
			and \cref{prop:parentsConnected} implies that $\pi_{2,1} \edge \alpha_m$,
			because $\alpha_m \in p(\gamma_{m-1})$.
			Thus it is possible to construct a shortest path of length $m-1$
			\begin{align*}
				P_{\alpha_0,\pi_{2,1}} = P_{\alpha_0,\gamma_{m-2}} \doubleplus \pi_{2,1}\,.
			\end{align*}
			This must be a shortest path, or else the distance between $\alpha_0$ and $\alpha_m$ would have been less than $m$.

			However, $P_{\alpha_0, \pi_{2,1}}$ now contains an edge
			\begin{equation}
				\{\gamma_{m-2}, \pi_{2,1}\} \in E_i\,,
			\end{equation}
			where $i = \ly(\gamma_{m-2}) = \ly(\gamma_{m-1}) > \ly(\alpha_0)$.
			But $i > \ly(\pi_{1,2})$, so $i > \max\{\ly(\pi_{2,1}), \ly(\alpha_0)\}$,
			thus contradicting the IH.
			Hence, we have shown that there is no shortest path from $\alpha_0$ to $\pi_{2,1}$
			which uses an edge on a higher layer than $\max \{\alpha_0,\pi_{2,1}\}$.
		\end{subproof}

		\noindent Thus we can conclude that the only paths of length $m-1$ are $P_{\alpha_0,\beta_{m-1}}$,
		for which we have already proven that the Theorem holds.
\end{fancyproof}

Now that we know higher layers are not necessary for routing,
we can start analysing the relation between parents of nodes and shortest paths.
We show that when routing from a vertex $\alpha \in V$ to a vertex $\beta \in V$, where $\ly(\beta) < \ly(\alpha)$,
there is always a shortest path that goes through some $\pi_\alpha \in p(\alpha)$.
And since the $(\pi_\alpha,\beta)$-path must also be shortest,
we can conclude that any vertex in that path must be on a lower or equal layer to $\min(\ly(\pi_\alpha), \ly(\beta))$
according to \cref{thm:noHigherEdge}.
Thus all nodes on an $(\alpha,\beta)$-path are on a lower layer than $\alpha$, except for $\alpha$ itself.
\begin{lemma}[Lower Layer Path]\label{lem:lowerLayerPath}
	Consider vertices
	\begin{equation}	
		\alpha_0, \alpha_m \in V : \ly(\alpha_1) < \ly(\alpha_m),
	\end{equation}
	with $d(\alpha_0,\alpha_m) = m$.
	Then, there exists a shortest path also of length $m$ that contains only vertices $\beta \in V_h$, with $ h < \ly(\alpha_0)$
	between $\alpha_0$ and $\alpha_m$, except $\alpha_0$.
\end{lemma}
\begin{fancyproof}
	We give a proof by induction over the distance $m$ from $\alpha_0$ to $\alpha_m$.

	\basis $m=1$:
		There is a direct hop from $\alpha_0$ to $\alpha_1$.
		This path satisfies the lemma.

	\ih
		If there exists a shortest path $P_{\alpha_0, \alpha_{\ell-1}}$ of length $m = \ell-1$ between vertices
		\begin{equation}	
			\alpha_0, \alpha_m \in V : \ly(\alpha_0) < \ly(\alpha_m),
		\end{equation}
		then there exists a shortest path also of length $\ell-1$ that contains only vertices $\beta \in V_h$, with $ h <\ly(\alpha_0)$.

	\is $m=\ell$:
		We distinguish two cases depending on the first hop in the path.
		\begin{enumerate}
			\item The first case is $\ly(\alpha_0) = \ly(\alpha_1)$.
				We can apply the Induction Hypothesis (IH) to a shortest path from $\alpha_1$ to $\alpha_\ell$ to get
				\begin{equation}
					P_{\alpha_1,\alpha_\ell} = \alpha_1, \beta_2, \ldots, \alpha_\ell\,.
				\end{equation}
				By prefixing $\alpha_0$ to this path we get another shortest path between $\alpha_0$ and $\alpha_\ell$
				that almost satisfies the lemma, except for $\alpha_1$.
				We know that $\beta_2 \in p(\alpha_1)$, because $\beta_2 \edge \alpha_1$ and $\ly(\beta_2) < \ly(\alpha_1)$.
				Furthermore, there is a common parent $\pi_{\alpha_0,\alpha_1}$ between $\alpha_0$ and $\alpha_1$ (\cref{prop:commonParent}).
				Parents are connected according to \cref{prop:parentsConnected} so $\pi_{\alpha_0, \alpha_1} \edge \beta_2$.
				We can thus make a shortest path from $\alpha_0$ to $\alpha_\ell$ by replacing $\alpha_1$ with $\pi_{\alpha_0,\alpha_1}$
				\begin{equation}
					P_{\alpha_0, \alpha_\ell} = \alpha_0, \pi_{\alpha_0,\alpha_1}, \beta_2, \ldots, \alpha_m\,.
				\end{equation}
				This path fulfills the lemma, because $\ly(\pi_{\alpha_0,\alpha_1}) < \ly(\alpha_0)$
				and the rest of the path already fulfilled it.
			\item Otherwise, it must be that $\ly(\alpha_0) > \ly(\alpha_1)$.
				The subpath from $\alpha_1$ to $\alpha_m$ must not use any edges of a layer
				higher than $\max(\ly(\alpha_1),\ly(\alpha_m))$ according to \cref{thm:noHigherEdge}.
				We know that $\ly(\alpha_0) > \ly(\alpha_1)$ and $\ly(\alpha_0) > \ly(\alpha_m)$,
				thus $\ly(\alpha_0) > \max(\ly(\alpha_1), \ly(\alpha_m))$.
				All vertices from $\alpha_1$ to $\alpha_m$ must thus be on a lower layer than $\alpha_0$,
				so the path already fulfills the lemma. \qedhere
		\end{enumerate}
\end{fancyproof}

A useful definition in the discussion of graph structures is that of a triangle.
\begin{definition}[Triangle]\label{def:triangle}
	A ``triangle'' is defined to be three distinct and pairwise adjacent vertices,
	$\beta_1,\beta_2,\beta_3$, and is noted down as $\{ \beta_1,\beta_2,\beta_3 \}$.
\end{definition}
The parents of two adjacent nodes form a triangle.
This triangle has two configurations, either one node is on a lower layer,
or all nodes are on the same layer.
We prove this in the following Proposition.
\begin{proposition}[Triangle Shape]\label{prop:triangleShape}
	Consider two vertices $\alpha_1,\alpha_2 \in L_k : k \in \{1,2,\ldots\}$
	and $\alpha_1 \edge \alpha_2$.
	Then parents $p(\alpha_1) \cup p(\alpha_2)$ form a triangle
	$\{\beta_1,\beta_2,\beta_3\}$,
	where $\ly(\beta_1) = \ly(\beta_2) \geq \ly(\beta_3)$.
\end{proposition}
\begin{fancyproof}
	First we show that the parents $p(\alpha_1) \cup p(\alpha_2)$ form a triangle.
	The vertices $\alpha_1$ and $\alpha_2$  are adjacent and on the same layer,
	so there must be a distinct common parent $\pi_{\alpha_1,\alpha_2}$
	according to \cref{prop:commonParent}.
	Let us call this common parent $\gamma_2 = \pi_{\alpha_1, \alpha_2}$.
	Furthermore, there are parents $\gamma_1 \in p(\alpha_1)$ and $\gamma_3 \in p(\alpha_2)$
	which are distinct from the common parent $\gamma_2$.

	We will show that $\gamma_1,\gamma_2$ and $\gamma_3$ are all pairwise adjacent so that they form a triangle.
	The edge $\alpha_1 \edge \alpha_2$ is created in \cref{alg:subdivisionSiblings}
	of the graph subdivision algorithm (\cref{alg:subdivisionIcosahedron}),
	because $\ly(\alpha_1) = \ly(\alpha_2)$.
	The construction algorithm step implies
	there is a triangle $\{\gamma_1,\gamma_2,\gamma_x\}$,
	where $\gamma_x$ is a parent of $\alpha_2$.
	Since $\alpha_2$ has only one other parent besides $\gamma_2$,
	we must have that $\gamma_x = \gamma_3$.
	Thus there exists a triangle $\{\gamma_1, \gamma_2, \gamma_3\}$
	which is equal up to permutation to $\{\beta_1, \beta_2, \beta_3\}$.

	We will now show that $\ly(\beta_1) = \ly(\beta_2) \geq \ly(\beta_3)$.
	Assume for a contradiction that
	\begin{equation}
		\ly(\beta_1) = \ly(\beta_2) < \ly(\beta_3)\,.\label{eq:higherLayerBeta}
	\end{equation}
	We distinguish two cases:
	\begin{enumerate}
		\item In the first case $\{\beta_1, \beta_2\} = p(\alpha_i)$ for $i \in \{1,2\}$.
			W.l.o.g.\ assume $\alpha_i = \alpha_1$.
			From \cref{prop:vertexLayer} we see that
			\begin{equation}
				\ly(\alpha_1) = \ly(\beta_1) -1 < \ly (\beta_3) - 1 = \ly(\alpha_2)\,,\label{eq:triangleShapeLayers}
			\end{equation}
			which contradicts the assumption $\ly(\alpha_1) = \ly(\alpha_2)$.

		\item Otherwise, $\beta_3$ is a common parent, so that $\beta_3 = \pi_{\alpha_1,\alpha_2}$.
			There must be some child $\alpha_3$ on the edge $\beta_1 \edge \beta_2$,
			where $\alpha_1 \ne \alpha_3 \ne \alpha_2$.
			By construction of the graph (\cref{alg:subdivisionSiblings} of \cref{alg:subdivisionIcosahedron})
			there must be edges $\alpha_1 \edge \alpha_3 \edge \alpha_2$.
			Similarly to \cref{eq:triangleShapeLayers}
			\begin{equation}
				\ly(\alpha_1) = \ly(\beta_3) - 1 < \ly(\beta_1) - 1 = \ly(\alpha_3)\,.
			\end{equation}
			This implies $\alpha_3 \in p(\alpha_1)$ since $\alpha_1 \edge \alpha_3$.
			However $\alpha_1$ already has two parents,
			and it is not possible for a vertex to have more than two parents.
			Thus this leads to a contradiction.
	\end{enumerate}
	In all cases the assumption leads to a contradiction,
	thus proving the Proposition.
\end{fancyproof}
Furthermore, the parents of a triangle on the same layer also form a triangle
with the same restrictions as in \cref{prop:triangleShape}.
\begin{corollary}[Triangle Into Triangle]\label{cor:triangleIntoTriangle}
	Consider vertices $\alpha_1,\alpha_2,\alpha_3 \in L_k : k\in \{1,2,\ldots\}$ that form a triangle $\{\alpha_1,\alpha_2, \alpha_3\}$.
	Then parents $p(\alpha_1) \cup p(\alpha_2) \cup p(\alpha_3)$ form a triangle
	$\{\beta_1,\beta_2,\beta_3\}$,
	where $\ly(\beta_1) = \ly(\beta_2) \geq \ly(\beta_3)$.
\end{corollary}
\begin{fancyproof}
	From \cref{prop:triangleShape} we know that the parents form a triangle
	$p(\alpha_1) \cup p(\alpha_2) = \{\beta_1, \beta_2, \beta_3\}$,
	where $\ly(\beta_1) = \ly(\beta_2) \geq \ly(\beta_3)$.
	It remains to show that $p(\beta_3) \subset \{\beta_1, \beta_2, \beta_3\}$.
	The triangle $\{\beta_1, \beta_2, \beta_3\}$ only contains one more vertex on the same layer
	that both $\alpha_1$ and $\alpha_2$ are connected to,
	since each of the three edges in the triangle is subdivided to one vertex
	in the subdivision algorithm (\cref{alg:subdivisionIcosahedron}).
	Thus that vertex must be $\alpha_3$ or it would not be adjacent to $\alpha_1$ and $\alpha_2$.
	This also means that both parents $p(\alpha_3)$ are in the triangle $\{\beta_1, \beta_2, \beta_3\}$,
	proving the Corollary.
\end{fancyproof}

We show that when two vertices $\alpha,\beta \in V$ are on the same layer and $d(\alpha,\beta) \geq 3$,
then there is a shortest path between $\alpha$ and $\beta$  that uses only lower layer vertices,
except $\alpha$ and $\beta$.
So when we can guarantee that two vertices are some distance apart,
we may assume that there is a parent in $\pi_\alpha \in p(\alpha)$ and a parent in $\pi_\beta \in p(\beta)$
that have $d(\pi_\alpha,\pi_\beta) = d(\alpha,\beta) -2$.
Thus there are some parents that are on a shortest path between $\alpha$ and $\beta$.
\begin{theorem}[Three Hops]
	\label{thm:threeHops}
	Consider a shortest path $P_{\alpha_0,\alpha_{m}}$ of length $m \geq 3$
	between two vertices on the same layer $\alpha_0, \alpha_{m} \in L_k, k>0$.
	Then there exists a path also of length $m$ that contains only vertices in
	$V_h, h < k$ between $\alpha_0$ and $\alpha_{m}$,
	except for $\alpha_0$ and $\alpha_{m}$.
\end{theorem}

\begin{figure}
	\centering
	\begin{tikzpicture}
		% Alpha
		\node (p12) at (0,0) {$\pi_{\alpha_1,\alpha_2}$};
		\node (p34) at (4,0) {$\pi_{\alpha_3,\alpha_4}$}
			edge [<-, bend left, red, densely dashed] (p12);
		\node (p23) at (2,3.5) {$\pi_{\alpha_2,\alpha_3}$}
			edge [bend right, mediumlightgray] (p12)
			edge [bend left, mediumlightgray] (p34);

		% Beta
		\node (a2) at (1,1.75)	{$\alpha_2$}
			edge (p12)
			edge (p23);
		\node (a5) at (2,0) [mediumlightgray]	{$\alpha_5$}
			edge [mediumlightgray] (p34)
			edge [mediumlightgray] (p12)
			edge [mediumlightgray] (a2);
		\node (a3) at (3,1.75) {$\alpha_3$}
			edge (p23)
			edge (p34)
			edge [mediumlightgray] (a5)
			edge [<-, blue] (a2);

		\node (a1) at (-1, 1.75) {$\alpha_1$}
			edge [->, red, densely dashed] (p12)
			edge [->, blue] (a2);
		\node (a4) at (5, 1.75) {$\alpha_4$}
			edge [<-, red, densely dashed] (p34)
			edge [<-, blue] (a3);
	\end{tikzpicture}
	\caption{An illustration for the proof of \cref{thm:threeHops}, in the case that $\ly(\alpha_2) = \ly(\alpha_3)$.
	There exists a triangle $\{\pi_{\alpha_1,\alpha_2}, \pi_{\alpha_2,\alpha_3}, \pi_{\alpha_3,\alpha_4}\}$
	such that there is an edge $\pi_{\alpha_1,\alpha_2} \edge \pi_{\alpha_3,\alpha_4}$.
	The original path $P_{\alpha_1,\alpha_4}$ (blue, solid arrow) can
	thus be rerouted through a lower layer $\hat P_{\alpha_1,\alpha_4}$ (red, dashed arrow).
}
	\label{fig:threeHops}
\end{figure}
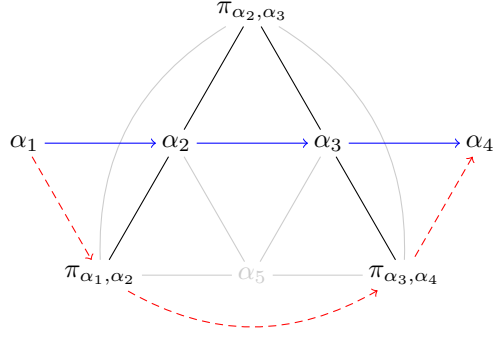
\begin{fancyproof}
	We give an inductive proof over $m$,
	the distance between a pair of vertices on the same layer $k$.

	\basis
	Let there be a shortest path of length $m=3$ given by $\alpha_0,\alpha_1, \alpha_2, \alpha_3$.
	We distinguish three cases depending on the layer of $\alpha_1$ and $\alpha_2$:
	\begin{enumerate}
		\item The first case is $\ly(\alpha_0) = \ly(\alpha_1) = \ly(\alpha_2) = \ly(\alpha_3)$.
			This case is illustrated in \cref{fig:threeHops}.
			According to \cref{prop:commonParent} there must be a unique common parent
			of $\alpha_0$ and $\alpha_1$ called $\pi_{\alpha_0, \alpha_1}$ which we will shorten to $\pi_{0,1}$.
			Furthermore, there are common parents $\pi_{1,2}$ and $\pi_{2,3}$,
			the common parents of $\alpha_1, \alpha_2$ and $\alpha_2,\alpha_3$ respectively.
			It must be the case that $\pi_{0,1} \ne \pi_{2,3}$ or there would exist a shorter path from $\alpha_0$ to $\alpha_3$.
			If either $\pi_{0,1} = \pi_{1,2}$ or $\pi_{1,2} = \pi_{2,3}$ then there exists a path
			\begin{equation}
				P_{\alpha_0, \alpha_3} = \alpha_0, \pi_{0,1}, \pi_{2,3}, \alpha_3\,.
			\end{equation}
			Otherwise $\pi_{0,1} \ne \pi_{1,2} \ne \pi_{2,3}$.
			According to \cref{prop:triangleShape} there must be a triangle $\{\beta_1, \beta_2, \beta_3\} = p(\alpha_1) \cup p(\alpha_2)$.
			Because all common parents are distinct, we can conclude that
			\begin{equation}
				p(\alpha_1) \cup p(\alpha_2) = \{\pi_{0,1}, \pi_{1,2}\} \cup \{\pi_{1,2}, \pi_{2,3}\} = \{\pi_{0,1}, \pi_{1,2}, \pi_{2,3}\} = \{\beta_1, \beta_2, \beta_3\}\,.
			\end{equation}
			Thus $\pi_{0,1} \edge \pi_{2,3}$ exists and we can create a path that fulfills the Theorem
			\begin{equation}
				P_{\alpha_0, \alpha_3} = \alpha_0, \pi_{0,1}, \pi_{2,3}, \alpha_3\,.
			\end{equation}
		\item The second case is when $\ly(\alpha_1) < \ly(\alpha_0)$ or $\ly(\alpha_2) < \ly(\alpha_0)$.
			W.l.o.g.\ assume that $\ly(\alpha_1) < \ly(\alpha_0)$.
			It must be the case that $\alpha_1 \in p(\alpha_2)$, because $\alpha_1 \edge \alpha_2$ 
			and $\ly(\alpha_1) < \ly(\alpha_2)$.
			The other parent of $\alpha_2$ must be the common parent with $\alpha_3$, $\pi_{2,3}$,
			or there would be a shorter path.
			Because parents are connected according to \cref{prop:parentsConnected} we can create
			a path that fulfills the theorem
			\begin{equation}
				P_{\alpha_1, \alpha_3} = \alpha_1, \alpha_2, \pi_{2,3}, \alpha_3\,.
			\end{equation}
		\item Otherwise, it must be that $\ly(\alpha_1) < \ly(\alpha_0)$ and $\ly(\alpha_2) < \ly(\alpha_0)$.
			In this case the path already fulfills the theorem.
	\end{enumerate}

	\ih
	Assume that for all $\alpha_0, \alpha_m \in L_k$ with distance $3 \leq m < \ell$
	there exists a shortest path that contains only vertices $\beta \in V_h, h < k$
	between $\alpha_0$ and $\alpha_{m}$, except for $\alpha_0$ and $\alpha_{m}$.

	\is $m=\ell$:
	Let there be a shortest path between $\alpha_0$ and $\alpha_\ell$ of length $\ell$
	\begin{equation}
		P_{\alpha_0,\alpha_{\ell}} = \alpha_0, \ldots, \alpha_{\ell-1}, \alpha_\ell\,.
	\end{equation}
	We distinguish cases by the relation between $\ly(\alpha_{\ell-1})$ and $\ly(\alpha_\ell)$.
	Note that $\ly(\alpha_{\ell-1}) \not> \ly(\alpha_\ell)$ because of \cref{thm:noHigherEdge}.
	\begin{enumerate}
		\item
			Consider the case that $\ly(\alpha_{\ell-1}) = \ly(\alpha_\ell)$ or its synonym $\ly(\alpha_0) = \ly(\alpha_1)$.
			Assume w.l.o.g.\ $\ly(\alpha_0) = \ly(\alpha_1)$.
			Then we can create a shortest path of length $\ell-1$ between $\alpha_1$ and $\alpha_{\ell}$
			using the Induction Hypothesis (IH)
			\begin{equation}
				P_{\alpha_1,\alpha_{\ell}} = \alpha_1,\beta_2, \ldots, \beta_{\ell-1},\alpha_{\ell}
			\end{equation}
			that uses edges in $E_h$, except the first and last hop.
			Extend $P_{\alpha_1,\alpha_{\ell}}$ with a hop to $\alpha_0$
			so that we get a new shortest path from $\alpha_0$ to $\alpha_\ell$ of length $\ell$
			\begin{equation}
				P'_{\alpha_0,\alpha_{\ell}} = \alpha_0, \alpha_1, \beta_2, \ldots, \beta_{\ell-1},\alpha_\ell\,.
				\label{eq:inductionHypothesisPath}
			\end{equation}
			We know that $\beta_2 \in p(\alpha_1)$,
			because it must be in a lower layer than $\alpha_2$ (IH).
			Additionally, there must be some common parent $\pi_{\alpha_0, \alpha_1}$ (\cref{prop:commonParent}).
			Parents are connected according to \cref{prop:parentsConnected}.
			Thus we can construct a shortest path of length $\ell$ which meets the criteria of the theorem by replacing
			$\alpha_1$ with $\pi_{\alpha_0,\alpha_1}$
			\[
				\hat P_{\alpha_0,\alpha_{\ell}} = \alpha_0, \pi_{\alpha_0,\alpha_1}, \beta_2, \ldots, \beta_{\ell-1},\alpha_\ell\,.
			\]

		\item\label{item:threeHopsLowerLayer}
			Otherwise, it holds that $\ly(\alpha_0) > \ly(\alpha_1)$
			and $\ly(\alpha_\ell) > \ly(\alpha_{\ell-1})$.
			According to \cref{thm:noHigherEdge} we
			know that the path between $\alpha_1$ and $\alpha_{\ell-1}$
			cannot contain edges to a layer higher than $\max(\ly(\alpha_1), \ly(\alpha_{\ell-1}))$, since it is a shortest path.
			We know that $\ly(\alpha_0) = \ly(\alpha_m)$,
			as such $\ly(\alpha_0) > \ly(\alpha_1)$ and $\ly(\alpha_0) > \ly(\alpha_{\ell-1})$.
			Thus $\ly(\alpha_0) = \ly(\alpha_m) > \max(\ly(\alpha_1), \ly(\alpha_{\ell-1}))$.
			All vertices from $\alpha_1$ to $\alpha_m$ must thus be on a lower layer than
			$\alpha_0$ and $\alpha_m$,
			so the path fulfills the theorem.
	\end{enumerate}
	In all cases it is possible to construct a path according to the theorem.
\end{fancyproof}

\noindent We use this Theorem and also \cref{lem:lowerLayerPath} extensively for our routing algorithm,
because they show that routing to the parents of nodes is always possible if
we can provide guarantees on the distance and relative layering.

\subsubsection{Routing Algorithm}\label{sec:technicalDetails}
In this section we provide the proofs of the theorems used in \cref{sec:routing}.
First we prove the optimality of the global routing algorithm (\ref{sec:proofOptimality}), then the optimality of
the local routing algorithm with labelling (\ref{sec:appendixLabelling})
and finally we provide the analysis of the resources needed for our local routing algorithm (\ref{sec:complexity}).

\paragraph{Global Routing Algorithm}\label{sec:proofOptimality}
In this section we prove the optimality of the routing algorithm given in \cref{alg:spherePath}.
Consider vertices $\alpha, \beta \in V$ so that $\ly(\alpha) \geq \ly(\beta)$.
We assume for a contradiction in \cref{lem:optimalChoicePath,lem:optimalChoice}
that when $d(\alpha, \beta) > 6$ the step the routing algorithm takes is to a vertex $\pi_1$,
for which it holds that
\begin{equation}
	d(\pi_1, \beta) > d(\alpha_, \beta) -1\,.
\end{equation}
In other words, we assume that the algorithm takes a suboptimal step.
From this assumption we get a contradiction, thus
showing that for $d(\alpha, \beta) >6$ the algorithm always makes an optimal choice.
Then, we show that the algorithm also makes an optimal choice if $\ly(\alpha) < \ly(\beta)$
and/or $d(\alpha, \beta) \leq 6$, thus showing it is always optimal (\cref{thm:sphereOptimal}).

\begin{lemma}[Graph Structure]\label{lem:optimalChoicePath}
	Consider two vertices $\alpha, \beta \in V$, so that $\ly(\alpha) \geq \ly(\beta)$ and $d(\alpha,\beta) > 6$.
	Assume that the routing algorithm (\cref{alg:spherePath}) starting at $\alpha$ chooses to go to $\pi_1$,
	where $d(\pi_1, \beta) > d(\alpha,\beta) -1$.
	Then there exists a shortest $(\alpha,\beta)$-path of the form
	\begin{equation}
		P_{\alpha,\beta} = \alpha, \pi_2, \gamma_2, \eta_2, \ldots, \beta\,,
	\end{equation}
	where $\pi_2 \in p(\alpha), \gamma_2 \in p(\pi_2)$.
	Moreover, if $\ly(\pi_1) = \ly(\pi_2)$ then $\eta_2 \in p(\gamma_2)$.
\end{lemma}

\begin{fancyproof}
	We show that the lemma holds by analysis of three different cases regarding the relation of the layers of $\alpha$ and $\beta$.
	The algorithm will choose a parent on the first hop of the path, since $d(\alpha,\beta) > 6$,
	and we assumed that the first hop is to $\pi_1$,
	so $\pi_1 \in p(\alpha)$.
	Furthermore, we know that $\ly(\alpha) \geq \ly(\beta)$ and $d(\alpha,\beta) > 6$
	so we can apply \cref{lem:lowerLayerPath} or \cref{thm:threeHops}.
	In these statements it is shown that there is a shortest $(\alpha,\beta)$-path
	that has a $\pi_2 \in p(\alpha)$ at the second position in the path,
	and in the case of \cref{thm:threeHops} also a $\pi \in p(\beta)$ in the second to last position.
	%These statements imply that there is a shortest path through some $\pi_2 \in p(\alpha)$.
	We will use the Lemma and Theorem as functions that give some $\pi_2$,
	and in the case of \cref{thm:threeHops}, $\pi$.
	We assumed that the algorithm takes a suboptimal choice, so $\pi_1 \ne \pi_2$.
	Since the algorithm does not choose $\pi_2$ it must be that $\ly(\pi_1) \leq \ly(\pi_2)$ according to the Parent Rule (\cref{def:parentRule}),
	thus
	\begin{equation}
		\ly(\pi_2) = \ly(\alpha) -1\,,\label{eq:optimalChoicePathPiLayer}
	\end{equation}
	according to \cref{prop:vertexLayer}.
	We define
	\begin{equation}
		\pi_\beta = \begin{cases}
			\pi & \text{if } \ly(\alpha) = \ly(\beta),\text{ then \cref{thm:threeHops} gives $\pi_2\in p(\alpha)$ and $\pi \in p(\beta)$ on a shortest path,}\\
			\beta & \text{otherwise } \ly(\alpha) > \ly(\beta) \text{ and we apply \cref{lem:lowerLayerPath} to get just $\pi_2 \in p(\alpha)$.}
		\end{cases}\label{eq:optimalChoicePathPiBeta}
	\end{equation}
	%We do this to prevent a case analysis for the cases of $\ly(\alpha) \geq \ly(\beta)$.
	Thus there exists a $\pi_\beta$ on a shortest $(\alpha,\beta)$-path
	according to either \cref{lem:lowerLayerPath} or \cref{thm:threeHops},
	where $d(\pi_2, \pi_\beta) > 4$.
	%\begin{equation}
	%	P_{\alpha,\beta} = \alpha,\pi_2, \ldots, \pi_\beta, \beta\,,\label{eq:graphStructureShortPath1}
	%\end{equation}
	%\todoi{not so nice}
	%where if $\pi_\beta = \beta$, the last hop can be ignored.
	We know that
	\begin{equation}
		\ly(\pi_2) \geq \ly(\pi_\beta)\,,
	\end{equation}
	because in the first case of \cref{eq:optimalChoicePathPiBeta} it holds that $\ly(\pi) \leq \ly(\beta) - 1$ (\cref{prop:vertexLayer}),
	and in the second case $\ly(\alpha) > \ly(\beta)$ so $\ly(\pi_2) \geq \ly(\beta)$.
	That means we can apply \cref{lem:lowerLayerPath} or \cref{thm:threeHops} 
	to a shortest $\pi_2,\pi_\beta$-path.
	We define
	\begin{equation}
		\gamma_\beta = \begin{cases}
			\gamma & \text{if } \ly(\pi_2) = \ly(\pi_\beta),
			\text{ then \cref{thm:threeHops} gives $\gamma_2 \in p(\pi_2)$ and $\gamma \in p(\pi_\beta)$ on a shortest path,}\\
			\pi_\beta & \text{otherwise } \ly(\pi_2) > \ly(\pi_\beta), \text{ and \cref{lem:lowerLayerPath} gives us just a $\gamma_2 \in p(\pi_2)$ on a shortest path.}
		\end{cases}
	\end{equation}
	Thus there exists a $\gamma_\beta$ on a shortest $\pi_2,\pi_\beta$-path
	according to \cref{lem:lowerLayerPath} or \cref{thm:threeHops},
	where $d(\gamma_2, \gamma_\beta) \geq 3$.
	All in all, there exists a shortest $\alpha,\beta$-path
	\begin{equation}
		P_{\alpha,\beta} = \alpha, \pi_2, \gamma_2, \eta_2 \dots, \beta\,,
	\end{equation}
	as stated in the Lemma.

	It remains to show that $\eta_2 \in p(\gamma_2)$ if $\ly(\pi_1) = \ly(\pi_2)$.
	Let the common parent of $\pi_1$ and $\pi_2$ be $\gamma_1 = \pi_{\pi_1,\pi_2}$ (\cref{prop:commonParent}),
	so that $\{\gamma_1, \gamma_3\} = p(\pi_1)$.
	Note that $\gamma_3 \ne \gamma_2$ because $\gamma_3 \not\in p(\pi_2)$.
	%and the other parent $\gamma_3 \in p(\pi_1)$ is distinct from $\gamma_1$ and $\gamma_2$.
	If $\gamma_2 = \gamma_1$ then there would exist an optimal path through $\pi_1$,
	since $\gamma_1$ is the common parent of $\pi_1$ and $\pi_2$.
	However, that would mean $d(\pi_1, \beta) = d(\alpha,\beta) - 1$, leading to a contradiction.

	Thus the interesting case is where $\gamma_2 \neq \gamma_1$.
	From the Grandparent Rule (\cref{def:grandparentRule})
	we know that $\ly(\gamma_3) \leq \ly(\gamma_2)$
	or the algorithm would have chosen $\pi_2$.
	Since $\ly(\pi_1) = \ly(\pi_2)$ and $\pi_1 \edge \pi_2$,
	their parents form a triangle $\{\gamma_x, \gamma_y, \gamma_z\}$,
	where $\ly(\gamma_x) = \ly(\gamma_y) \geq \ly(\gamma_z)$ (\cref{prop:triangleShape}).	
	If only one vertex can be on a lower layer, it must either be $\gamma_3$ or $\gamma_1$,
	because $\ly(\gamma_3) \leq \ly(\gamma_2)$.
	Thus $\ly(\gamma_1) \leq \ly(\gamma_2)$, so
	\begin{equation}
		\ly(\gamma_2) = \ly(\pi_2) -1  \geq \ly(\pi_\beta) -1 \geq \ly(\gamma_\beta)\,.\label{eq:optimalChoiceGammaLayer}
	\end{equation}
	Furthermore, $d(\gamma_2, \gamma_\beta) \geq 3$,
	so we can apply either \cref{thm:threeHops} or \cref{lem:lowerLayerPath}
	to get a shortest path which contains $\eta_2 \in p(\gamma_2)$
	\begin{equation}
		P = \alpha,\pi_2, \gamma_2, \eta_2, \ldots, \gamma_\beta, \pi_\beta, \beta\,,
	\end{equation}
	proving the Lemma.
\end{fancyproof}

Now we know that there is some shortest path from $\alpha$ to $\beta$
that passes through $\pi_2$, $\gamma_2$, and also $\eta_2$ if $\ly(\pi_1) = \ly(\pi_2)$.
Using this information we can show that the routing algorithm makes an optimal choice at $\alpha$ when routing to $\beta$.

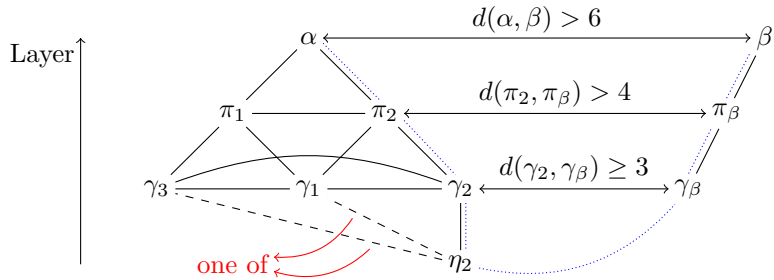
\begin{figure}
	\centering
	\begin{tikzpicture}
		\node (a) at (0,0) {$\alpha$};
		\node (b) at (6,0) {$\beta$};

		\node (pi1) at ($(a) + (-1,-1)$) {$\pi_1$}
			edge (a);
		\node (pi2) at ($(a) + (1,-1)$) {$\pi_2$}
			edge (a)
			edge (pi1);
		\node (pibeta) at ($(b) + (-0.5,-1)$) {$\pi_\beta$}
			edge (b);
		
		\node (g1) at ($(pi1)!0.5!(pi2) + (0,-1)$) {$\gamma_1$}
			edge (pi1)
			edge (pi2);
		\node (g2) at ($(pi2) + (1,-1)$) {$\gamma_2$}
			edge (pi2)
			edge (g1);
		\node (g3) at ($(pi1) + (-1,-1)$) {$\gamma_3$}
			edge (pi1)
			edge (g1)
			edge[bend left=20] (g2);
		\node (gammabeta) at ($(pibeta) + (-0.5,-1)$) {$\gamma_\beta$}
			edge (pibeta);

		\node (e2) at ($(g2) + (0,-1)$) {$\eta_2$}
			edge (g2);

		\draw[dashed] (e2) -- (g1) node[near end] (g1edge) {};
		\draw[dashed] (e2) -- (g3) node[near start] (g3edge) {};

		\node[red] (either) at ($(g3)!0.5!(g1) + (0,-1)$) {one of};
		\draw[red, ->] (g1edge) to[bend left] (either.10);
		\draw[red, ->] (g3edge) to[bend left] (either.-10);

		\draw[<->] (a) -- (b) node[midway, above] (abmid) {$d(\alpha,\beta) > 6$};
		\draw[<->] (pi2) -- (pibeta) node[midway, above] (pimid) {$d(\pi_2,\pi_\beta) > 4$};
		\draw[<->] (g2) -- (gammabeta) node[midway, above] (gammamid) {$d(\gamma_2, \gamma_\beta) \geq 3$};

		%\draw[mediumlightgray] (b) -- ++(0.5,1) -- (pibeta);
		%\draw[mediumlightgray] (pibeta) -- ++(0.5,1) -- (gammabeta);

		\draw[densely dotted, blue] (a.-20) -- (pi2.-250);
		\draw[densely dotted, blue] (pi2.-20) -- (g2.-250);
		\draw[densely dotted, blue] (g2.-70) -- (e2.-290);
		\draw[densely dotted, blue] (e2) to[bend right] (gammabeta);
		\draw[densely dotted, blue] (gammabeta.-270) -- (pibeta.-135);
		\draw[densely dotted, blue] (pibeta.-270) -- (b.-135);

		\draw[<-] ($(a) + (-3,0)$) node[anchor=north east]() {Layer} -- ++(0,-3);
	\end{tikzpicture}
	\caption{
		The graph structure when routing from $\alpha$ to $\beta$
		when $\ly(\alpha) \leq \ly(\beta)$, $\ly(\pi_1) = \ly(\pi_2)$,
		and $\ly(\pi_2) = \ly(\pi_\beta)$.
		Either $\gamma_1$ or $\gamma_3$ are connected to $\eta_2$ (dashed).
		The assumed shortest path is also given (dotted blue),
		where between $\eta_2$ and $\gamma_\beta$ there could be many vertices.
		There is either an edge $\gamma_1 \edge \eta_1$ or $\gamma_3 \edge \eta_2$,
		which shows that if the algorithm chooses $\pi_1$,
		it is still on an optimal path.
	}
	\label{fig:graphStructure}
\end{figure}

%\todoi{Make symmetric}
\begin{lemma}[Optimal Choice]\label{lem:optimalChoice}
	Consider vertices $\alpha,\beta \in V$ where $d(\alpha,\beta) > 6$ and $\ly(\alpha) \geq \ly(\beta)$.
	Then the routing algorithm (\cref{alg:spherePath}) chooses a parent $\pi_1 \in p(\alpha)$ so that $d(\pi_1, \beta) = d(\alpha,\beta)-1$.
\end{lemma}

\begin{fancyproof}
	We know that $\ly(\alpha) \leq \ly(\beta)$ and $d(\alpha,\beta) > 6$
	so it is possible to apply \cref{lem:lowerLayerPath} or \cref{thm:threeHops}
	to see that there must be a $\pi_2 \in p(\alpha)$ for which $d(\pi_2, \beta) = d(\alpha,\beta) -1$.
	The routing algorithm will also choose to go to a parent $\pi_1 \in p(\alpha)$,
	since $\beta \not\in N^\alpha$.
	If $\pi_1 = \pi_2$ then the routing algorithm made an optimal choice.
	Otherwise, $\pi_1 \ne \pi_2$ and we will show that $d(\pi_1,\beta) = d(\alpha,\beta)$
	through a proof by contradiction.

	Assume for a contradiction that $d(\pi_1, \beta) > d(\alpha, \beta) -1$.
	From \cref{lem:optimalChoicePath} we know that there exists
	a shortest $(\alpha,\beta)$-path of the form
	\begin{equation}
		P_{\alpha,\beta} = \alpha, \pi_2, \gamma_2, \eta_2, \dots, \beta\,,\label{eq:optimalChoicePath}
	\end{equation}
	where $\pi_2 \in p(\alpha)$ and $\gamma_2 \in p(\pi_2)$.
	The algorithm will always choose the parent $\pi_1$
	with $\ly(\pi_1) \leq \ly(\pi_2)$.
	We distinguish two cases regarding the relation between the layers of $\pi_1$ and $\pi_2$.
	\begin{enumerate}
		\item The first case is $\ly(\pi_1) < \ly(\pi_2)$.
			We know that $\pi_1 \in p(\pi_2)$ since parents are connected
			$\pi_1 \edge \pi_2$ (\cref{prop:parentsConnected}).
			It must be that $\gamma_2 \ne \pi_1$,
			or else \cref{eq:optimalChoicePath} would not be a shortest path.
			However, we know that parents are connected so $\pi_1 \edge \gamma_2$ exists,
			thus there must exist an $(\alpha,\gamma_2)$-path
			$P_{\alpha,\gamma_2} = \alpha, \pi_1, \gamma_2$
			that is equally long as $P'_{\alpha,\gamma} = \alpha, \pi_2, \gamma_2$.

		\item
			Since $\ly(\pi_1) \leq \ly(\pi_2)$ the only remaining case is $\ly(\pi_1) = \ly(\pi_2)$.
			We know that $\pi_1 \edge \pi_2$,
			so there must exist a common parent $\gamma_1$ (\cref{prop:commonParent}).
			If $\gamma_1 = \gamma_2$,
			then $d(\pi_1, \beta) = d(\alpha, \beta) -1$,
			because there is a $(\alpha,\gamma_2)$-path
			\begin{equation}
				P_{\alpha, \gamma_2} = \alpha,\pi_1,\gamma_2
			\end{equation}
			that is equal in length to a subsection of \cref{eq:optimalChoicePath}.
			But that would contradict the assumption $d(\pi_1, \beta) > d(\alpha,\beta) -1$,
			so we can assume that $\gamma_1 \ne \gamma_2$.
			That implies $p(\pi_2) = \{\gamma_1, \gamma_2\}$
			and $\gamma_2 \not\in p(\pi_1)$.
			Let $p(\pi_1) = \{\gamma_1, \gamma_3\}$.
			From \cref{prop:triangleShape} we know that $\{\gamma_1, \gamma_2, \gamma_3\}$ form a triangle,
			and that only one $\gamma_i$ can be on a lower layer than the other $\gamma_j$'s.
			We also know that the algorithm will apply the Grandparent Rule,
			thus $\ly(\gamma_3) \leq \ly(\gamma_2)$
			or the algorithm would have chosen $\pi_2$ instead of $\pi_1$.
			Moreover, we know that there is an $\eta_2 \in p(\gamma_2)$
			on a shortest path (\cref{lem:optimalChoicePath}).
			We can then distinguish two cases:
			\begin{enumerate}
				\item If $\ly(\gamma_1) < \ly(\gamma_2) = \ly(\gamma_3)$ or $\ly(\gamma_3) < \ly(\gamma_1) = \ly(\gamma_2)$,
					then assume w.l.o.g.\ $\gamma_1$ is on the lowest layer.
					We then know that $\gamma_1 \in p(\gamma_2)$,
					but $\eta_2 \neq \gamma_1$,
					or the $\alpha,\eta_2$-subpath of \cref{eq:optimalChoicePath}
					would not be shortest.
					However, parents are connected so $\gamma_1 \edge \eta_2$ exists,
					so there exists a path
					\begin{equation}
						P_{\alpha, \eta_2} = \alpha, \pi_1, \gamma_1, \eta_2
					\end{equation}
					that is equal in length as the $(\alpha,\eta_2)$-subpath of \cref{eq:optimalChoicePath}.
					A similar proof holds for $\gamma_3$ on the lowest layer,
					by replacing $\gamma_1$ with $\gamma_3$.
				%\item If $\ly(\gamma_3) < \ly(\gamma_2) = \ly(\gamma_1)$,
				%	then $\gamma_3 \in p(\gamma_2)$
				%	because they are connected.
				%	It must be that $\eta_2 \neq \gamma_3$ or it would not be a shortest path.
				%	But then we can construct a shortest path through $\pi_1$ to $\eta_2$ 
				%	because parents are connected
				%	\begin{equation}
				%		P_{\alpha, \eta_2} = \alpha, \pi_1, \gamma_3, \eta_2\,.
				%	\end{equation}

				\item Otherwise $\ly(\gamma_1) = \ly(\gamma_2) = \ly(\gamma_3)$ holds,
					because $\ly(\gamma_3) \leq \ly(\gamma_2)$
					and the restrictions on layers for the triangle according to \cref{prop:triangleShape}.
					This situation is also depicted in \cref{fig:graphStructure},
					when ignoring the $\beta, \pi_\beta$ and $\gamma_\beta$.
					Then $\{\gamma_1, \gamma_2, \gamma_3\}$ form a triangle on the same layer,
					so the parents also form a triangle $\{\eta_1, \eta_2, \eta_3\}$ according to \cref{cor:triangleIntoTriangle}.
					Every $\gamma_i$ must have two distinct parents,
					but there are only 3 distinct combinations of parents:
					\begin{equation}
						\{\eta_1, \eta_2\}, \{\eta_2, \eta_3\}, \{\eta_3, \eta_1\}\,.
					\end{equation}
					Thus there must be another vertex $\gamma_i \ne \gamma_2$
					so that $\gamma_i \edge \eta_2$.
					We can conclude that there exists a shortest path that passes through $\pi_1$
					\begin{equation}
						P_{\alpha, \beta} = \alpha, \pi_1, \gamma_i, \eta_2, \ldots, \beta\,,
					\end{equation}
					since for both $\gamma_i$, $\gamma_1 \edge \pi_1$ and $\gamma_3 \edge \pi_1$.
			\end{enumerate}
	\end{enumerate}
	In all cases it is possible to create an $(\alpha,\eta_2)$-path
	through $\pi_1$ that is the same length as the $(\alpha,\eta_2)$-subpath
	in \cref{eq:optimalChoicePath}.
	Thus it must be that $d(\pi_1, \beta) = d(\pi_2, \beta) = d(\alpha,\beta) -1$,
	proving that the parent that the algorithm chooses is on a shortest $(\alpha,\beta)$-path.

\end{fancyproof}

\begin{theorem}[Sphere Routing Optimal]\label{thm:sphereOptimal}
	Consider $\alpha,\beta \in V$, where $\alpha$ is the sender and $\beta$ the receiver.
	Then the routing algorithm (\cref{alg:spherePath}) chooses a vertex $\pi_1 \in N(\alpha)$ so that $d(\pi_1, \beta) = d(\alpha,\beta) - 1$,
	or a vertex $\pi_2 \in N(\beta)$ so that $d(\alpha, \pi_1) = d(\alpha, \beta) -1$.
\end{theorem}
\begin{fancyproof}
	We have shown in \cref{lem:optimalChoice} that the algorithm makes
	an optimal choice if $d(\alpha,\beta) > 6$ and $\ly(\alpha) \geq \ly(\beta)$.
	If $\ly(\beta) > \ly(\alpha)$ then the algorithm instead takes a step from $\beta$.
	We can invert the positions of $\alpha$ and $\beta$
	so that algorithm can perform the exact same step for $d(\beta,\alpha) > 6$ and $\ly(\beta) \geq \ly(\alpha)$,
	of which we have already shown that it is optimal (\cref{lem:optimalChoice}).
	Thus if $d(\alpha, \beta) > 6$,
	then the algorithm performs an optimal step no matter if $\ly(\alpha) \geq \ly(\beta)$
	or $\ly(\beta) > \ly(\alpha)$.

	We will show that the algorithm also makes an optimal choice when $d(\alpha,\beta) \leq 6$.
	In this case, we use a known routing algorithm, Dijkstra's algorithm~\cite{kleinberg2006algorithm}, to finds a path between $\alpha$ and $\beta$.
	Searching in the 6-hop neighbourhood is sufficient, because $d(\alpha,\beta) \leq 6$
	and a shortest path of length 6 or less cannot use vertices outside this neighbourhood.
	Since it is known that Dijkstra's algorithm finds a shortest path,
	the base case does so as well if $d(\alpha, \beta) \leq 6$.
	Thus the algorithm makes an optimal choice no matter if $d(\alpha,\beta) \leq 6$ or
	$d(\alpha,\beta) > 6$.
\end{fancyproof}

\paragraph{Labelling}\label{sec:appendixLabelling}
	The labelling is defined in \cref{eq:label} as $\la(\alpha)$.
	Very important is the fact that $\left|\la(\alpha)_i\right| \leq 3$, for any $i$.
	This greatly limits the size of the labelling,

	\begin{lemma}[Label Bound]\label{lem:labelBound}
		The label size is bounded as $\abs{ \la(\alpha)_\ell } \in \{1,2,3\}$,
		for any $\alpha \in V$ and $\ell \in \mathbb N$.
		Additionally, for any $\beta, \gamma \in \la(\alpha)_\ell : \beta \ne \gamma$
		it holds that $\beta \edge \gamma$ and $\ly(\beta) = \ly(\gamma)$.
	\end{lemma}
	\begin{fancyproof}
		We will prove this lemma using induction over the label index $\la(\alpha)_i$,
		using a constrained set of cases.

		\basis $i=1$.
			We know that the label starts with only one node $\alpha$, at $\la(\alpha)_1$.
			Since there are no pairs of vertices, all pairs of vertices are adjacent
			and of the same layer.

		\ih
			Assume that $\left|\la(\alpha)_{\ell-1}\right| \in \{1,2,3\}$,
			and that all vertices in $\la(\alpha)_{\ell-1}$ are pairwise adjacent
			and of the same layer.
			The label is thus in one of the following configurations:
			\begin{enumerate}
				\item \label{item:boundedNonSimpleIH1}
					One vertex.
				\item \label{item:boundedNonSimpleIH2}
					Two adjacent vertices of the same layer.
				\item \label{item:boundedNonSimpleIH3}
					Three pairwise adjacent vertices of the same layer.
			\end{enumerate}
			We will show that the induction step will only result in one of these cases.

		\is $i=\ell$.
			According to the Induction Hypothesis (IH) $\la(\alpha)_{\ell-1}$ is in one of three configurations,
			we will handle each case individually.
			\begin{enumerate}
				\item \label{item:boundedNonSimple1}
					The number of vertices is 1 for $\la(\alpha)_{\ell-1}$.
					Let $\{\beta_1, \beta_2\} = p(\alpha)$.
					This can result in:
					\begin{enumerate}
						\item \label{item:oneNonSimpleOne}
							The first case is $\ly(\beta_1) < \ly(\beta_2)$.
							Only $\beta_1$ is added to $\la(\alpha)_\ell$
							in accordance with the Parent Rule (\cref{def:parentRule}).
						\item \label{item:oneNonSimpleTwo}
							Otherwise $\ly(\beta_1) = \ly(\beta_2)$ holds.
							Then the label element $\la(\alpha)_\ell = \{\beta_1, \beta_2\}$.
							Furthermore, $\beta_1 \edge \beta_2$, because parents are adjacent (\cref{prop:parentsConnected}).							
					\end{enumerate}

				\item \label{item:boundedNonSimple2}
					There are 2 vertices in the label.
					Let the two vertices be called $\alpha_1,\alpha_2 \in \la(\alpha)_{\ell-1}$,
					According to \cref{prop:triangleShape} the parents of $\alpha_1$ and $\alpha_2$ form a triangle
					$\{\beta_1, \beta_2, \beta_3\}$,
					where $\ly(\beta_1) = \ly(\beta_2) \geq \ly(\beta_3)$.
					For each vertex in $\la(\alpha)_\ell$ must still hold that
					they are on the same layer,
					or they would be unfavored by the Parent Rule (\cref{def:parentRule}).
					Furthermore, the Grandparent Rule (\cref{def:grandparentRule}) could remove a parent 
					if the grandparents are on a higher layer.
					Thus this can result in the following cases for $\la(\alpha)_\ell$,
					depending on the relative layering of the $\beta_i$ and the Grandparent Rule:
					\begin{enumerate}
						\item \label{item:twoNonSimpleOne}
							One vertex.
						\item \label{item:twoNonSimpleTwo}
							Two adjacent vertices of the same layer.
						\item \label{item:twoNonSimpleThree}
							Three vertices that are all pairwise adjacent and of the same layer.
					\end{enumerate}

				\item \label{item:boundedNonSimple3}
					There are 3 vertices in the label.
					Let $\{\alpha_1,\alpha_2,\alpha_3\} = \la(\alpha)_{\ell-1}$,
					where all vertices are pairwise adjacent and of the same layer.
					Because these vertices are all pairwise adjacent, they form a triangle.
					A triangle $\{\alpha_1,\alpha_2,\alpha_3\}$ with
					$\ly(\alpha_1) = \ly(\alpha_2) = \ly(\alpha_3)$
					is only formed by three parents in a triangle $\{\beta_1,\beta_2,\beta_3\}$,
					according to \cref{cor:triangleIntoTriangle}.
					Again, only vertices of the lowest layer are added
					because of the Parent Rule,
					so that all vertices in $\la(\alpha)_\ell$ are of the same layer.
					Furthermore, the Grandparent Rule can remove certain parents.
					This can result in the following cases for $\la(\alpha)_\ell$:
					\begin{enumerate}
						\item One vertex.
						\item Two adjacent vertices of the same layer.
						\item Three vertices that are pairwise adjacent and of the same layer.
					\end{enumerate}
			\end{enumerate}
			So for all given cases the lemma holds, and each case results in one given in the IH.
	\end{fancyproof}

\paragraph{ Analysis of the Local Routing algorithm}\label{sec:complexity}
	As is common, we assume that the ID of a node uses constant space, $O(1)$,
	in accordance with the uniform cost model.
	The ID is saved as a unique binary number in the graph.
	We will first analyse the size of the data stored in every vertex.
	Using that information we are able to analyse the running time of routing per vertex. Let $N$ be the number of nodes in the network and let the number of layers be defined as~\eqref{eq:nRelatedToK}
	\begin{equation}
		k = \frac{1}{2}\log_2\left( \frac{N-2}{10} \right) \,.
	\end{equation}

	\begin{theorem}[Memory size]\label{thm:spaceComplexityRouting}
		The classical memory size per node for the local routing algorithm (\cref{alg:localRoute}) is $O(\log^6 N)$.
	\end{theorem}

	\begin{figure}
		\centering
		\begin{tikzpicture}
			% Center node
			\node (bi) at (0,0)	{$\beta_i$};

			% Parents
			\node (ax) at (1,2) {$\alpha_x$}
				edge (bi);
			\node (ay) at (-1,2) {$\alpha_y$}
				edge (bi)
				edge [mediumlightgray] (ax);

			% Neighbours
			\node (bi1) at (-2,0.5)		{$\beta_{i-2}$}
				edge (bi)
				edge [mediumlightgray] (ay);
			\node (bi2) at (-2,-0.5)	{$\beta_{i-1}$}
				edge (bi)
				edge [mediumlightgray] (bi1)
				edge [mediumlightgray, bend left=20] (ax);
			\node (bi3) at (2,0.5)		{$\beta_{i+1}$}
				edge (bi)
				edge [mediumlightgray] (ax);
			\node (bi4) at (2,-0.5)		{$\beta_{i+2}$}
				edge (bi)
				edge [mediumlightgray] (bi3)
				edge [mediumlightgray, bend right=20] (ay);

			% Children
			\node (g1) at (-1,-2) {$\gamma_1$}
				edge (bi);
			\node (g2) at (-0.25, -2) {$\gamma_2$}
				edge (bi);
			\node (gj) at (1,-2) {$\gamma_j$}
				edge (bi);
			\draw [dotted] (0, -1.5) -- (0.65,-1.5);

			% Text
			\node (l1) at (-5, 2)	{Parents, $L_{< \ell}$};
			\node (l2) at (-5, 0)	{Siblings, $L_\ell$};
			\node (l3) at (-5, -2)	{Children, $L_{> \ell}$};
			\draw [decorate,decoration={brace,amplitude=10pt}] (-3.5,-1) -- (-3.5,1);
		\end{tikzpicture}
		\caption{The neighbourhood of a node, which is directly related to the size complexity.
			Some edges of the neighbours are drawn using a light gray line.
			The node $\beta_i$ stores the IDs of $2$ parents, $4$ siblings,
			and $O(k)$ children, where $k$ is the number of layers.
		}
		\label{fig:spaceComplexityNode}
	\end{figure}
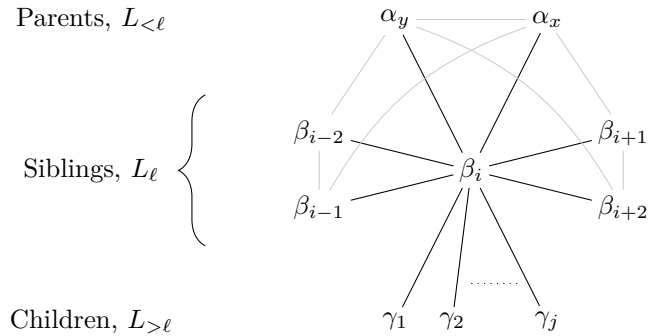%

	\begin{fancyproof}
		We first analyze the contribution to the memory size of the labelling,
		followed by that of $N^\alpha$.
		We know that $\left|\la(\alpha)_i\right| \leq 3$ for every
		$i\in\{1, \ldots, \abs{\la(\alpha)} \}$ (\cref{lem:labelBound}) resulting in
		\begin{equation}
			\left| \la(\alpha) \right| \leq 3\cdot \ly(\alpha) = O(k)\,.\label{eq:labelSize}
		\end{equation}
		A vertex must store its own label, and receive a label of the target node $\beta$,
		which also has a space complexity of $O(k)$.

		We assume that the data $N^\alpha$ for Dijkstra's algorithm is stored in the network node.
		Alternatively, it may request it over the network; trading off space with time.
		The neighbourhood of a node is illustrated in \cref{fig:spaceComplexityNode}.
		The number of children of a node is (\ref{sec:appShortestPathStructure})
		\begin{equation}
			\sum_{i=\ly(\alpha) + 1}^{k} 6 = 6(k - \ly(\alpha)) = O(k)\,.
		\end{equation}
		The number of vertices in the neighbourhood of a node is thus
		\begin{equation}
			\left| N(\alpha) \right| = 2 + 4 + 6(k-\ly(\alpha)) = O(k)\,.
		\end{equation}
		We take the 6th order neighbourhood, resulting in a space complexity of
		\begin{equation}
			|N^\alpha| = O(k^6)\,.\label{eq:spaceComplexityNeighbourhood}
		\end{equation}
		Thus, in total, the space complexity per vertex is
		\begin{equation*}
			O(k^2 + k^6) = O(k^6) = O(\log^6 N)\,.\qedhere
		\end{equation*}
	\end{fancyproof}

	\noindent Now that we know the memory size,
	we can use that information in the running time analysis.
	Given two sets $\mathbb A$ and $\mathbb B$,
	where the size is $|\mathbb A| = O(x)$ and $|\mathbb B| = O(y)$.
	Then it holds for the time complexity of the $\cap$ operation that
	\begin{equation}
		T(\mathbb A \cap \mathbb B) = O(\min(x,y))\,.\label{eq:setIntersection}
	\end{equation}
	We can achieve this by taking each element from the smallest set,
	and checking presence in the other set.
	We assume each element of the sets has size $O(1)$ in accordance with the uniform cost model.
	Then hashing it will also take $O(1)$.
	Finding the element in a hash set is amortized $O(1)$ with an appropriately sized hashing table.
	Let the smallest set be of size $O(n) = O(\min(x,y))$.
	To find each element of the smallest set we iterate through all elements of the smallest set in $O(n)$
	and check its presence in the larger set in $O(1)$.
	As such this operation can be completed in $O(n)$ time.

	\begin{theorem}[Local Running Time] \label{thm:timeComplexityRouting}
		The running time per node of the local routing algorithm (\cref{alg:localRoute}) is $O(\log N)$.
	\end{theorem}
	\begin{fancyproof}
		First we investigate the time complexity of the base case where Dijkstra's algorithm is used,
		going from top to bottom.
		First is a check $\alpha = \beta$ which is $O(1)$.
		Then comes a check if $\alpha \in \la(\beta)_i$ for some $i$.
		There are $\abs{\la(\beta)}$ entries to check, which is upper bounded by $O(k)$.

		It is assumed that $N^\alpha$ is already stored in the network node.
		The time complexity of the intersection $N^\alpha \cap L_\beta$
		is $O(\min(k^6, k)= O(k)$ (\cref{eq:spaceComplexityNeighbourhood,eq:labelSize,eq:setIntersection}).
		Once an intersection has been found we know the set of possible destinations
		\begin{equation}
			\Gamma = N^\alpha \cap L_\beta\,.
		\end{equation}
		We know that $\Gamma \subseteq L_\beta$, so $|\Gamma| = O(k)$.
		For every $\gamma \in N^\alpha$ we assume the implementation has stored a shortest $(\alpha,\gamma)$-path
		as given by Dijkstra's algorithm.
		Since the path is at most of size 6, this has a size of $O(1)$.
		Thus we can calculate $d(\alpha,\gamma)$ in $O(1)$.
		Mapping $\forall\gamma \in \Gamma \to d(\alpha,\gamma)$ takes $O(k)$ time complexity.
		Finding the minimum in this data structure can be done through a linear traversal in $O(k)$,
		because the data is of size $O(k)$.
		Thus it is possible to calculate
		\begin{equation}
			\gamma^{\text{min}} = \argmin_{\gamma \in N_\alpha \cap L_\beta} \{ |P_{\alpha,\gamma}| : P_{\alpha,\gamma} = \text{dijkstra}(\alpha,\gamma)\}
		\end{equation}
		in $O(k)$.
		As seen before, the path from $\alpha$ to $\gamma^\text{min}$ is readily available
		for a given $\gamma$ in $O(1)$ by storing the $(\alpha,\gamma^\text{min})$-path given by Dijkstra's algorithm.
		Thus the total running time of the base case is $O(k)$.
		
		Finding an element from $\la(\alpha)_2$ is only $O(1)$ because $\la(\alpha)_2 \leq 2$.
		The neighbourhood search that computes
		\begin{equation}
			\la(\beta)_{i-1} \cap N(\alpha)
		\end{equation}
		is $O(1)$ because $\abs{\la(\beta)_{i-1}} \leq 3$.
		Thus the time complexity of the base case overshadows that of inductive step,
		resulting in a total time complexity of
		\begin{equation}
			O(k) = O(\log N)
		\end{equation}
		proving the theorem.
	\end{fancyproof}

	\noindent Since the diameter of the graph $D(G) = 2k +3 = O(k)$ (\cref{prop:diameter}) and every vertex takes $O(k)$,
	the routing of any packet will take at most $O(k \cdot k) = O(\log^2 N)$ total time.

\subsection{Technical Details Replenishing Entanglement}
In this section we prove some properties on the number of time steps required to establish
entanglement according to our proposed network structures.
We have made some assumptions in \cref{sec:replenishing} about the operations and their the time required.
Our first assumption is that there are two operations:
creation steps, which create entanglement between network nodes that have a physical quantum link;
and entanglement swaps, which establish entanglement across larger distances.
We furthermore assume that each of these operations takes exactly one time step.
Second, each edge may only be operated upon by one operation at each time step.
The reasoning is that it should not be possible to start entanglement swapping with entanglement that must still be established.
Third, we restrict each node to one entanglement swapping per time step.
It is possible that operations can only be performed on a subset of the qubits that are stored in the quantum memory.
Here we assume that this subset is of size $2$.
%Note that if we have entanglement on the edges $\{a,b\},\{b,c\} \in E$
%and we perform an entanglement swap at $b$,
%then $a$ and $c$ are still free to perform their own entanglement swaps in this time step.
%We can justify this by looking at the resources affected by the swapping at $b$.
%At $a$ and $c$ we can regard the entanglement with $b$ as occupied.
%However, when $a$ or $c$ performs an entanglement swap to create an edge on a lower layer,
%they will never use the entanglement with $b$,
%since $b$ is on a higher layer than $a$ and $c$.

Let us first introduce notation that unifies the ring and sphere graph
since we are proving that the same statements hold for both.
We will assume the sphere notation of $V_k$ and $E_k$ to indicate vertices and edges generated in iteration $k$,
where $k=0$ is the basis.
When referring to the ring notation we will do so explicitly in the superscript by using `$\circ$', e.g. $E_k^\circ$.
We define the subdivided ring similarly to the sphere as $G_k = (V_k, \cup_{i \in \{1,\dots,k\}} E_i)$.
The set of vertices is almost the same, $V_k = V_{k-1}^\circ$ for all $k>0$;
similarly to the sphere, the set of vertices in layer $k$ is defined as
\begin{equation}
	L_k = \begin{cases}
		V_1^\circ & \text{if } k = 0,\\
		V_{k+1}^\circ \setminus V_k^\circ & \text{otherwise.}\\
	\end{cases}
\end{equation}
However, $E_k^\circ$ also includes edges of previous iterations, so we define
\begin{equation}
	E_k = \begin{cases}
		E_1^\circ & \text{if } k = 0,\\
		E_{k+1}^\circ \setminus E_k^\circ & \text{otherwise,}\\
	\end{cases}
\end{equation}
which completes our definition of $G_k$ for the ring.
This also allows us to use familiar functions using the same definitions for the sphere as for the ring graph
such as the parent function $p(\alpha)$ given in \cref{eq:directParent}.
Additionally, let $T: E \to \mathbb N$ be the function that gives the number of time steps required to create entanglement along $E$.
We first look at the time required to establish entanglement from scratch.
\begin{theorem}[Entangling Time]\label{thm:entanglingTime}
	Consider the subdivided graph $G_k = (V_k, \cup_{i \in \{1,\dots,k\}} E_i)$
	with no entanglement distributed.
	Then the number of time steps $T(E)$ required to establish entanglement along $E$ is
	\begin{equation*}
		T(E) = 2k + 1 = O(\log N)\,.
	\end{equation*}
\end{theorem}
\begin{fancyproof}
	Assume that the network is given by $G_k$, so $k$ subdivisions were performed.
	We will show with a direct proof that given entanglement along all edges above a given layer $m \in \{1,\dots, k\}$,
	\begin{equation}
		F_{m+1} = \bigcup_{i \in \{m+1, \dots, k\}} E_i\,,
	\end{equation}
	it is possible to construct entanglement along all edges on layer $m$ and above given by
	\begin{equation}
		F_{m} = \bigcup_{i \in \{m, \dots, k\}} E_i
	\end{equation}
	in two time steps.
	Using this statement we can prove the theorem for $T(E)$.

	For a given $\ell$ and given entanglement distributed along $F_{\ell+1}$
	we give an algorithm to distribute entanglement along $F_\ell$ in two time steps.
		Informally, we will perform an entanglement swap at all nodes in $L_{\ell+1}$
	%\[
	%		L_{\ell+1} = V_{\ell+1} \setminus V_{\ell+2}\,,
	%\]
	(\cref{eq:layerVertices}) to create entanglement along the edges $E_{\ell}$.
	%\todoi{This is sphere-specific, we need to work out some common notation.}
	At the same time we will recreate entanglement along edges $E_{\ell+1}$ (that were just used for $E_\ell$).
	We do this by performing entanglement swaps at most nodes in $L_{i}$ to create entanglement along
	edges $E_{i-1}$ for all $\ell < i < k$,
	thereby moving entanglement from $E_k \to E_{k-1} \to \dots \to E_{\ell+1}$.
	Now a large part of the entanglement at $E_k$ has been used to create entanglement on lower layers,
	but this is remedied by performing an entanglement creation operation in a second time step
	for most of $E_k$ simultaneously.

	Formally, the first step is that all vertices in
	\begin{equation}
		\mathbb L = L_{\ell+1} \cup
			\left\{
				\alpha \in \bigcup_{i \in \{\ell+2, \dots, k\}} L_i
				: \ly(\pi_1) \ne \ly(\pi_2), \text{where } \{\pi_1,\pi_2\} = p(\alpha)
			\right\}
	\end{equation}
	perform an entanglement swap to create entanglement between
	the parents of $\alpha \in \mathbb L$,
	i.e.\ $\{\pi_1,\pi_2\} \in E_{\ly(\alpha)-1} = p(\alpha)$,
	using entanglement along $\{\alpha,\pi_1\}, \{\alpha,\pi_2\} \in E_{\ly(\alpha)}$.
	The only exception is the ring
	where following this procedure would lead to two entangled pairs being created for $E_1^\circ$.
	This can be remedied at no cost by only performing an entanglement swapping at one side of the ring.

	It is possible to perform an entanglement swap at all vertices $\alpha \in \mathbb L$ because there is entanglement on the edge $\{\alpha,\beta\} \in E$
	for some $\beta \in V$,
	since $\{\alpha,\beta\} \in F_{\ell+1}$ as $\ly(\alpha) > \ell$.
	Furthermore, note that each node is only performing one entanglement swap in this time step,
	thus this does not violate our restriction that a node can perform only one entanglement swap in a time step.
	To see why this is the case, assume that there are three nodes $a,b,c$ where there is entanglement along $a \edge b \edge c$
	and $b$ performs an entanglement swap so that $a$ becomes entangled with $c$.
	Now nodes $a$ and $c$ have not used their qubits that are entangled with $b$ in this time step,
	since $b$ is a child of $a$ and $c$, and nodes do not create entanglement for higher layers.
	The result is entanglement distributed along
	\begin{equation}
		F'_\ell = F_\ell \setminus \{ \{\gamma_1,\gamma_2\} \in E_k : \ly(\gamma_2) < k \}\,,
	\end{equation}
	except if $\ell = k-1$ where $F'_\ell = F_\ell \setminus E_k$.
	This results from the fact that each vertex in $L_{\ell+1}$ is a child of an edge in $E_\ell$,
	and that the remaining vertices in $\mathbb L$ recreate the entanglement along $L_{\ell+1}$,
	where only entanglement along edges $\{\alpha,\beta\} \in E_{\ell +1}$ with
	$\ly(\alpha) \ne \ly(\beta)$ was used.

	In the second time step we then establish entanglement along $F_\ell \setminus F'_\ell \subseteq E_k$
	using concurrent creation operations, thus establishing entanglement along all edges in $F_\ell$.

	With this algorithm we can prove the time required to create entanglement along all edges in $G_k$.
	To create entanglement along $E$ (or equivalently, $F_0$), we need entanglement along $F_1$ plus two time steps.
	Then to create entanglement along $F_1$ we need entanglement along $F_2$ plus two time steps.
	Continuing this recursively we get
	\begin{equation}
		T(E) = T(F_0) = T(F_1) + 2 = \dots = T(F_k) + 2k = 2k + 1 = O(\log N)\,,
	\end{equation}
	proving the Theorem.
\end{fancyproof}

\noindent We also give an upper bound on the establishing of entanglement after some operation or decoherence.
This is useful to upper bound the time steps necessary after performing a routing operation.
\begin{theorem}[Edge Entangling]
	Consider a ring or sphere graph $G_k = (V, E)$ where entanglement has been distributed along
	all edges $E$.
	If at any point in time the entanglement along the edges $S \subseteq E$ has been consumed, 
	then it takes at most $2 \abs{S}$ time steps to establish once again entanglement along all edges $E$.
\end{theorem}
\begin{fancyproof}
	We will first look at the case where $\abs{S} = 1$ and expand that to any number of edges.
	The process to recreate entanglement if $\{e\} = S$ can be summarized as follows:
	Unless $e \in E_k$, let $\{\beta_1,\beta_2\} = e$, 
	\[
		\alpha \in V : \{\beta_1,\beta_2\} \in p(\alpha)\,,
	\]
	and perform an entanglement swap on $\alpha$ to create entanglement along $e$,
	and recursively also on the edges that $\alpha$ uses to perform this swap until the base layer is reached
	(see also the proof of \cref{thm:entanglingTime}).
	%Let $\text{child}(e): E \to E$, where $e=\{\beta_1,\beta_2\}$, be a non-unique mapping defined as
	%\begin{equation}
	%	\text{child}(e) =
	%	\{ \{\beta_1, \alpha\} , \{\beta_2,\alpha\} : \alpha \in V,  e = p(\alpha)\}.
	%\end{equation}
	%Then we recursively perform an entanglement swapping along all edges in the set
	%\begin{equation}
	%	\text{swap}(S) = \begin{cases}
	%		\emptyset & \text{if } S = \emptyset\,,\\
	%		S \cup \text{swap}(\{\text{child}(e) : e \in S, e \not\in E_k \}) & \text{otherwise.}
	%	\end{cases}
	%\end{equation}
	The second step is performing a creation step on all affected edges in $E_k$
	to recreate the entanglement on the base layer.
	Using this procedure it is possible to restore entanglement for $\abs{S}=1$ in two time steps.

	For any set of edges $S$, some edge $e_1 \in S$ may, however,
	require an entanglement swapping using another edge $e_2 \in S$.
	We define an ordering based on a layering of edges
	\begin{gather}
		\ly \colon E \to \mathbb N\,,\\
		\ly(e) = \max( \ly(\alpha_1), \ly(\alpha_2) ) : \{\alpha_1, \alpha_2\} = e\,.
	\end{gather}
	Using this ordering we know that $\ly(e_1) < \ly(e_2)$,
	or otherwise $e_1$ would not require the entanglement swapping along $e_2$,
	since $e_2$ would only involve nodes on the same or lower layer.
	And we know nodes only perform a swap to create entanglement for strictly lower layers.
	We modify the procedure to first recreate entanglement for the edges on the highest layer
	\begin{equation}
		S^{\max} = \argmax \{ \ly(e) : e \in S \}\,.
	\end{equation}
	Now for any $e'_1 \in S^{\max}$ there cannot be an $e'_2 \in S$ that it depends on,
	since $\ly(e'_1) \geq \ly(e'_2)$.
	We can then concurrently restore entanglement along all edges in $S^{\max}$ in two time steps,
	according to the procedure for restoring a single edge.
	Now we iterate this procedure for $S \setminus S^{\max}$,
	until all edges have been restored.
	Since $\abs{S^{\max}} \geq 1$ the size of $S$ strictly decreases on every iteration by at least one,
	so the procedure takes at most $2\abs{S}$ time steps as specified in the Theorem.
\end{fancyproof}
Note that by combining the above two theorems, we have that replenishing the entanglement in the entire graph
can take time at most 
\begin{equation}
	\min(2|S|, 2k+1) = O(\log N)\,.
\end{equation}

\end{document}